\documentclass[12pt]{article}
\usepackage{amsfonts,amsmath,upgreek,bm}
\usepackage[paper=letterpaper,margin=.85in]{geometry}
\usepackage{graphicx}

\usepackage[dvipsnames]{xcolor}
\definecolor{darkblue}{rgb}{0.1,0.1,.7}
\usepackage[colorlinks, linkcolor=darkblue, citecolor=darkblue, urlcolor=darkblue, linktocpage]{hyperref} 
\usepackage[square, comma, compress,numbers]{natbib}
\usepackage{physics}
\usepackage{footmisc} % Preserve the existing footnote layout.
\usepackage{tikz}
\usepackage{caption} % Preserve ordinary caption formatting without subcaption.

\usepackage{lmodern}
\usetikzlibrary{arrows.meta}

\definecolor{ink}{HTML}{233344}
\definecolor{blue}{HTML}{246B93}
\definecolor{pale}{HTML}{EDF4F8}
\definecolor{muted}{HTML}{637181}
\definecolor{line}{HTML}{CDD6DE}

\numberwithin{equation}{section}
\newcommand{\be}{\begin{equation}}
\newcommand{\ee}{\end{equation}}

\newcommand*{\boxcolor}{black}
\makeatletter
\renewcommand{\boxed}[1]{\textcolor{\boxcolor}{%
\tikz[baseline={([yshift=-1ex]current bounding box.center)}] \node [rectangle, minimum width=1ex,rounded corners,draw] {\normalcolor\m@th$\;\,\displaystyle#1\;\,$};}}
 \makeatother

\newcommand*{\boxfill}{gray!10} % light gray background
\makeatletter
\newcommand{\boxedGrey}[1]{%
  \tikz[baseline={([yshift=-1ex]current bounding box.center)}]
    \node[
      rectangle,
      minimum width=1ex,
      rounded corners,
      draw=\boxcolor,
      fill=\boxfill
    ]{\m@th$\;\,\displaystyle#1\;\,$};%
}
\makeatother

\newcommand{\btau}{\boldsymbol\tau}
\newcommand{\bnu}{\boldsymbol\nu}
\newcommand{\bomega}{\boldsymbol\omega}

\def\w{\omega}
\def\sl{\text{SL}(2,\mathbb{Z})}
\def\erf{\text{Erf}}
\def\mpos{>0}
\def\sGen{{\mathfrak{s}}}
\def\sGenb{{\bm{\mathfrak s}}}
\def\Re{\text{Re}}
\def\Im{\text{Im}}

\def\la{\langle}
\def\ra{\rangle}
\newcommand{\es}[2] {\begin{equation} \label{#1} \begin{split} #2 \end{split} \end{equation}}
\newcommand {\beq} {\begin {equation}}
\newcommand {\eeq} {\end {equation}}

\DeclareMathOperator*{\sumint}{%
\mathchoice%
  {\ooalign{$\displaystyle\sum$\cr\hidewidth$\displaystyle\int$\hidewidth\cr}}
  {\ooalign{\raisebox{.14\height}{\scalebox{.7}{$\textstyle\sum$}}\cr\hidewidth$\textstyle\int$\hidewidth\cr}}
  {\ooalign{\raisebox{.2\height}{\scalebox{.6}{$\scriptstyle\sum$}}\cr$\scriptstyle\int$\cr}}
  {\ooalign{\raisebox{.2\height}{\scalebox{.6}{$\scriptstyle\sum$}}\cr$\scriptstyle\int$\cr}}
}

\begin{document}

\thispagestyle{empty}
\vspace*{0.2cm}
\begin{center}
{\bf{\LARGE RMT${}_2$: \\ Random Spectral Correlations for CFT${}_2$ \\{\vskip .05 in}and AdS${}_3$ Gravity}\\
}
\bigskip\rm\bigskip
 Jan Boruch,${}^a$ Gabriele Di Ubaldo,${}^{b,c}$ Felix M.\ Haehl,${}^d$ Eric Perlmutter,${}^{e,f}$ Moshe Rozali${}^g$\\
\bigskip\rm\bigskip
{\small
${}^a$ Leinweber Institute for Theoretical Physics at Stanford,
382 Via Pueblo, Stanford, CA 94305, U.S.A.
\\\vspace{2mm}
${}^b$ Leinweber Institute for Theoretical Physics and Department of Physics,\\ University of California, Berkeley, CA 94720, U.S.A.
\\\vspace{2mm}
${}^c$ RIKEN iTHEMS Center for Interdisciplinary Theoretical and Mathematical Sciences,\\ 2-1 Hirosawa, Wako, Saitama 351-0198 Japan
\\\vspace{2mm}
${}^d$ School of Mathematical Sciences \& STAG Research Centre, University of Southampton,\\ Highfield Campus, SO17 1BJ, Southampton, U.K.
\\\vspace{2mm}
${}^e$ Universit\'{e} Paris-Saclay, CNRS, CEA, Institut de Physique Th\'{e}orique, 91191, Gif-sur-Yvette, France
\\\vspace{2mm}
${}^f$ Institut des Hautes Études Scientifiques, 
35 Route de Chartres, 91440, Bures-sur-Yvette, France
\\\vspace{2mm}
${}^g$ Department of Physics and Astronomy, University of British Columbia,\\ Vancouver, V6T 1Z1, Canada\\
}\normalsize
\bigskip\rm\bigskip
{\tt jboruch@stanford.edu, gdiubaldo@berkeley.edu, f.m.haehl@soton.ac.uk, perl@ipht.fr, rozali@phas.ubc.ca}

\vspace{0.75cm}
{\bf Abstract}

\end{center}
\noindent We present the full framework of RMT${}_2$, a consistent unification of random matrix theory and modular invariance. The systematic prescription, which augments the existing construction by incorporating spin, lifts random matrix correlators to $\text{SL}(2,\mathbb{Z})$ Poincar\'e sums of their Mellin transforms, in a manner which enforces random matrix correlations near extremality in every spinning primary sector. The resulting RMT${}_2$ correlators are further expressible in terms of an orthogonal basis of Eisenstein series and Maass cusp forms. We demonstrate the construction by lifting the two-point correlators at arbitrary genus of the Airy model, in both unitary and orthogonal symmetry classes, exhibiting the modular-invariant topological expansion of RMT${}_2$; a resummation of the genus expansion shows that the modular-invariant spectral form factor preserves the ramp-plateau structure. Holographically, RMT${}_2$ provides a concrete benchmark and explicit candidate expressions for comparison with off-shell wormhole amplitudes in AdS${}_3$ gravity, with any number of torus boundaries. We explicitly compute the universal RMT${}_2$ three-point correlator, corresponding to the putative three-boundary wormhole.

\setcounter{page}{0}
\setcounter{tocdepth}{2}
\setcounter{footnote}{0}
\newpage
%%%%%%%%%%%%%%%%%%%%%%%%%%%%%%%%
%%%%%%%%%%%%%%%%%%%%%%%%%%%%%%%%

\tableofcontents

%%%%%%%%%%%%%%%%%%%%%%%%%%%%%%%%%%%%%%%%%%%%%%%%
%%%%%%%%%%%%%%%%%%%%%%%%%%%%%%%%%%%%%%%%%%%%%%%%
\section{Introduction}
\label{sec:intro}
%%%%%%%%%%%%%%%%%%%%%%%%%%%%%%%%%%%%%%%%%%%%%%%%
%%%%%%%%%%%%%%%%%%%%%%%%%%%%%%%%%%%%%%%%%%%%%%%%

Random matrix theory (RMT) has proven to be a powerful tool across multiple branches of physics \cite{Bohigas:1983er,Haake:2010fgh}. There are two particular perspectives on RMT that we will be concerned with in this paper. The first is that of random matrix universality, which may be broadly summarized as the statement that in semiclassical regimes, the Hamiltonian of a chaotic quantum mechanical system can be accurately modeled by a random matrix. Another perspective arises in the context of AdS/CFT and holographic duality, where double-scaled matrix integrals are dual to two-dimensional theories of quantum gravity \cite{Cotler:2016fpe,Saad:2018bqo,Saad:2019lba}.

An important avenue of current research is to extend these insights to cases where the chaotic system is described by a local quantum field theory instead of a discrete quantum mechanical system. One widely relevant setting for this investigation is that of two-dimensional conformal field theory (CFT), where locality and modular invariance impose non-trivial global constraints on the spectrum, relating low- and high-energy spectra as well as spectra across different spin sectors. The natural expectation, supported from myriad points of view, is that upon properly accounting for symmetries, high-energy spectral data of CFTs exhibit the hallmark features of random matrix universality.

The most basic question is how it is possible to satisfy the global spectral constraints due to both modular invariance {\it and} random matrix universality simultaneously. An answer was provided by the RMT$_2$ framework for spectral statistics of CFT${}_2$, originally proposed in \cite{Haehl:2023tkr,DiUbaldo:2023qli,Haehl:2023xys,Haehl:2023mhf} and developed further in \cite{Boruch:2025ilr,Perlmutter:2025ngj,Boruch:2026hbr}. In analogy with the two perspectives on RMT described above, the goal of RMT$_2$ is two-fold. Its primary goal is to understand the universal spectral correlations compatible with the symmetries of a CFT${}_2$ and random matrix statistics in the requisite spectral regime; as we explain further below, this is the `near-extremal' part of the spectrum, where the lightest states of a given spin are known to exhibit universal behavior \cite{Ghosh:2019rcj,Pal:2023cgk}. A secondary goal is to investigate to what extent AdS$_3$ gravitational observables are simply determined by random matrix statistics combined with modular invariance -- that is to say, to what extent RMT${}_2$ itself describes universal gravitational behavior. See \cite{Dyer:2016pou,Cotler:2020ugk, Belin:2020hea, Maxfield:2020ale, Chandra:2022bqq, Belin:2023efa,Collier:2023fwi,Collier:2024mgv,Jafferis:2025vyp,deBoer:2025oge, Hartman:2025ula,Jafferis:2026gzn,  Belin:2026pko, Ambrosini:2026int, Perlmutter:2026ujr} for a selection of pertinent recent works on universality and/or randomness in AdS$_3$/CFT$_2$.

One motivation for the gravitational inquiry comes from the discovery of \cite{DiUbaldo:2023qli} (see also \cite{Haehl:2023mhf}), where it was shown that the two-boundary AdS${}_3$ torus wormhole of Cotler and Jensen \cite{Cotler:2020ugk}  follows exactly from a minimal implementation of random matrix statistics and modular invariance. This points to the simplicity of the torus wormhole and raises the question of whether more complicated wormhole amplitudes, with more boundaries or higher genus, contain information about the CFT spectrum that transcends symmetry and universality. In this context, the goal of RMT$_2$ is to provide the simplest candidate wormhole amplitudes, consistent with symmetries and RMT universality, against which any gravitational amplitude can be compared: that is, RMT${}_2$ should be viewed as the benchmark for any modular-invariant form factor that may or may not have additional, intrinsic structure beyond the minimal requirements. 

The mathematical backbone of the RMT$_2$ framework is harmonic analysis on the fundamental domain ${\cal F}=\mathbb{H}/\text{SL}(2,\mathbb{Z})$ \cite{rankin0, selberg0,terras2013harmonic, Iwaniec2021AnalyticNT}, first applied systematically in the context of CFT$_2$ partition functions in \cite{Benjamin:2021ygh} (see also \cite{Keller:2014xba}). The fluctuating part of the primary state-counting partition function is a square-integrable modular-invariant function, and therefore admits an expansion in the harmonic basis of real-analytic Eisenstein series and Maass cusp forms. Rather than writing the partition function as a standard state sum, in which the modularity constraint on energy levels is not manifest, expanding it in the harmonic basis amounts to solving the constraints of modular invariance once and for all. This allows one to directly impose random matrix statistics as a condition on the spectral overlaps (after which one can extract the energy eigenspectrum). This is how the torus wormhole was rederived in \cite{DiUbaldo:2023qli,Haehl:2023mhf}, and how the scalar component of RMT$_2$ form factors was constructed in \cite{Boruch:2025ilr}.

A challenge that previous studies of RMT${}_2$ did not address in generality is that of the spinning sectors of the partition function, where `spin' here refers to the quantum number $j$ labeling two-dimensional CFT local primary operators.  Let us explain the origin of the challenge. Maass cusp forms do not have a scalar component: this implies that the near-extremal statistics of the scalar spectrum can be imposed through Eisenstein overlaps alone.
The latter are computable because the Fourier coefficients of Eisenstein series are known analytically in closed form. For spinning sectors the situation is qualitatively different: near the cusp, no single cusp form dominates, and therefore all cusp forms contribute to the near-extremal behavior of every spin sector. Ensuring that random matrix universality is realized in all spin sectors simultaneously is therefore a highly non-trivial task. This is further complicated by the fact that the Fourier coefficients of the cusp forms are erratic and not known analytically, obstructing a  systematic implementation of the RMT$_2$ procedure in spinning sectors. 

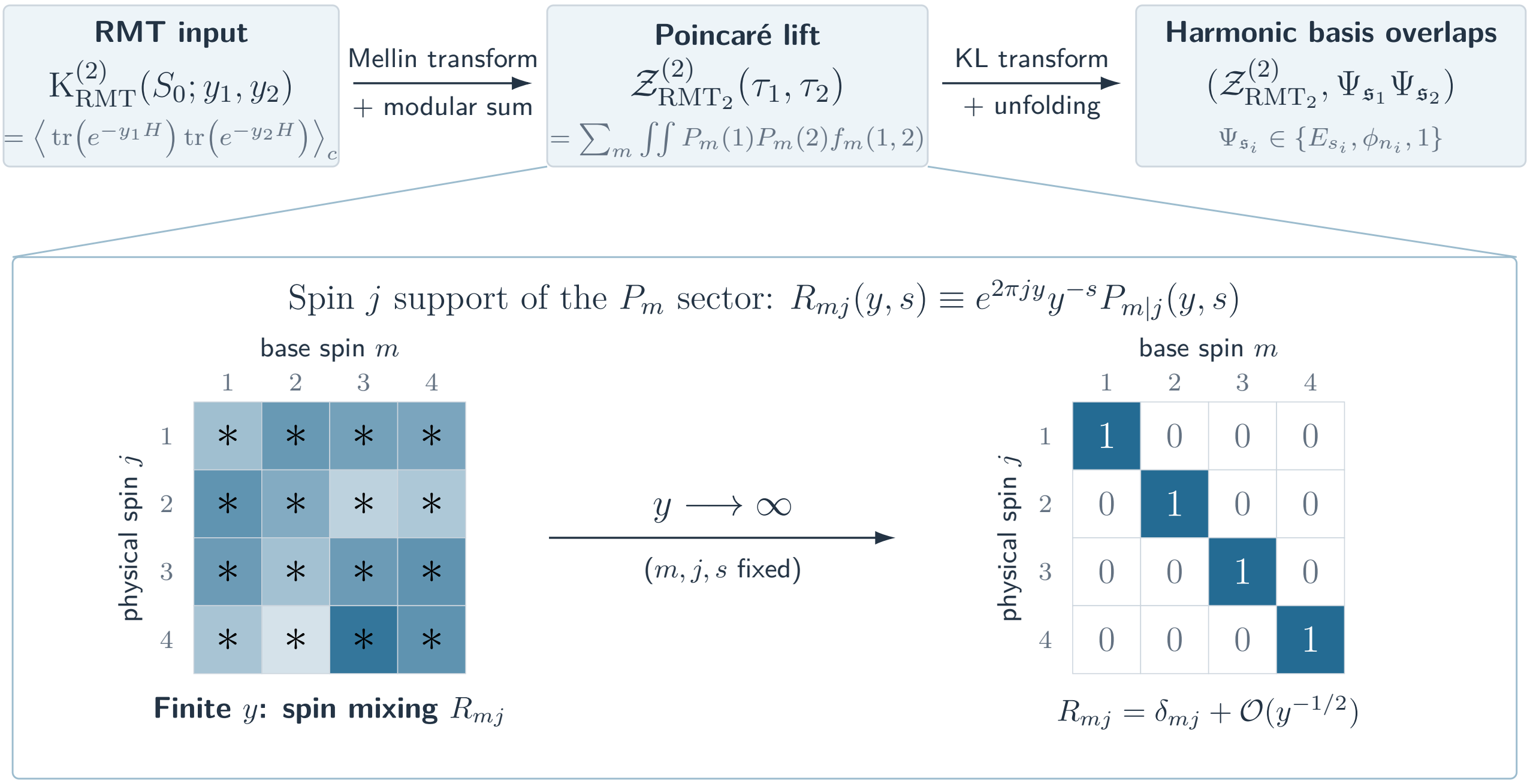
\begin{figure}[t]
\begin{tikzpicture}[x=1cm,y=1cm,
  every node/.style={text=ink,inner sep=0pt},
  arrow/.style={-{Latex[length=2.3mm,width=1.5mm]},draw=ink,line width=.65pt},
  small/.style={font=\sffamily\fontsize{8.5}{10}\selectfont},
  note/.style={font=\sffamily\fontsize{8}{9.5}\selectfont,text=muted}]

\path[use as bounding box] (-.05,-.02) rectangle (16.85,8.95);

% Three steps of the construction.
\draw[fill=pale,draw=line,line width=.55pt,rounded corners=2pt]
  (0,7.1) rectangle (3.7,8.88);
\draw[fill=pale,draw=line,line width=.55pt,rounded corners=2pt]
  (6.0,7.1) rectangle (10.2,8.88);
\draw[fill=pale,draw=line,line width=.55pt,rounded corners=2pt]
  (12.5,7.1) rectangle (16.8,8.88);

\node[small,font=\sffamily\bfseries\fontsize{9.5}{11}\selectfont]
  at (1.85,8.56) {RMT input};
\node[font=\fontsize{11}{13}\selectfont] at (1.85,8.02)
  {$\mathrm K_{\rm RMT}^{(2)}(S_0;y_1,y_2)$};
\node[note] at (1.85,7.4) {$=\big\langle \tr  (e^{-y_1H})\tr ( e^{-y_2H}) \big\rangle_c$};

\node[small,font=\sffamily\bfseries\fontsize{9.5}{11}\selectfont]
  at (8.1,8.56) {Poincar\'e lift};
\node[font=\fontsize{12}{14}\selectfont] at (8.1,8.02)
  {$\mathcal Z_{\rm RMT_2}^{(2)}(\tau_1,\tau_2)$};
\node[note] at (8.1,7.4)
  {$= \sum_{m}  \iint P_m(1) P_m(2) f_m(1,2)$};

\node[small,font=\sffamily\bfseries\fontsize{9.5}{11}\selectfont]
  at (14.65,8.56) {Harmonic basis overlaps};
\node[font=\fontsize{11}{13}\selectfont] at (14.65,8.02)
  {$(\mathcal Z^{(2)}_{\text{RMT}_2},\Psi_{\sGen_1}\Psi_{\sGen_2})$};
\node[note] at (14.65,7.4)
  {$\Psi_{\sGen_i}\in\{E_{s_i},\phi_{n_i},1\}$};

\draw[arrow] (3.86,8.02) -- (5.84,8.02);
\node[small] at (4.85,8.30) {Mellin transform};
\node[small] at (4.85,7.76) {+ modular sum};
\draw[arrow] (10.36,8.02) -- (12.34,8.02);
\node[small] at (11.35,8.30) {KL transform};
\node[small] at (11.35,7.76) {+ unfolding};

% Expanding lines connect the Poincare lift to its detailed illustration.
\draw[blue!45,line width=.55pt] (6.0,7.1) -- (.1,6.1);
\draw[blue!45,line width=.55pt] (10.2,7.1) -- (16.7,6.1);
\draw[blue!45,line width=.55pt,rounded corners=2pt]
  (.1,.34) rectangle (16.7,6.1);
\node[font=\fontsize{11}{13}\selectfont] at (8.4,5.62)
  {Spin $j$ support of the $P_m$ sector: $R_{mj}(y,s)\equiv e^{2\pi jy}y^{-s}P_{m|j}(y,s)$};

% Bottom graphic
\foreach \xzero in {2.1,11.8}{
  \node[small] at ({\xzero+1.5},5.08) {base spin $m$};
  \node[small,rotate=90] at ({\xzero-.7},3.00) {physical spin $j$};
  \foreach \i in {1,2,3,4}{
    \node[note] at ({\xzero+(.75)*(\i-.5)},4.72) {$\i$};
    \node[note] at ({\xzero-.30},{4.5-(.75)*(\i-.5)}) {$\i$};
  }
}

\pgfmathsetseed{2026}
\foreach \j in {1,2,3,4}{
  \foreach \m in {1,2,3,4}{
    \pgfmathsetmacro{\ytop}{4.5-.75*(\j-1)}
    \pgfmathsetmacro{\xl}{2.1+.75*(\m-1)}
    \pgfmathsetmacro{\xr}{11.8+.75*(\m-1)}
    \pgfmathrandominteger{\cellshade}{35}{100}
    \pgfmathsetmacro{\cellopacity}{0.50+0.50*rnd}
    \fill[blue!\cellshade!white,fill opacity=\cellopacity]
      (\xl,\ytop) rectangle ++(.75,-.75);
    \def\starcolor{black}
    \node[text=\starcolor,font=\fontsize{15}{17}\selectfont]
      at ({\xl+.375},{\ytop-.375}) {$\ast$};
    \ifnum\j=\m
      \fill[blue] (\xr,\ytop) rectangle ++(.75,-.75);
      \node[text=white,font=\fontsize{12}{14}\selectfont]
        at ({\xr+.375},{\ytop-.375}) {$1$};
    \else
      \fill[white] (\xr,\ytop) rectangle ++(.75,-.75);
      \node[text=muted,font=\fontsize{11}{13}\selectfont]
        at ({\xr+.375},{\ytop-.375}) {$0$};
    \fi
    \draw[draw=line,line width=.3pt]
      (\xl,\ytop) rectangle ++(.75,-.75);
    \draw[draw=line,line width=.3pt]
      (\xr,\ytop) rectangle ++(.75,-.75);
  }
}

\draw[arrow] (6.02,3.00) -- (9.85,3.00);
\node[font=\fontsize{12}{14}\selectfont] at (7.94,3.3) {$y\longrightarrow\infty$};
\node[small] at (7.94,2.64) {($m,j,s$ fixed)};

\node[small,font=\sffamily\bfseries\fontsize{9.5}{11}\selectfont]
  at (3.6,1.08) {Finite $y$: spin mixing $R_{mj}$};
\node[font=\fontsize{9.5}{11}\selectfont] at (13.3,1.08)
{$R_{mj}=\delta_{mj}+{\cal O}(y^{-1/2})$};
\end{tikzpicture}
\caption{
Overview of the RMT$_2$ construction, demonstrated for $n=2$. {\it Top row:} The Mellin transform $f_m(\bar s_1,\bar s_2)$ of a prescribed matrix model correlator determines the weight for products of Poincar\'{e} series, producing a modular-invariant amplitude. Unfolding expresses its Eisenstein, cuspidal, and mixed harmonic overlaps through Kontorovich-Lebedev integral transforms. {\it Bottom:} Each Poincar\'{e} series has a fixed base spin label $m$ but support over all physical spins $j$. After removing the Boltzmann factor and seed power, the positive-spin Fourier
coefficients become diagonal in $(m,j)$ in the near-extremal limit $y\rightarrow \infty$.}
\label{fig:summary_figure}
\end{figure}

In this paper, we solve this technical problem and present a general construction of  RMT$_2$, completing the construction of \cite{Boruch:2025ilr} by incorporating random matrix dynamics in sectors of arbitrary spin. We summarize the paper in the subsection below, but let us give a compact explanation here. The key idea is the use of an additional family of Poincar\'e sums, which span the same space of functions as Eisenstein series and Maass cusp forms, but have more convenient analytic properties. The role of these Poincar\'e sums, labeled by an integer $m$ referred to as ``base spin", is to effectively resum all the Maass cusp forms relevant for a given CFT spin sector $j$ near extremality. More precisely, in the CFT spin $j$ spectrum, the leading contribution is captured simply by a {\it single} Poincar\'e sum with base spin $m=j$ (see Figure \ref{fig:summary_figure}). This property allows us to efficiently determine the explicit coefficients of the Poincar\'e expansion of the partition function in terms of the near-extremal data of every CFT spin sector. The resulting expression gives, by construction, a modular-invariant partition function which in the near-extremal limit reproduces the imposed near-extremal data. It is straightforward to re-express the obtained partition function in terms of Eisenstein series and Maass cusp forms, with explicit coefficients. 

We begin with a compact summary of, and guide to, the remainder of this paper.

%%%%%%%%%%%%%%%%%%%%%%%%%%%%%%%%%%%%%%%%%%%%%%%%
%%%%%%%%%%%%%%%%%%%%%%%%%%%%%%%%%%%%%%%%%%%%%%%%
\subsection{Summary: AdS$_3$/RMT$_2$ with spin }
\label{sec:summary}
%%%%%%%%%%%%%%%%%%%%%%%%%%%%%%%%%%%%%%%%%%%%%%%%
%%%%%%%%%%%%%%%%%%%%%%%%%%%%%%%%%%%%%%%%%%%%%%%%

The central CFT object of interest is the fluctuating part of the primary state-counting torus partition function, i.e., the part that is sensitive only to Virasoro primary states above the black hole threshold. This is a normalizable function on the fundamental domain $\cal F$. We start by briefly summarizing different ways to spectrally decompose $L^2({\cal F})$-normalizable modular-invariant functions. The most obvious way to do so is by employing a spectral decomposition in terms of orthogonal eigenfunctions of the Laplacian on the fundamental domain \cite{Benjamin:2021ygh}: these consist of real-analytic Eisenstein series, Maass cusp forms and the constant function.

A crucial theme in this work will be the imposition of random matrix statistics in operator spectra with fixed, {\it arbitrary} integer spin; as explained earlier, the properties of Eisenstein series and Maass cusp forms make this difficult a priori. Solving this problem is the main contribution of this work, which we now explain.\footnote{In what follows, $s\in\frac12+i\mathbb{R}$ labels the critical line unless otherwise noted, and $\tau=x+iy$ is the modulus of the torus (i.e. $y$ is an inverse temperature and $x$ is conjugate to spin). The {\it near-extremal limit} is $y\to\infty$ at fixed spin.}

\paragraph{Spectral decomposition and Poincar\'e sums (Section \ref{sec:background}).}

The $\text{SL}(2,\mathbb{Z})$ spectral eigenbasis spanned by Eisenstein series $E_{s}$ and Maass cusp forms $\phi_n$ is a canonical basis for three reasons: it is an honest orthogonal basis, so overlaps are unique and there are no null vectors; it is the basis in which the CFT trace formula is formulated \cite{DiUbaldo:2023qli,Haehl:2023tkr}; and it diagonalizes the torus wormhole amplitude, in the same way that Berry's diagonal approximation diagonalizes the linear ramp in the periodic-orbit basis \cite{DiUbaldo:2023qli}, as explained further below. Conceptually, it is useful to think of these as forming what is analogous to a {\it plane wave basis}. 

On the other hand, we will discuss a basis -- more precisely, an overcomplete frame -- of Poincar\'{e} sums $P_m(\tau,s)$, which is analogous to {\it coherent states}, i.e. localized superpositions of spectral eigenstates. Explicitly, the (parity-even) Poincar\'e series is
\begin{equation}
\label{eq:summaryPoincare}
 P_m(\tau,s)=d_m\!\sum_{\gamma\in\Gamma_\infty\backslash \text{SL}(2,\mathbb Z)}
 \Im(\gamma\tau)^s \, e^{-2\pi m\Im(\gamma\tau)}
 \cos\!\big(2\pi m\Re(\gamma\tau)\big) \,,
 \qquad d_m=2-\delta_{m0}\,.
\end{equation}
Here $m\geq0$ and $\Gamma_\infty$ is the subgroup of translations: the sum adds the inequivalent modular images of a seed of definite spin. It converges for $\Re(s)>1$, but we will find a convenient analytic continuation to the critical line $\text{Re}(s)=\frac12$.  These functions form an overcomplete frame, analogous to wavepackets of harmonic eigenfunctions. Indeed, for $m=0$ the Poincar\'e sums are precisely Eisenstein series, $P_0(\tau,s)=E_s(\tau)$. The usefulness of the Poincar\'e frame stems from its cuspidal (large-$y$) behavior: the spin $j>0$ Fourier component of $P_m$, call it $P_{m|j}$, exhibits the following leading asymptotics:
\begin{equation}
\label{eq:summaryCusp}
 P_{m|j}(y,s)
 = e^{-2\pi j y} \left[\delta_{mj}\,y^s\, +{\cal O}(1) \right] \,, \qquad s=\frac12+i\omega\,,
\end{equation}
at fixed $m$ and $\omega$. Thus the seed dominates precisely in the spin it was designed to describe, $j=m$. The \emph{base spin} $m$ labels an entire modular-invariant function, with support on all physical spins $j$; the two labels coincide precisely for the leading asymptotic term.

\paragraph{The RMT${}_2$ construction (Section \ref{sec:rmt2general}).} In Section \ref{sec:rmt2general}, we present the general construction for $n$-point correlations in \eqref{eq:mellinBasic} and \eqref{eq:RMT2defPmRep}. For simplicity of presentation, let us here consider $n=2$, with an input matrix model correlator $\text{K}^{(2)}_{\rm RMT}(S_0;y_1,y_2)=\langle \tr  (e^{-y_1H})\tr ( e^{-y_2H}) \rangle_c$. The latter will typically describe a double-scaled matrix model with a near-extremal entropy $S_0$ that organizes the genus expansion.
The goal of the RMT${}_2$ construction is to provide a (doubly) modular-invariant function ${\cal Z}^{(2)}_{\text{RMT}_2}(\tau_1,\tau_2)$ which reduces to $\text{K}^{(2)}_{\rm RMT}(S_0(j);y_1,y_2)$ in the limit $y_i\rightarrow\infty$ for every fixed spin $j$ sector. 
We achieve this by computing the Mellin transform of $\text{K}^{(2)}_{\rm RMT}$ and using it as the coefficient in a (double) expansion in Poincar\'e sums:
\begin{equation}
\label{eq:summaryLift}
\begin{split}
 f^{(2)}_{\rm RMT}(S_0;\bar s_1,\bar s_2)
 &=\int_0^\infty\!\!\int_0^\infty
 \frac{dy_1\,dy_2}{y_1^{1+i\omega_1}y_2^{1+i\omega_2}}
 \,\text{K}^{(2)}_{\rm RMT}(S_0;y_1,y_2),\\
 {\cal Z}^{(2)}_{{\rm RMT}_2}(\tau_1,\tau_2)
 &\equiv\sum_{m\geq0}\frac1{d_m}
 \int_{{\cal C}_{\rm crit}^2}\frac{ds_1\,ds_2}{(2\pi i)^2}
 \,P_m(\tau_1,s_1)P_m(\tau_2,s_2)\,
 f^{(2)}_{\rm RMT}(S_0(m);\bar s_1,\bar s_2)\,.
\end{split}
\end{equation}
Modular invariance of this construction is manifest. Further, near the cusp, thanks to \eqref{eq:summaryCusp} the Poincar\'e series $P_m$ are dominated by their seeds for $j=m$, thus turning the integrals back into inverse Mellin transforms: this implies, for the spin-$(j_1,j_2)$ Fourier component of the RMT${}_2$ amplitude,
\begin{equation}
\label{eq:summaryNearExtremal}
 {\cal Z}^{(2)}_{{\rm RMT}_2|j_1,j_2}(y_1,y_2)  \approx \frac{\delta_{j_1j_2}}{d_{j_1}} e^{-2\pi(j_1y_1+j_2y_2)}\sqrt{y_1y_2}\,  \text{K}^{(2)}_{\rm RMT}(S_0(j_1);y_1,y_2)\,, \qquad (y_i\rightarrow\infty)\,.
\end{equation}
The exponential accounts for the spin-dependent ground-state energy, while the factor $\sqrt{y_1y_2}$ arises from stripping off Virasoro descendants while preserving modular invariance. The required near-extremal reduction to the fixed-spin RMT input is therefore established.

Taking \eqref{eq:summaryLift} as the definition of the RMT${}_2$ lift (for $n=2$), the expansion in the harmonic eigenbasis follows by unfolding the Poincar\'e sums: writing collectively $\Psi_\sGen$ to mean either Eisenstein series or cusp forms with Fourier coefficients denoted as ${{\tt c}_m^{(\sGen)}}$, we find that the overlaps take the form 
\begin{equation}
\label{eq:summaryKL}
\begin{split}
 &\left({\cal Z}^{(2)}_{{\rm RMT}_2},\Psi_{\sGen_1}\Psi_{\sGen_2}\right)_{\mpos}=\sum_{m>0}d_m\,{\bar{\tt c}_m^{(\sGen_1)}}\,
 {\bar {\tt c}_m^{(\sGen_2)}}
  \prod_{i=1}^2\left[\int_0^\infty\frac{dy_i}{y_i}\,e^{-2\pi my_i}
 K_{\sGen_i-\frac12}(2\pi my_i)\right]
 \text{K}^{(2)}_{\rm RMT}(S_0(m);y_1,y_2)
\end{split}
\end{equation}
where the subscript ``$>0$'' indicates that we consider the $m>0$ contributions. 
The square brackets denote a universal integration kernel that arises from the modular inner product with measure $dx\,dy/y^2$ on each torus. The integrals \eqref{eq:summaryKL} are Kontorovich-Lebedev (KL) transforms: a Bessel-$K$ transform replacing the Mellin transform in the harmonic basis. These transforms determine Eisenstein, cusp, and mixed overlaps via a uniform integral transform. As we will expand on later, the sum over Fourier coefficients ${\bar{\tt c}_m^{(\sGen_1)}}$ naturally leads to the appearance of Dirichlet-like series and L-functions associated with Eisenstein series and cusp forms.

\paragraph{Explicit worked example: Airy RMT${}_2$ (Section \ref{sec:airyRMT2}).}

In Section~\ref{sec:airyRMT2}, we present a worked example of our construction for the RMT${}_2$ lift of the Airy model, in both GUE and GOE ensembles. We work at $n=2$ and arbitrary genus $g$. We focus especially on the so-called $\uptau$-scaling limit, which allows us to zoom in on spectral statistics on energy scales of the same order as the average level spacing: working at complex time $y_{1,2}=\beta\pm iT$, this is the regime of $T,e^{S_0}\to\infty$ and $\uptau=Te^{-S_0}$ fixed. In the unitary class, the result has a rather compact expression:
\begin{equation}
\begin{split}
    &{\cal Z}^{(2)}_{\uptau\text{-Airy}}(\tau_1,\tau_2) 
    = {\cal Z}_{\text{RMT}_2}^{(0,2)}(\tau_1,\tau_2)
    \\
    &\quad + \sum_{m\geq 0} d_m^{-1} \sum_{g\geq 1}a_g\, e^{-2g S_0(m)}\sum_{k= 0}^{g-1} \binom{g-1}{k} \left[ P_m\left(\tau_1,2g-k\right)P_m\left(\tau_2,g+k+1\right) \right]_\text{sym} 
\end{split}
\end{equation}
where $ {\cal Z}^{(0,2)}_{{\rm RMT}_2}(\tau_1,\tau_2)$ is the universal RMT${}_2$ correlator defined immediately below, the combinatorial coefficients $a_g$ are defined in \eqref{eq:tauAiryGUE2}, $d_m = 2-\delta_{m0}$, and `sym' denotes symmetrization in $\tau_1$ and $\tau_2$. Each positive-genus term of the corresponding RMT${}_2$ is a finite sum of products of Poincar\'e series. The orthogonal class additionally includes half-integer genus and logarithmic terms, both associated with unorientable crosscap geometries. These examples allow us to validate the ramp-plateau transition for both the scalar and the spinning sectors of RMT${}_2$.

\paragraph{Universal RMT${}_2$ correlations at $n=2,3$ and AdS${}_3$ torus wormholes (Section \ref{sec:ramp}).}
The most basic input of RMT, which is independent of the choice of spectral curve, is the two-point function.\footnote{At late times this produces the famous `linear ramp'. Here we impose the RMT two-point function at fixed $y_{1,2}$, not only its late time limit.} In our variables, this takes the form 
\begin{equation}
\text{K}^{(0,2)}_{\rm RMT}(y_1,y_2)=\frac{{\tt C}_{\rm RMT}}{2\pi}\frac{\sqrt{y_1y_2}}{y_1+y_2}\,,
\end{equation}
where ${\tt C}_{\rm RMT}=1,2,\frac12$ for the unitary, orthogonal, and symplectic classes. Its RMT${}_2$ lift \cite{DiUbaldo:2023qli} reproduces the Cotler-Jensen AdS$_3$ wormhole amplitude \cite{Cotler:2020ugk} and has two equivalent representations, which follow from the general prescription above: in the Poincar\'e frame,
\es{eq:summaryCJ}{
 {\cal Z}^{(0,2)}_{{\rm RMT}_2}
 (\tau_1,\tau_2)=\sum_{m\geq0}\frac1{d_m}\int_{\mathbb R}\frac{d\omega}{2\pi}\,
 h(\omega)P_m(\tau_1,\tfrac12+i\omega)P_m(\tau_2,\tfrac12-i\omega)\,,\quad\; h(\omega)=\frac{{\tt C}_{\rm RMT}}{2\cosh(\pi\omega)}}
Passage to the harmonic basis of \cite{DiUbaldo:2023qli} is straightforward; further, the coefficient function $h(\omega)$ is the same in both the Poincar\'e frame and the harmonic basis, a remarkable simplicity special to the $n=2$ correlator at genus zero. 

The RMT amplitude at three points and genus zero is also universal: 
\begin{equation}
\text{K}^{(0,3)}_{\rm RMT}(S_0;y_1,y_2,y_3)= \frac{e^{-S_0}}{(2\pi\gamma)^{3/2}} \, \sqrt{y_1y_2y_3} \,,
\end{equation}
with $\gamma$ fixing the energy normalization. The RMT${}_2$ lift is particularly compact in the Poincar\'e frame:
\begin{equation}
\label{eq:summaryThree}
 {\cal Z}^{(0,3)}_{{\rm RMT}_2}(\tau_1,\tau_2,\tau_3)
 =\frac1{(2\pi\gamma)^{3/2}}
 \left[e^{-S_0(0)}\prod_{i=1}^3\widehat E_1(\tau_i)
 +\frac14\sum_{m>0}e^{-S_0(m)}\prod_{i=1}^3P_m(\tau_i,1)\right].
\end{equation}
Here $\widehat E_1$ is the finite part of $E_s$ after subtracting its pole at $s=1$ (see \eqref{e1hatdef}). The remaining Poincar\'e sums have a finite analytic continuation at $s=1$. Unlike \eqref{eq:summaryCJ}, this amplitude has non-diagonal harmonic overlaps and mixed Eisenstein-cusp form correlations. The $m=0$ term supplies the scalar limit, entirely determined by Eisenstein series. The $m>0$ terms contain further Eisenstein contributions, which are a feature of our particular construction and are subleading near extremality. We find for the spectral decomposition the following triple overlaps with the $\delta$-normalized harmonic basis vectors:
\begin{equation}
    \left(  {\cal Z}^{(0,3)}_{\text{RMT}_2},\,\Psi_{\sGen_1}\Psi_{\sGen_2} \Psi_{\sGen_3} \right) \;=\; \frac{\widetilde{D}_{\Psi_{\bar\sGen_1}\times\Psi_{\bar\sGen_2}\times \Psi_{\bar\sGen_3}}\left(S_0;\frac{3}{2}\right)}{32(2\pi \gamma)^{3/2}} \,
    \prod_{i=1}^3 \Gamma(\sGen_i)\, \Gamma(1-\sGen_i) + \textsf{[$m=0$ term]} 
\label{eq:Z03resultIntro}
\end{equation}
with the generalized (entropy-weighted) triple product Dirichlet series
\begin{equation}
\widetilde{D}_{\Psi_{\sGen_1}\times\Psi_{\sGen_2}\times \Psi_{\sGen_3}}(S_0;t) \equiv 8 \sum_{m\geq 1} \frac{{\tt c}^{(\sGen_1)}_m {\tt c}^{(\sGen_2)}_m {\tt c}^{(\sGen_3)}_m}{m^t} \, e^{-\frac{2}{3} t\,S_0(m)} \,.
\end{equation}
The triple-cusp form contribution supplies the leading terms in the spinning spectrum. This is a candidate three-boundary amplitude in AdS$_3$ pure gravity, awaiting an independent bulk calculation. The result \eqref{eq:Z03resultIntro} exemplifies a general feature of our construction: some overlaps (such as those with three Eisenstein series for $m\geq 1$) produce terms that are modular-invariant by themselves but are subleading near extremality. This reflects the non-uniqueness of the RMT${}_2$ construction away from extremality beyond the $(g,n)=(0,2)$ case. Understanding what choice is selected by AdS$_3$ pure gravity is an interesting open problem, which we comment on in Section \ref{ssec:minimality}.

{\bf Several appendices} collect various complementary results and checks: random matrix theory in Mellin space (Appendix~\ref{app:RMTmellin}), properties of Poincar\'e series (Appendix~\ref{app:poincare}), a gravity interpretation of the Poincar\'e and spectral decompositions (Appendix~\ref{app:wdw}), the parity-odd sector (Appendix~\ref{app:parity}), zero modes (Appendix~\ref{app:zeromodes}), the Dirichlet series and L-functions that appear in our results (Appendix~\ref{app:Lfunctions}), alternative presentations of the RMT${}_2$ correlators (Appendix~\ref{app:alt}), numerical checks (Appendix~\ref{app:numerics}), a basis change for $n=2$ (Appendix~\ref{sec:CJnewderivation}) and the calculations for the GOE Airy model (Appendix~\ref{app:GOE}). We highlight Appendix \ref{app:RMTmellin}, which stands alone as an alternative, Mellin presentation of ordinary random matrix theory which may be useful for other purposes.

%%%%%%%%%%%%%%%%%%%%%%%%%%%%%%%%%%%%%%%%%%%%%%%%
%%%%%%%%%%%%%%%%%%%%%%%%%%%%%%%%%%%%%%%%%%%%%%%%
\section{Spectral decomposition and Poincar\'e sums}
\label{sec:background}
%%%%%%%%%%%%%%%%%%%%%%%%%%%%%%%%%%%%%%%%%%%%%%%%
%%%%%%%%%%%%%%%%%%%%%%%%%%%%%%%%%%%%%%%%%%%%%%%%

In this section we lay the mathematical foundation for the construction of spinning RMT${}_2$ form factors. In particular, this involves a review of the $\text{SL}(2,\mathbb{Z})$ spectral decomposition and of Poincar\'e sums relevant for our construction. For more detailed reviews, we refer the reader to the previous works cited.

%%%%%%%%%%%%%%%%%%%%%%%%%%%%%%%%%%%%%%%%%%%%%%%%
\subsection{The CFT trace formula and $\text{SL}(2,\mathbb{Z})$ spectral decomposition}
%%%%%%%%%%%%%%%%%%%%%%%%%%%%%%%%%%%%%%%%%%%%%%%%

For the standard $\text{SL}(2,\mathbb{Z})$ spectral decomposition in terms of real-analytic Eisenstein series and Maass cusp forms, we use the collective notation
\be 
\label{eq:PsiDef}
\Psi_\sGen (\tau) \equiv \sum_{j\geq 0} d_j \cos(2\pi j x)\, \Psi_{\sGen|j}(y) = \sum_{j\geq 0} d_j \cos(2\pi j x) \, {\tt c}_j^{(\sGen)} \, \sqrt{y} K_{\sGen - \frac{1}{2}} (2\pi y j) \,,\qquad d_j \equiv 2-\delta_{j0}\,,
\ee
where $\sGen  \in {\cal C}_\text{crit} = \frac{1}{2}+i \mathbb{R}$ is either continuous or discrete, depending on whether $\Psi_\sGen$ is an Eisenstein series ($\sGen \equiv s \equiv  \frac{1}{2}+i\omega$), or a Maass cusp form ($\sGen \equiv s_n \equiv \frac{1}{2}+i\omega_n$). The Fourier coefficients in the two cases are:\footnote{For spin $j=0$, one has $E_{s|0}(y) = y^s+\varphi(s)y^{1-s}$ with $\varphi(s)$ as in \eqref{eq:reflection}.}
\be 
\label{eq:PsiDef2}
 \begin{split}
   \Psi_{\sGen} \equiv E_{\frac{1}{2}+i\omega} \, : \qquad\;\;\, {\tt c}_j^{(\sGen)} &=   {\tt a}_j^{(s)}  = \frac{2\sigma_{2i\omega}(j)}{j^{i\omega} \Lambda(-i\omega)}  \,,\\
    \Psi_\sGen \equiv \phi_n^{+}\, : \qquad\;\;\, {\tt c}_j^{(\sGen)} & =  {\tt b}_j^{(n,+)} \,,\\
 \end{split}
\ee
with $\Lambda(s)= \pi^{-s} \Gamma(s)\zeta(2s)$ and ${\tt b}_j^{(n,+)}$ the standard erratic Fourier coefficients of even-parity Maass cusp forms normalized such that the Petersson norm is $|\!|\phi_n|\!|=1$ (and ${\tt b}_1^{(n,+)}>0$).\footnote{Our Eisenstein normalization and Mellin-transform convention agree with \cite{Boruch:2025ilr}. Relative to the completed Eisenstein series $E_s^*=\Lambda(s)E_s$ used in \cite{DiUbaldo:2023qli}, the Fourier coefficients satisfy ${\sf a}_{j\text{\cite{DiUbaldo:2023qli}}}^{(s)} =\Lambda(s){\tt a}_{j\mathrm{[here]}}^{(s)}$, while the unit-normalized cusp coefficients obey ${\sf b}_{j \text{\cite{DiUbaldo:2023qli}}}^{(n)} ={\tt b}_{j \mathrm{[here]}}^{(n)}$. For the Hecke-normalized forms $\nu_{n,+}$ of \cite{Haehl:2023xys}, our conventions are $\phi_n=\nu_{n,+}/|\!|\nu_{n,+}|\!|$ and ${\tt b}_{j \mathrm{[here]}}^{(n)} =a_{j \text{\cite{Haehl:2023xys}}}^{(n,+)} /(2|\!|\nu_{n,+}|\!|)$, with $\omega_n=R_n^+$.} In general, there also exist odd-parity Maass cusp forms $\phi_n^-$ with a Fourier sine-decomposition. However, in this work, we assume for simplicity that the CFT under consideration has a parity-even spectrum. This simplifies the notation, while generalizations are straightforward. In the following we omit the `$+$' label. We discuss some  generalizations in Appendix \ref{app:parity}. 

We note in passing that one could instead work with the completed Eisenstein series $E_s^*(\tau) \equiv \Lambda(s) E_s(\tau)$, which are real and satisfy $E_s^*(\tau) = E_{1-s}^*(\tau)$. We prefer to work with the (complex-valued) Eisenstein series as defined by \eqref{eq:PsiDef} and \eqref{eq:PsiDef2} as these have a more transparent near-extremal limit. They satisfy the functional equation 
\begin{equation}
\label{eq:reflection}
    E_{1-s}(\tau) = \varphi(1-s) E_{s}(\tau) \,,\qquad \varphi(s) \equiv \frac{\Lambda(1-s)}{\Lambda(s)} \,.
\end{equation}

The Eisenstein series and Maass cusp forms furnish an orthogonal basis for modular-invariant, square-integrable functions. We abbreviate the spectral decomposition of an $n$-fold modular-invariant, square-integrable function as follows:
\begin{equation}
\label{eq:specD}
    \begin{split}
    {\cal Z}^{(n)}(\btau) \equiv {\cal Z}^{(n)}(\tau_1,\ldots,\tau_n) = \prod_{k=1}^n \left( \sumint_{\sGen_k}  \, \Psi_{\sGen_k}(\tau_k) \right) \, \left( {\cal Z}^{(n)},\,\Psi_{\sGen_1} \cdots \Psi_{\sGen_n}\right) \,,
    \end{split}
\end{equation}
where $\sumint_\sGen \Psi_{\sGen}(\tau)$ stands for sums over normalized cusp forms ($\sum_{n=1}^\infty  \phi_n(\tau)$), plus integrals over Eisenstein series ($\int_{{\cal C}_\text{crit}} \frac{ds}{4\pi i}\,E_s(\tau)$), plus overlaps with the constant function $1$ (with ``measure'' $\frac{3}{\pi}$). The inner product is the nested Petersson inner product over the fundamental domain ${\cal F} = \mathbb{H}/\text{SL}(2,\mathbb{Z})$,
\begin{equation}
     \left( {\cal Z}^{(n)},\,\Psi_{\sGen_1} \cdots \Psi_{\sGen_n}\right)
     \equiv \int_{{\cal F}^n}\frac{dx_1 dy_1}{y_1^2} \cdots \frac{dx_n dy_n}{y_n^2} \, {\cal Z}^{(n)}(\tau_1,\ldots,\tau_n) \,\overline{\Psi_{\sGen_1}(\tau_1)} \cdots \overline{\Psi_{\sGen_n}(\tau_n)} \,,
\end{equation}
where $\tau_k = x_k + i y_k$.\footnote{Note that complex conjugation of $\Psi_\sGen$ amounts to complex conjugation of the Fourier coefficients ${\tt c}_j^{(\sGen)}$. For Eisenstein series on the critical line, this amounts to $\overline{E_s} = E_{\bar s}$. The Maass cusp forms are real-valued, $\overline{\phi_n} = \phi_n$.}
The orthogonality relations are:
\begin{equation}
    \begin{split}
    (E_{s_1},E_{s_2}) &= 2\pi \big( \delta(\omega_1-\omega_2) + \varphi(1-s_2) \delta(\omega_1+\omega_2) \big) \qquad (s_i = \tfrac{1}{2}+i\omega_i) \,,\\
    (\phi_{n_1},\phi_{n_2}) &=  \delta_{n_1n_2} \,,\\
    (E_s,\phi_n) &= 0 \,.
    \end{split}
\end{equation}
For $n=1$, this is the ordinary $L^2(\mathcal{F})$ spectral decomposition of subtracted CFT partition functions \cite{Benjamin:2021ygh}, and the Eisenstein overlap defines (up to gamma factors) a Rankin-Selberg L-function \cite{Perlmutter:2025ngj}. We will mainly be concerned here with $n>1$.

%%%%%%%%%%%%%%%%%%%%%%%%%%%%%%%%%%%%%%%%%%%%%%%%
\subsection{Poincar\'e decomposition}
\label{s32}
%%%%%%%%%%%%%%%%%%%%%%%%%%%%%%%%%%%%%%%%%%%%%%%%

While the Eisenstein series and cusp forms form an orthogonal basis over which to expand modular-invariant partition functions and form factors, it will prove useful to consider a different set of modular-invariant functions, more adapted to the near-extremal limit.
We define the spin-$m$ Poincar\'{e} sum as
\begin{equation}
{\cal P}_m (\tau , s) = \sum_{\gamma \in \Gamma_\infty \setminus \text{SL}(2,\mathbb{Z})} 
\Im(\gamma \tau)^s \,
e^{2\pi i m \cdot \gamma \tau} \qquad\quad (\tau \in \mathbb{H},\; \text{Re}(s)>1) \,,
\end{equation}
which is normalizable in the $\tau=x+iy$ fundamental domain for $m> 0$. The sum converges only for $\text{Re}(s)>1$, and the functions are defined elsewhere by analytic continuation. Since we will always assume a parity-even spectrum, it is convenient to define the corresponding parity-even Poincar\'e sum\footnote{Parity conjugation acts as $\tau \mapsto -\bar{\tau}$. See Appendix \ref{app:parity} for more details and a discussion of the parity-odd sector.}
\begin{equation}
\label{eq:PmDef0}
    P_m(\tau,s) \equiv \frac{d_m}{2}\left( {\cal P}_m(\tau,s) + {\cal P}_m(-\bar{\tau},s) \right)\,,\qquad d_m =2-\delta_{m0} \,.
\end{equation}

The Eisenstein series are spin-0 Poincar\'e series:
\begin{equation}
P_0(\tau,s)=E_s(\tau)\,.
\end{equation}
More generally, for any $m$ and $s$ the spectral decomposition of $P_m$ exists:
\be 
\label{eq:Pintegration}
P_m(\tau , s) = \sumint_\sGen \left( P_m(\,\cdot\,,s),\,\Psi_\sGen\right) \Psi_\sGen(\tau) 
\ee
with explicit overlaps (see, for example, \cite{Goldfeld1979,motohashi1997}, and Appendix \ref{app:poincare})
\begin{equation}
\label{eq:goldfeldOverlaps}
\begin{split}
\left( P_m (\,\cdot\, ,s) , \Psi_\sGen \right) &= 
\frac{2\,\bar{\tt c}^{(\sGen)}_m  \, \sqrt{\pi}}{(4\pi m)^{s-\frac{1}{2}}}
\frac{\Gamma(s-\sGen) \Gamma(s+\sGen-1)}{\Gamma(s)} \qquad (m>0)\,.
\end{split}
\end{equation}
Note that this implies that $P_m(\tau,s)$ has poles along the critical line. In all integrals along the critical line (such as \eqref{eq:Pintegration}), we take ${\cal C}_\text{crit}=\frac12+0^++i\mathbb{R}$ (i.e., $\omega \in \mathbb{R} - i 0^+$). We leave this implicit except when it is important.

Our study will crucially rely on the Fourier decomposition of the Poincar\'e series, which takes the following form: 
\be 
\label{eq:PmFourier}
\begin{split}
P_{m} (\tau ,s) &\equiv  
\sum_{j \geq 0} d_j\,\cos(2\pi j x) \, P_{m|j}(y,s)  \\
&= 
\sum_{j \geq 0} d_j\,\cos(2\pi j x) \left( 
y^s \,  \delta_{m,j} \, e^{-2\pi m y}
+y^{1-s} \,G_{m|j}(y,s)  \right) \,,
\end{split}
\ee
where the spin-off-diagonal term $G_{m|j}$ is given in terms of Kloosterman sums and Whittaker functions, see \eqref{eq:KloosterG}. The motivation for working with $P_m$ in our construction of RMT${}_2$ is the following property:

\noindent{\bf Cuspidal limit:} 
{\it In the cuspidal limit ($y\rightarrow \infty$), for fixed $m$ and any fixed $j>0$, the `remainder' terms $G_{m|j}$ are subleading in \eqref{eq:PmFourier}; on ${\cal C}_\text{crit}$, the relative suppression is ${\cal O}(y^{-1/2})$.}

\begin{figure}
\begin{center}
\includegraphics[width=.5\textwidth]{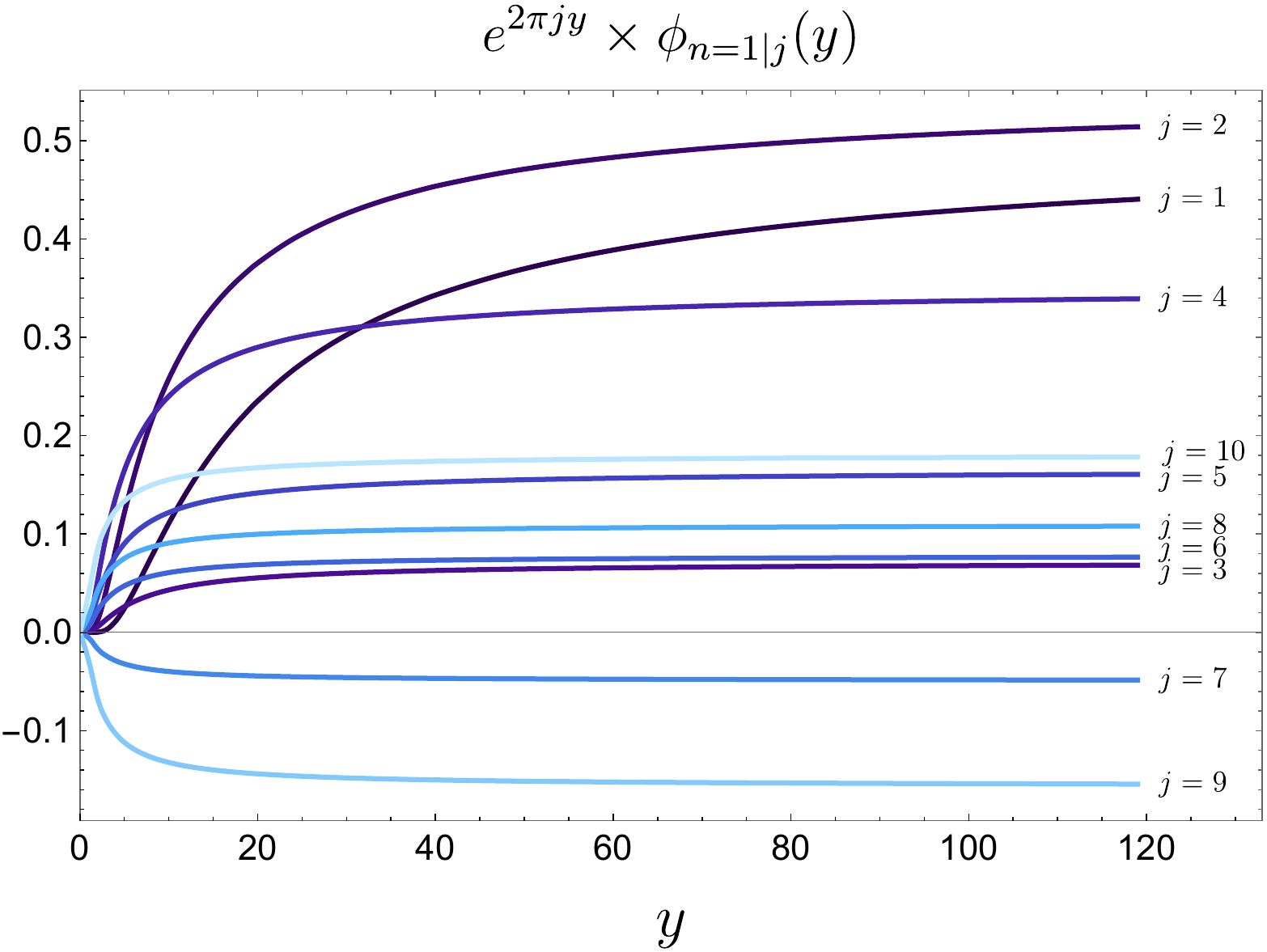}$\,$\includegraphics[width=.48\textwidth]{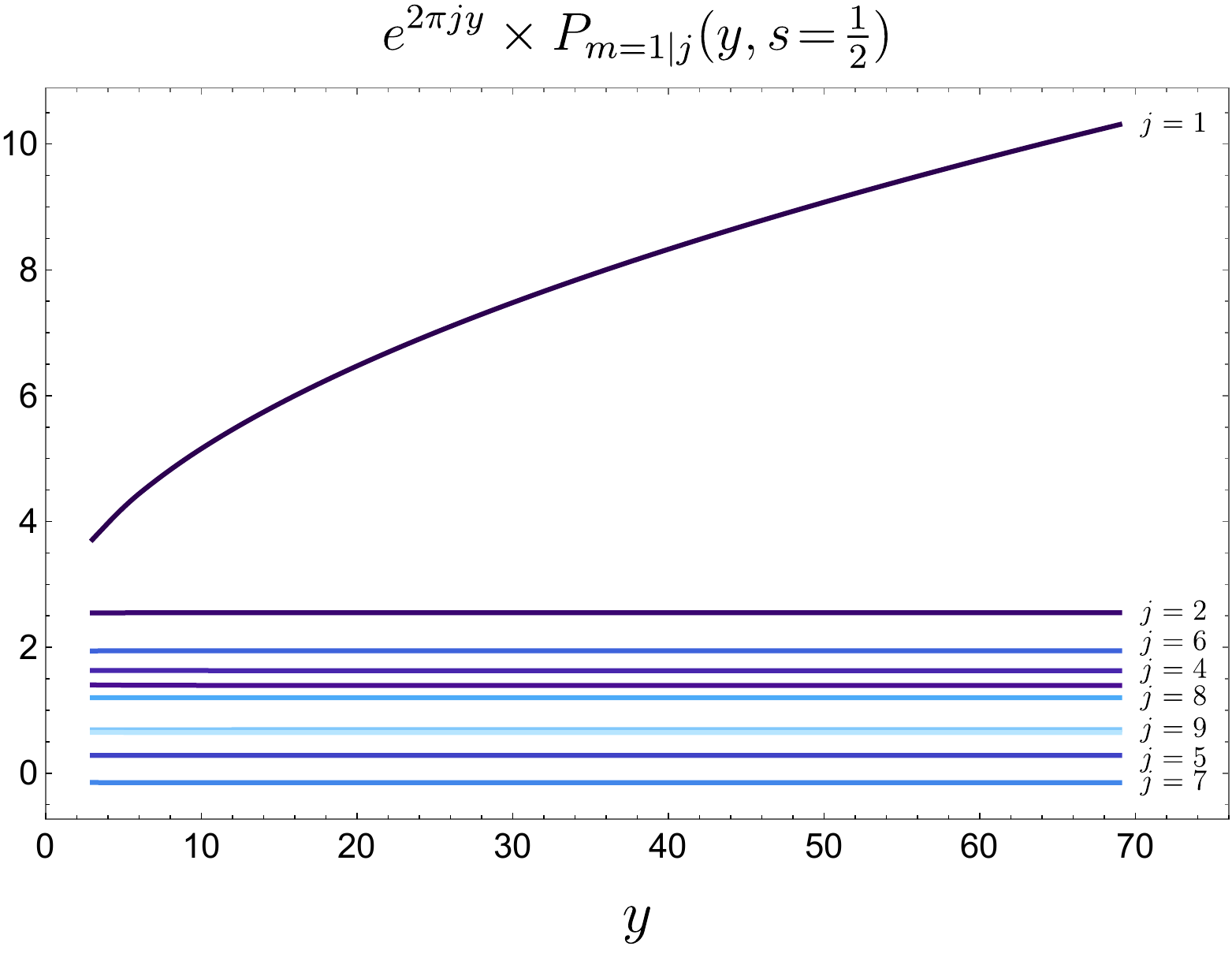}
\end{center}
\caption{Plots of fixed-spin components of the $n=1$ Maass cusp form (left) and the $m=1$ Poincar\'e sum (right). In both cases we rescale by the universal factor $e^{2\pi j y}$, i.e., we renormalize every spin sector by its ground state energy. For large $y$, the cusp form becomes an irregular superposition of different spin components; the Poincar\'e sum is dominated by $j=m$.}
\label{fig:compare}
\end{figure}

This statement is illustrated in Fig.\ \ref{fig:compare}. The right panel shows the square-root-growth of the $j=m$ sector for Poincar\'e sums with all $j\neq m$ components being almost perfectly constant. On the other hand, the left panel shows how cusp forms become democratic superpositions of all their spin components (i.e. frequencies) at the cusp: in the cuspidal limit, all cusp forms contribute to all spin sectors $j\geq 1$. Further illustration can be found in Appendix \ref{app:poincare}.

From the Fourier decomposition, we see that it is useful to think of $P_m$ as a generalization of the Eisenstein series with a fixed dominant cuspidal behavior in the spin-$m$ sector. More precisely (details in Appendix \ref{app:poincare}),
\begin{equation}
\label{eq:PmNE}
    \begin{split}
       m> 0: \qquad  P_{m|j}(y ,s) \; &\stackrel{y\rightarrow\infty}{\approx} \; \delta_{mj} \, y^s \, e^{-2\pi m y} +
       \delta_{0j}  \, \frac{2\sqrt{\pi}\,\Gamma(s-\frac12) \sigma_{1-2s}(m)}{\Gamma(s)\zeta(2s)}\, y^{1-s}
       +\ldots \,, \\ 
       m=0: \qquad\;\;\, P_0 (\tau ,s) \; &\stackrel{y\rightarrow\infty}{\approx} \; y^s + \varphi(s) \, y^{1-s}+ \ldots \,.
    \end{split}
\end{equation}
Conceptually, we can think of the Poincar\'e sums as similar to a coherent state basis, with $P_m$ built to support the $j=m$ spectrum near the cusp.

We note, however, that the set of $\{P_m(\tau,s)\}$ is not actually a complete basis of $L^2({\cal F})$, but rather an {\it overcomplete frame} \cite{Christensen2016Frames,Ali2014CoherentStates}: the Poincar\'e sums are not orthogonal and do not admit an orthogonal resolution of the identity. Their value lies in the fact that they form `wavepackets' of Eisenstein series and cusp forms that localize on fixed spin $m=j>0$ sectors at the cusp (much like Eisenstein series localize on the $j=0$ sector). A further interpretation of the harmonic basis and Poincar\'e frame in terms of quantization of Wheeler-DeWitt wavefunctions of three-dimensional gravity is given in Appendix \ref{app:wdw}.

%%%%%%%%%%%%%%%%%%%%%%%%%%%%%%%%%%%%%%%%%%%%%%%%
%%%%%%%%%%%%%%%%%%%%%%%%%%%%%%%%%%%%%%%%%%%%%%%%
\section{General construction of RMT${}_2$}
\label{sec:rmt2general}
%%%%%%%%%%%%%%%%%%%%%%%%%%%%%%%%%%%%%%%%%%%%%%%%
%%%%%%%%%%%%%%%%%%%%%%%%%%%%%%%%%%%%%%%%%%%%%%%%

We now present our general construction of RMT${}_2$ $n$-point correlators ${\cal Z}_{\text{RMT}_2}^{(n)}(\btau)$. These admit a decomposition into spin sectors:
\begin{equation}
    {\cal Z}_{\text{RMT}_2}^{(n)}(\btau) = \prod_{i=1}^n \left( \sum_{j_i\geq 0} d_{j_i}\, \cos(2\pi j_i x_i) \right)  {\cal Z}_{\text{RMT}_2|{\bf j}}^{(n)}({\bf y}) \,,
    \label{eq:ZFourierDecomp}
\end{equation}
where ${\cal Z}_{\text{RMT}_2|{\bf j}}^{(n)}({\bf y})$ describes $n$-point spectral correlations of operators with spins ${\bf j}=(j_1,\ldots,j_n)$ on $n$ tori with modular parameters $\btau=(\tau_1,\ldots,\tau_n) = (x_1+iy_1,\ldots,x_n+iy_n) = {\bf x} + i{\bf y}$.

We begin in Section \ref{sec:axioms} with a list of minimal requirements for ${\cal Z}_{\text{RMT}_2}^{(n)}$ to satisfy. Then, we present a systematic way to construct a solution to these axioms.

We will use as input for the construction of RMT$_2$ the correlation functions of standard RMT ensembles. We can consider arbitrary single-trace double-scaled matrix models in any of the GUE, GOE, GSE classes:\footnote{There is no obstruction to constructing RMT$_2$ for standard, non-double-scaled matrix models such as a modular-invariant Gaussian matrix model.  These are  interesting in their own right but we focus on double-scaled models for their applications to quantum gravity and CFTs. A particularly relevant example to have in mind is Jackiw-Teitelboim (JT) gravity, for which $\rho^\text{JT}(E) = e^{S_0} \rho^\text{JT}_0(E)$ with $\rho^\text{JT}_0(E) =\text{sinh}  (2\pi \sqrt{E})/(2\pi)^2$. The emergence of a JT gravity sector in the fixed (large) spin sector of CFT${}_2$ has been previously observed and derived for other observables \cite{Ghosh:2019rcj,Pal:2023cgk}, so the lift of the correlators of the JT gravity matrix model is a natural instantiation of RMT${}_2$.}
\es{}{
\la \;\cdot\;\ra= \frac{1}{\text{Z}_\text{RMT}}\int dH \, (\,\cdot\,) \, e^{-N\,\tr(V(H))}\,.
}
We will denote the correlation functions of eigenvalue densities as
\es{}{
\rho^{(n)}(E_1,\dots, E_n)=\la \rho(E_1)\dots\rho(E_n)\ra \,,
}
and the connected correlation function of partition functions 
as
\es{}{
\text{K}_{\rm RMT}^{(n)}(S_0;{\bf y}) = \la\tr(e^{-y_1H})\dots \tr(e^{-y_n H}) \ra_c \,.
}
These can be decomposed in a genus expansion in $e^{-S_0}$,
\begin{equation}
  \text{K}_\text{RMT}^{(n)}(S_0;{\bf y}) = \sum_{g\geq 0} \text{K}_\text{RMT}^{(g,n)}({\bf y}) \, e^{(2-n-2g) S_0}\,. 
\end{equation}
For the GUE universality class, $g\in\mathbb{Z}_{\geq0}$, while in the GOE and GSE ensembles, $g$ also takes positive half-integer values.

%%%%%%%%%%%%%%%%%%%%%%%%%%%%%%%%%%%%%%%%%%%%%%%%
\subsection{Axioms}
\label{sec:axioms}
%%%%%%%%%%%%%%%%%%%%%%%%%%%%%%%%%%%%%%%%%%%%%%%%

As proposed in \cite{Boruch:2025ilr}, an RMT${}_2$ $n$-point connected form factor ${\cal Z}_{\text{RMT}_2}^{(n)}(\btau)$ is the minimal lift of a given input random matrix amplitude $\text{K}^{(n)}_\text{RMT}({\bf y})$, defined by the requirement to obey the following consistency conditions:

\begin{enumerate}
  \item {\bf Modular invariance.} The RMT${}_2$ amplitude should be modular-invariant on all $n$ tori,
  \begin{equation}
      \boxedGrey{
      {\cal Z}_{\text{RMT}_2}^{(n)}(\tau_1,\ldots,\tau_n) = {\cal Z}_{\text{RMT}_2}^{(n)}(\gamma_1\tau_1,\ldots,\gamma_n\tau_n) \,,\qquad \gamma_i \in \text{SL}(2,\mathbb{Z}) \text{ for } i=1,\ldots,n\,.
      }
  \end{equation}
and of moderate growth at the cusps.\footnote{Specifically, $|{\cal Z}_{\text{RMT}_2}^{(n)}(\tau_1,\ldots,\tau_n)| \leq C \prod_{i=1}^n y_i^A$ for $y_i\geq 1$ and some $A$. The spectral overlaps then exist as distributions in the spectral parameters, using Zagier’s renormalization of the Rankin-Selberg integral where needed \cite{zbMATH03796039}.}
\item {\bf Reduction to RMT in the near-extremal limit.} The RMT${}_2$ amplitude admits a {\it near-extremal limit} \cite{Ghosh:2019rcj,Pal:2023cgk} for every fixed spin sector ${\bf j}=(j_1,\dots, j_n)$ where all $y_i \rightarrow \infty$ uniformly (i.e., $y_i \rightarrow \gamma y_i$ and $\gamma \rightarrow \infty$). In this limit, we recover the reference RMT input in every spin sector. The component of the RMT${}_2$ amplitude with spins ${\bf j}$ on the $n$ tori obeys:
\begin{equation}
\boxedGrey{
\begin{aligned}
  {\cal Z}_{\text{RMT}_2|{\bf j}}^{(n)}({\bf y}) 
  \stackrel{\substack{\text{near-}\\ \text{extremal}}}{\longrightarrow} \; &d_{j_1}^{1-n} \, e^{-2\pi(j_1y_1+j_2y_2+\ldots)}\sqrt{y_1 \cdots y_n} \, \text{K}_\text{RMT}^{(n)}(S_0(j_1);{\bf y})  \, \delta_{j_1,j_2,\dots,j_n}\\
  &+ \text{subleading}\!
\end{aligned}
}
\label{eq:NErequirement}
\end{equation}
Let us explain the right-hand side:
The factor $d_j^{1-n}$ is simply our convention for defining the spin-${\bf j}$ component of the amplitude. The exponential factor accounts for the spinning ground state energy. The factor $\sqrt{y_1 \cdots y_n}$ is present because we are constructing an object that counts Virasoro primaries, and stripping off Virasoro descendants in the standard modular-invariant fashion (by appending a factor of $\sqrt{y_i}|\eta(\tau_i)|^2$ for each torus) introduces such a factor \cite{Benjamin:2021ygh}. The parameter $S_0(j)$ is the near-extremal entropy that generally depends on the spin $j$ and organizes the topological expansion of the underlying matrix model.
\end{enumerate}

These are the minimal constraints on RMT${}_2$ amplitudes, solely based on the required symmetry properties and physical near-extremal limit. It is a priori not obvious that it is possible to satisfy them, i.e., that a modular-invariant lift of arbitrary RMT correlators should even exist. Our first task will thus be to provide a construction of ${\cal Z}_{\text{RMT}_2}^{(n)}$ that satisfies these axioms. 

Before writing down ${\cal Z}_{\text{RMT}_2}^{(n)}$, let us make two remarks. 

First, the reader may already see why the Poincar\'e sums $P_m$ will be useful in satisfying Axiom 2, particularly for the spinning sectors ${\bf j \neq 0}$. Whereas the latter are challenging in the {\it harmonic} basis for the reasons discussed in Section \ref{s32}, the cuspidal property \eqref{eq:PmNE} of the Poincar\'e sums $P_m$ is perfectly tailored to solve that problem.

Second, let us record a further property of RMT$_2$, which turns out to follow non-trivially from Axiom 2 in our construction (see Section \ref{sec:plateau}), pertaining to the two-point spectral correlator $\text{K}^{(2)}_\text{RMT}(y_1,y_2)$, analytically continued to complex time $y_{1,2} \rightarrow \beta \pm i T$, in the limit $T \gg \beta$. Universal information about the sine kernel, including the transition from the initial linear growth (``ramp'') to the asymptotic constant ``plateau'' can be extracted perturbatively in the so-called $\uptau$-scaling limit \cite{Saad:2022kfe} (see also \cite{Haake:2010fgh,Okuyama:2020ncd,Blommaert:2022lbh,Weber:2024ieq,DiUbaldo:2025cmg}), which focuses on frequencies of the order of the level spacing:
\begin{equation}\label{tauscdef}
    T, \, e^{S_0} \rightarrow \infty\,,  \qquad \uptau = T\, e^{-S_0} \text{ fixed.}
\end{equation}
We denote the $\uptau$-scaled spectral form factor (SFF) as 
\begin{equation}
    {\cal K}^{(2)}_{\uptau\text{-RMT}}(S_0;\beta,\uptau) \equiv \lim_{\substack{T,e^{S_0}\rightarrow\infty\\ \uptau=Te^{-S_0} \text{ fix}}}  \, e^{-S_0} \, \text{K}^{(2)}_\text{RMT}\big(S_0;y_1=\beta+iT,y_2=\beta-iT\big)\,.
\end{equation}
In this limit, we will demand -- and indeed, show -- that the RMT${}_2$ two-point function reproduces the same $\uptau$-scaling limit as its underlying RMT in every spin sector:
\begin{equation}
\label{eq:axiom3}
\lim_{\substack{T,e^{S_0}\rightarrow\infty\\ \uptau=Te^{-S_0} \text{ fix}}}  \; \frac{e^{-S_0(j)}}{\sqrt{y_1y_2}} \,{\cal Z}^{(2)}_{\text{RMT}_2|j,j}(y_1,y_2)\Big|_{y_1=\beta+iT,y_2=\beta-iT}  = d_j^{-1} \, e^{-4\pi j \beta}\, {\cal K}^{(2)}_{\uptau\text{-RMT}}(S_0(j);\beta,\uptau)\,.
\end{equation}

%%%%%%%%%%%%%%%%%%%%%%%%%%%%%%%%%%%%%%%%%%%%%%%%
\subsection{Construction}
\label{sec:construction}
%%%%%%%%%%%%%%%%%%%%%%%%%%%%%%%%%%%%%%%%%%%%%%%%

We now turn to our construction of ${\cal Z}^{(n)}_{\text{RMT}_2}$ consistent with the requirements outlined above. In brief, our procedure for defining minimal RMT${}_2$ correlators is as follows. First, we determine their expansion coefficients in the Poincar\'e decomposition, called $f_{\text{RMT}|{\bf m}}^{(n)}(\bar{\bf s})$, by {\it demanding the emergence of RMT correlators in each spin sector in the near-extremal limit}. We then {\it lift the result to a modular-invariant function} by forming the corresponding linear combination of Poincar\'e series, which preserves the near-extremal structure. The expansion in the harmonic basis of Eisenstein series and cusp forms follows from a simple basis rotation. Indeed, the first axiom implies that the RMT$_2$ amplitude admits a decomposition in terms of real-analytic Eisenstein series and Maass cusp forms:
\begin{equation}
    \begin{split}
    {\cal Z}_{\text{RMT}_2}^{(n)}(\btau)  = \prod_{k=1}^n \left( \sumint_{\sGen_k}  \, \Psi_{\sGen_k}(\tau_k) \right) \, \left( {\cal Z}_{\text{RMT}_2}^{(n)},\,\Psi_{\sGen_1} \cdots \Psi_{\sGen_n}\right) , \qquad \Psi_{\sGen_i} \in \left\{ E_{\frac{1}{2}+i\omega_i},\, \phi_{n_i},\,1\right\} ,
    \end{split}
\end{equation}
with well-defined (generally distributional) spectral overlaps $\left( {\cal Z}_{\text{RMT}_2}^{(n)},\,\Psi_{\sGen_1} \cdots \Psi_{\sGen_n}\right)$. 

As preparation, and without loss of generality, let us decompose the RMT${}_2$ amplitude into different {\it base spin sectors} labeled by ${\bf m}=(m_1,\ldots,m_n)\in\mathbb{Z}^n_{\geq0}$:
\begin{equation}
\label{eq:Zdecompm}
   {\cal Z}_{\text{RMT}_2}^{(n)}(\btau) = \sum_{{\bf m \geq 0}}   {\cal Z}_{\text{RMT}_2}^{(n)}(\btau | {\bf m} ).
\end{equation}
The decomposition \eqref{eq:Zdecompm} is {\it not} the same as the Fourier decomposition \eqref{eq:ZFourierDecomp} of ${\cal Z}_{\text{RMT}_2}^{(n)}(\btau)$ into different spinning sectors ${\cal Z}_{\text{RMT}_2|{\bf j}}^{(n)}({\bf y})$ on the $n$ tori: every fixed ${\bf m}$ component ${\cal Z}_{\text{RMT}_2}^{(n)}(\btau | {\bf m} )$ is by itself modular-invariant and therefore has support over all spins ${\bf j}$. The significance of the decomposition \eqref{eq:Zdecompm} is the near-extremal limit: we will construct the RMT${}_2$ amplitude such that the near-extremal limit in the spin ${\bf j}$ spectrum is dominated by the $\bf{m}=\bf{j}$ component ${\cal Z}_{\text{RMT}_2}^{(n)}(\btau | {\bf m= j})$. Furthermore, ${\cal Z}_{\text{RMT}_2}^{(n)}(\btau | {\bf m} )$ depends on $S_0({\bf m} )$ which organizes the topological expansion of the near-extremal matrix model correlators. This  can be interpreted as the near-extremal entropy in the spectrum with spins $j_i=m_i$.\footnote{With a view towards more general applicability, our convention for distributing $e^{-S_0({\bf m} )}$ factors in \eqref{eq:Zdecompm} is slightly different from \cite{Boruch:2025ilr}.} 

In detail, the aforementioned procedure is implemented as follows:
\begin{enumerate}
  \item Start with the RMT input, i.e., a connected $n$-point RMT form factor, which is the basic input that we wish to recover near extremality,
        \begin{equation}
        \text{K}_\text{RMT}^{(n)}(S_0;{\bf y}) \equiv \left\langle \tr (e^{-y_1H}) \cdots \tr (e^{-y_n H}) \right\rangle_c \,.
        \end{equation}
  Mellin transform this quantity to get the {\it Mellin space RMT amplitude}
  \begin{equation}
  \label{eq:mellinBasic}
  \boxedGrey{
  f_{\text{RMT}}^{(n)}(S_0;\bar{\bf s} ) = 
 \int_0^\infty \left( \prod_{i=1}^n \frac{dy_i}{y_i^{1+i\omega_i}} \right) \text{K}_\text{RMT}^{(n)}(S_0;{\bf y})
 \equiv {\cal M}_{{\bf y}} \left[ \text{K}_\text{RMT}^{(n)}(S_0;{\bf y}) \right] (-i\bomega ) 
  }
  \end{equation}
  where ${\bf s}=(s_1,\ldots,s_n)$ and $\bar s_i = \tfrac{1}{2}-i\omega_i$.\footnote{When analytically continuing off the critical line, it is understood that $\bar s_i \equiv 1- s_i$.}
  \item In the Poincar\'e decomposition, we take the modular-invariant RMT${}_2$ lift to be {\it diagonal in base spin ${\bf m}$}. We define the lift by using $f_{\text{RMT}}^{(n)}(S_0;\bar{\bf s} )$ as the overlap coefficient for every sector of base spins ${\bf m}$, where the near-extremal entropy $S_0=S_0({\bf m})$ is the only parameter carrying dependence on ${\bf m}$:
  \begin{equation}
    f_{\text{RMT}|{\bf m} }^{(n)}(\bar{\bf s} ) \equiv d_{m_1}^{1-n}\,f^{(n)}_\text{RMT}(S_0(m_1);\bar{\bf s} ) \,\delta_{m_1m_2} \delta_{m_2m_3} \cdots \delta_{m_{n-1}m_n}  \,.
  \end{equation}
  The lift is then defined as 
  \begin{equation}
  \begin{split}
   {\cal Z}^{(n)}_{\text{RMT}_2} (\btau) &\equiv \sum_{{\bf m \geq 0}} \; {\cal Z}^{(n)}_{\text{RMT}_2} (\btau|{\bf m} ) \\
&\equiv \sum_{{\bf m \geq 0}}\; \int_{{\cal C}_\text{crit}^n} \left( \prod_{i=1}^n \frac{ds_i}{2\pi i} \, P_{m_i}(\tau_i,s_i) \right)_\text{sym} \, f_{\text{RMT}|{\bf m} }^{(n)}(\bar{\bf s} )  \; + \;  \text{\tt[zero modes]}\,.
\end{split}
\end{equation}
The $\text{\tt [zero modes]}$ are overlaps with the trivial basis function $\Psi_\sGen(\tau) = 1$, and ``sym.'' indicates a sum over all permutations (with combinatorial coefficient $\frac{1}{n!}$). The zero modes can be determined iteratively from $f^{(n)}_\text{RMT}$ by imposing appropriate asymptotics. In order not to distract from our main points, we discuss these in Appendix \ref{app:zeromodes}.

In short, making the base spin diagonality explicit:
    \begin{equation}
  \boxedGrey{
   {\cal Z}^{(n)}_{\text{RMT}_2} (\btau) \equiv \sum_{m\geq 0}d_m^{1-n} \int_{{\cal C}_\text{crit}^n} \left( \prod_{i=1}^n \frac{ds_i}{2\pi i} \, P_{m}(\tau_i,s_i) \right)_\text{sym} \, f^{(n)}_\text{RMT}(S_0(m);\bar{\bf s} )  \; + \;  \text{\tt[zero modes]}
   }
   \label{eq:RMT2defPmRep}
  \end{equation}
This construction ensures that the input amplitude $\text{K}_\text{RMT}^{(n)}$ is recovered near extremality for {\it every} spin: the Poincar\'e sums in this limit simply invert the Mellin transform. Indeed, the term involving $P_{m}\cdots P_{m}$ dominates the near-extremal behavior in the spin sector with spins $j_1=\ldots=j_n=m$ and it naturally exhibits the correct power of $e^{-S_0({\bf m} )}$. We check this in more detail below.
\end{enumerate}

Having found one solution to the RMT${}_2$ axioms, it is immediately clear that this solution cannot be unique: there are infinitely many terms one could add to \eqref{eq:RMT2defPmRep} that are modular-invariant but subleading in the near-extremal limit. This is to be expected, in analogy to the statement in quantum mechanics that the early-time behavior of the spectral form factor of any particular system is not universal, even if the late-time behavior is described by RMT. We will focus on the prescription \eqref{eq:RMT2defPmRep}, which is `minimal' in the Poincar\'e sum basis, and comment further on the associated ambiguities in Section \ref{ssec:minimality}.

%%%%%%%%%%%%%%%%%%%%%%%%%%%%%%%%%%%%%%%%%%%%%%%%
\subsubsection{Passage to harmonic basis}
%%%%%%%%%%%%%%%%%%%%%%%%%%%%%%%%%%%%%%%%%%%%%%%%
Having defined the spinning RMT${}_2$ amplitude this way, we can work out the overlaps in the harmonic basis of Eisenstein series and cusp forms -- collectively denoted as $\Psi_\sGen \in \{E_{\frac{1}{2}+i\omega},\phi_n,1\}$.  For the ${\bf m=0}$ sector, the definition \eqref{eq:RMT2defPmRep} and the fact that $P_{0}(\tau,s)=E_s(\tau)$ immediately give
\begin{equation}
    \boxedGrey{
    \left(  {\cal Z}_{\text{RMT}_2}^{(n)} ,\, E_{s_1} \ldots E_{s_n}  \right)_{m=0}
     =   f^{(n)}_\text{RMT}(S_0(0);\bar{\bf s} )  
     + [\text{reflections } s_i \!\rightarrow\! 1-s_i]
     }
     \label{eq:m0sector}
\end{equation}
and vanishing overlaps with cusp forms. There are a total of $2^n$ terms, accounting for all ``reflections'' $s_i\rightarrow 1-s_i$ with coefficients $\varphi(1-s_i)$, which ensure that the overlap is consistent with the functional equation $E_{1-s}(\tau) = \varphi(1-s) E_{s}(\tau)$.

For ${\bf m\neq 0}$, we find that the definition \eqref{eq:RMT2defPmRep} has overlaps with harmonic basis functions given via {\it (multiple) Kontorovich-Lebedev (KL) transforms:}\footnote{Note that, for every $\Psi_{\sGen_i}$ that is an Eisenstein series $E_{s_i}$, the expression \eqref{eq:spec11} satisfies the functional equation $E_{1-s_i}(\tau) = \varphi(1-s_i) E_{s_i}(\tau)$. The reason is that the Eisenstein Fourier coefficients satisfy $\bar{\tt a}_m^{(\sGen_i)}=\varphi(1-\sGen_i) \bar{\tt a}_m^{(1-\sGen_i)}$.}
\begin{equation}
\label{eq:spec11}
\left(  {\cal Z}_{\text{RMT}_2}^{(n)} ,\, \Psi_{\sGen_1} \ldots \Psi_{\sGen_n}  \right)_{\mpos}
  = \sum_{{\bf m >0}} d_{m_1}^{1-n}\, \bar{\tt c}_{m_1}^{(\sGen_1)}\cdots\bar{\tt c}_{m_n}^{(\sGen_n)}\, 
   h_{\text{RMT}|{\bf m} }^{(n)}(\sGenb )
\end{equation}
with
\begin{equation}
h_{\text{RMT}|{\bf m} }^{(n)}(\sGenb )  = d_{m_1}^n\,
  \delta_{m_1 \dots m_n} 
 \int_0^\infty \left( \prod_{i=1}^n  \frac{dy_i}{y_i}\, e^{-2\pi m_i y_i}\, K_{\sGen_i-\frac{1}{2}}(2\pi m_iy_i) \right)\text{K}_\text{RMT}^{(n)}(S_0(m_1);{\bf y}) 
 \,.
\label{eq:KLgeneral}
\end{equation}
Equivalently, for every diagonal base spin $m>0$, we get an $n$-fold KL transform of the RMT input: 
\begin{equation}
\label{eq:spec11Rep}
\boxedGrey{
\begin{aligned}
  &\left(  {\cal Z}_{\text{RMT}_2}^{(n)} ,\, \Psi_{\sGen_1} \ldots \Psi_{\sGen_n}  \right)_{\mpos}
  \\
  &\quad = \sum_{m > 0} d_m\,\bar{\tt{c}}_{m}^{(\sGen_1)}\cdots \bar{\tt c}_{m}^{(\sGen_n)} \int_0^\infty \left( \prod_{i=1}^n  \frac{dy_i}{y_i}\, e^{-2\pi m y_i}\, K_{\sGen_i-\frac{1}{2}}(2\pi my_i) \right)\text{K}_\text{RMT}^{(n)}(S_0(m);{\bf y})
\end{aligned}
}
\end{equation}
These overlaps also fully determine ${\cal Z}^{(n)}_{\text{RMT}_2}(\btau)$ via the spectral decomposition \eqref{eq:specD}. We recall the overlaps of the Poincar\'e series with the harmonic basis elements given in \eqref{eq:goldfeldOverlaps}.

%%%%%%%%%%%%%%%%%%%%%%%%%%%%%%%%%%%%%%%%%%%%%%%%
\subsection{Further explanation and general checks}
%%%%%%%%%%%%%%%%%%%%%%%%%%%%%%%%%%%%%%%%%%%%%%%%

We will now explain the construction from the previous section in more detail. We discuss ${\bf m=0}$ and ${\bf m \neq 0}$ separately.

\paragraph{Scalar RMT${}_2$ (${\bf m=0}$).}
In \cite{Boruch:2025ilr}, the ${\bf m=0}$ sector \eqref{eq:m0sector} was already described. We refer to this as {\it scalar RMT${}_2$} because it is determined by the scalar near-extremal behavior. However, the scalar RMT${}_2$ amplitude of course has support in all spins, as required by modular invariance. In short, the scalar amplitude is of the form\footnote{The measure here is $\frac{1}{2\pi i}$ instead of the $\frac{1}{4\pi i}$ that appears in the usual spectral decomposition. The reason is that the Mellin amplitude $f^{(n)}_\text{RMT}(S_0;\bar{{\bf s}})$ is not reflection symmetric under $s_i \rightarrow 1-s_i$. When integrated against Eisenstein series, each of the $2^n$ reflections in \eqref{eq:m0sector} contributes equally to \eqref{eq:scalarRMT2def}.}
\begin{equation}
\begin{split}
{\cal Z}^{(n)}_{\text{RMT}_2}(\btau | {\bf m=0}) &= \int_{{\cal C}_\text{crit}^n}\left( \prod_{i=1}^n\frac{ds_i}{2\pi i} \, E_{s_i}(\tau_i) \right)_\text{sym} f^{(n)}_\text{RMT}(S_0(0);\bar{\bf s} ) + \text{\tt [zero modes]} \,,
\end{split}
\label{eq:scalarRMT2def}
\end{equation}
where `sym' denotes symmetrization in the moduli $\btau$, and the coefficients $f^{(n)}_\text{RMT}$ are Mellin transforms of the RMT input with respect to the inverse temperatures ${\bf y}$. 

Let us briefly give an intuitive argument why this definition satisfies the near-extremal limit \eqref{eq:NErequirement}. By definition, the Eisenstein series can be written as Poincar\'e sums, whose near-extremal limit was discussed in \eqref{eq:PmNE}:
\begin{equation}
\begin{split}
 E_{s_i}(\tau_i) \equiv P_{m=0}(\tau_i,s_i) \, &\stackrel{\substack{\text{near-}\\\text{extremal}}}{\longrightarrow} \,  y_i^{s_i} + \varphi(s_i) \, y_i^{1-s_i}  + \ldots
 \\
 &\quad\;\;\; = \sqrt{y_i} \left( y_i^{i\omega_i} + \varphi(s_i) y_i^{-i\omega_i} \right) + \ldots 
 \end{split}
 \end{equation}
 where `$\ldots$' denotes subleading terms in the large $y_i$ limit.
This simplified form turns \eqref{eq:scalarRMT2def} into an inverse Mellin transform, which thus reproduces the RMT input by definition:
\begin{equation}
    \begin{split}
    &{\cal Z}^{(n)}_{\text{RMT}_2}(\btau | {\bf m = 0}) 
    \\
    &\quad \stackrel{\substack{\text{near-}\\\text{extremal}}}{\longrightarrow} \, 
    \sqrt{y_1\cdots y_n}\int_{\mathbb{R}^n} \left( \prod_{i=1}^n \frac{d\omega_i}{2\pi} \, \left( y_i^{i\omega_i} + \varphi(s_i) y_i^{-i\omega_i} + \ldots \right)\right)_\text{sym} f^{(n)}_\text{RMT}(S_0(0);\bar{\bf s}=\tfrac{1}{2}-i\bomega)
    \\
    &\qquad\;\;\; = \sqrt{y_1\cdots y_n} \, \text{K}_\text{RMT}^{(n)}(S_0(0);{\bf y}) + \ldots \,.
    \end{split}
\end{equation}
For reasonable overlaps $f^{(n)}_\text{RMT}$, the `reflected' term involving $\varphi(s_i)$ oscillates rapidly along the integration contour and is typically subleading, thus producing the last line.\footnote{A more precise argument is as follows (see, e.g., \cite{Collier:2022emf}): the $\omega_i$ integrals can be performed by contour deformation, picking up poles of $f^{(n)}_\text{RMT}$. The reflected term involving $\varphi(s_i)=\frac{\Lambda(1-s_i)}{\Lambda(s_i)} = \pi^{2s_i-1} \, \frac{\Gamma(1-s_i)\zeta(2-2s_i)}{\Gamma(s_i)\zeta(2s_i)}$ also contributes to this procedure, but it leads to powerlaw corrections to the leading term, involving negative powers of $y$. These are important to ensure modular-invariance, but they are subleading near extremality.} The kernel involving factors $y_i^{i\omega_i}$ is then nothing but the kernel for an inverse Mellin transform. In special cases products of reflected terms can give enhanced contributions, see, in particular, \eqref{eq:plus2ramp}.

\paragraph{Spinning RMT${}_2$ (${\bf m>0}$), explicit case of $n=1$.}
Because the spin-$m$ Poincar\'e sum $P_m$ dominates over all other Poincar\'e sums $P_{m' \neq m}$ as we take the near-extremal limit in the spin $j=m$ sector, the Poincar\'e sums furnish an orthogonal basis {\it in the near extremal limit}, and therefore their overlaps can be determined unambiguously in that limit. Further, the power-law scaling in \eqref{eq:PmNE} shows that the integral along the critical line of Poincar\'e series $P_m$ in the near-extremal limit inverts the Mellin transform, completely analogous to the discussion of Eisenstein series above.

To illustrate the construction explicitly, let us consider the simplest case $n=1$. Combining all elements of the basis, our proposal amounts to a decomposition of the form
\begin{equation}
\begin{split}
  {\cal Z}^{(1)}_{\text{RMT}_2}(\tau)
  &= \sum_{m\geq 0} \int_{{\cal C}_\text{crit}} \frac{ds}{2\pi i}\, P_m(\tau,s)  \, f_{\text{RMT}|m}^{(1)}(\bar s) 
  \\
  &= \int_{{\cal C}_\text{crit}} \frac{ds}{2\pi i} \left\{  E_{s}(\tau)\,f^{(1)}_{\text{RMT}|0}(\bar s) + \sum_{m\geq 1}  P_{m}(\tau,s)\,f^{(1)}_{\text{RMT}|m}(\bar s)
  \right\}\,.
  \end{split}
  \label{eq:Z1example}
\end{equation}
We already reviewed the coefficients $f^{(1)}_{\text{RMT}|0}(\bar s)$: they are simply Mellin transforms of the RMT input correlator with a complex conjugate argument:
\begin{equation}
 f^{(1)}_{\text{RMT}|0}\left(\bar s=\tfrac{1}{2}-i\omega\right) = {\cal M}_y \left[ \text{K}_\text{RMT}^{(1)}(S_0;y) \right] (-i\omega) \equiv f_\text{RMT}^{(1)}(S_0;\bar{s})  \,.
\end{equation}
We will see that the amplitudes $f_\text{RMT}^{(1)}(\bar{s})$ summarize the RMT input in an interesting way, thus leading to the structure of {\it `RMT in Mellin space'}, which we expand on in Appendix \ref{app:RMTmellin}.

Consider the near-extremal limit $y\rightarrow \infty$ of \eqref{eq:Z1example} in the spin $j>0$ sector. As shown in \eqref{eq:PmNE}, this limit is dominated by $P_{m=j}(\tau,s)$ with all other terms being subleading:
\begin{equation}
    {\cal Z}^{(1)}_{\text{RMT}_2|j>0}(y) \stackrel{\substack{\text{near-}\\\text{extremal}}}{\longrightarrow} \int_{{\cal C}_\text{crit}} \frac{ds}{2\pi i} \left\{ y^s \, e^{-2\pi j y} \, f^{(1)}_{\text{RMT}|j}(\bar s) + \text{subleading} \right\} \,.
    \label{eq:Zjgtr0}
\end{equation}
Since $s\in\frac{1}{2}+i\mathbb{R}$, this is nothing but ($\sqrt{y}$ times) the inverse Mellin transform, so we recover the RMT input:
\begin{equation}
\begin{split}
    \eqref{eq:Zjgtr0} \; &=  e^{-2\pi j y} \, \sqrt{y}\, {\cal M}^{-1}_{-i\omega}\left[ f^{(1)}_{\text{RMT}|j}(\bar s) \right](y) + \text{subleading}
    \\
    &= e^{-2\pi j y} \, \sqrt{y} \, \text{K}^{(1)}_\text{RMT}(y) + \text{subleading} \,.
    \end{split}
\end{equation}
The near-extremal limit is thus reproduced correctly for all spins.

Finally, in the harmonic basis, expanding ${\cal Z}^{(1)}_{\text{RMT}_2}$ in Poincar\'e sums and using the formula for the overlaps, we have for fixed $m>0$
\begin{align}
  \left( {\cal Z}^{(1)}_{\text{RMT}_2}(\,\cdot \, | m),\, \Psi_\sGen\right)
  &= \int_{{\cal C}_\text{crit}} \frac{ds}{2\pi i} \,  f_{\text{RMT}|m}^{(1)}(\bar s)\,\left( P_{m}(\,\cdot\,,s) ,\,\Psi_\sGen\right) \notag\\
  &= \bar{\tt c}_m^{(\sGen)}\int_{{\cal C}_\text{crit}} \frac{ds}{2\pi i} \, f_{\text{RMT}|m}^{(1)}(\bar s)\, \frac{2\sqrt{\pi}}{(4\pi m)^{s-\frac{1}{2}}}\;
     \frac{\Gamma\!\left(s-\sGen\right)\Gamma\!\left(s+\sGen-1\right)}{\Gamma(s)} \,.
\end{align}
We then use an integral representation for the Gamma functions (see \eqref{eq:convolutionKernel} below) to get:
\begin{equation}
  \left( {\cal Z}^{(1)}_{\text{RMT}_2}(\,\cdot \, | m),\,\Psi_{\sGen}\right)
  = 2\,\bar{\tt c}_m^{(\sGen)}\int_{{\cal C}_\text{crit}} \frac{ds}{2\pi i}\,f^{(1)}_{\text{RMT}|m}(\bar s)
    \int_0^\infty dy\; y^{s-\frac{3}{2}}\, e^{-2\pi m y}\, K_{\sGen-\frac{1}{2}}(2\pi m y)\,.
\end{equation}
The integral over $s$ can be done first, and it is an inverse Mellin transform, provided the critical line ${\cal C}_\text{crit} = \frac{1}{2} + i\mathbb{R}$ lies within the fundamental strip of the Mellin transform. Using the standard definition for the inverse Mellin transform,
\begin{equation}
  \int_{{\cal C}_\text{crit}} \frac{ds}{2\pi i}\; f^{(1)}_{\text{RMT}|m}\left( \bar s \right)\, y^{s-\frac{1}{2}}
  =  {\cal M}^{-1}_{-i\omega}\left[ f_{\text{RMT}|m}^{(1)} \right](y) 
  = \text{K}^{(1)}_{\text{RMT}}(y)
   \,,
\end{equation}
where $\text{K}^{(1)}_{\text{RMT}}(y) = \langle \tr (e^{-y H}) \rangle$ is the RMT one-point function, i.e., the partition function.
Using this, we find:
\begin{equation}\label{eq:UF}
  \left( {\cal Z}^{(1)}_{\text{RMT}_2}(\,\cdot \, | m)\,,\,\Psi_{\sGen}\right)
  =  2\,\bar{\tt c}_m^{(\sGen)} \int_0^\infty \frac{dy}{y}\,
    e^{-2\pi my}\, K_{\sGen-\frac{1}{2}}(2\pi my)\;\text{K}^{(1)}_{\text{RMT}}(y)\,.
\end{equation}
This formula (Kontorovich-Lebedev transform) provides the relation between the spectral decomposition and the expansion of the RMT${}_2$ amplitude in terms of Poincar\'e sums.  It immediately generalizes to $n>1$, which we summarized in \eqref{eq:spec11Rep}.

\paragraph{Relation between Mellin and Kontorovich-Lebedev transform.}
From the above derivation, it is clear that the Mellin amplitudes \eqref{eq:mellinBasic} and the KL transform \eqref{eq:KLgeneral} are closely related. Note that we can write the latter in terms of the former via a convolution integral:
\begin{equation}
 h^{(n)}_{\text{RMT}|{\bf m}} (\sGenb=\tfrac{1}{2}+i\bomega) = d_{m_1}^n \delta_{m_1m_2} \cdots \delta_{m_{n-1}m_n} \int_{(\mathbb{R}-i0)^n} \left(\prod_{i=1}^n \frac{d\nu_i}{2\pi} \, {\cal B}_{m_i}(\omega_i,\nu_i)\right) f^{(n)}_{\text{RMT}}(S_0(m_1);\bar{{\bf s}} = \tfrac{1}{2}-i\bnu)  
\end{equation}
where the universal convolution kernel is given by
\begin{equation}
\label{eq:convolutionKernel}
    \begin{split}
        {\cal B}_m(\omega,\nu) &= \int_0^\infty \frac{dy}{y}\, y^{i\nu} \, e^{-2\pi m y} \, K_{i\omega}(2\pi m y) 
        = \frac{\sqrt{\pi}}{(4\pi m)^{i\nu}} \frac{\Gamma(i(\nu-\omega))\Gamma(i(\nu+\omega))}{\Gamma(\frac{1}{2}+i\nu)} \,.
    \end{split}
\end{equation}
This shows that the ${\bf m>0}$ overlaps (which include all cusp form overlaps) are obtained from the ${\bf m=0}$ Eisenstein base sector by an integral convolution of the Mellin amplitude. We can hence view the Mellin space correlator $f^{(n)}_{\text{RMT}}(S_0;\bar{\bf s})$ as the fundamental building block, required to build the RMT${}_2$ lift.  In Appendix \ref{app:RMTmellin} we develop the Mellin space formalism in detail for general RMT correlators.

%%%%%%%%%%%%%%%%%%%%%%%%%%%%%%%%%%%%%%%%%%%%%%%%
%%%%%%%%%%%%%%%%%%%%%%%%%%%%%%%%%%%%%%%%%%%%%%%%
\section{Airy RMT${}_2$ and higher-genus amplitudes}
\label{sec:airyRMT2}
%%%%%%%%%%%%%%%%%%%%%%%%%%%%%%%%%%%%%%%%%%%%%%%%
%%%%%%%%%%%%%%%%%%%%%%%%%%%%%%%%%%%%%%%%%%%%%%%%

In this section we discuss the topological expansion of the simplest RMT${}_2$: the uplift of the Airy model with spectral curve
\begin{equation}
\label{eq:rhoAiry}
    \rho^{\text{Airy}}(E) = \rho_0^{\text{Airy}}(E) \, e^{S_0} \,,\qquad \rho_0^{\text{Airy}}(E) = \frac{1}{2\pi} \, \sqrt{E} \,.
\end{equation}
We discuss both the GUE and the GOE universality classes, which present qualitatively different features.\footnote{The GSE case is very similar to GOE and presents no qualitatively new ingredients.}

%%%%%%%%%%%%%%%%%%%%%%%%%%%%%%%%%%%%%%%%%%%%%%%%
\subsection{GUE Airy RMT${}_2$: general structure}
\label{sec:GUEairyRMT2}
%%%%%%%%%%%%%%%%%%%%%%%%%%%%%%%%%%%%%%%%%%%%%%%%

The two-point RMT spectral form factor for the Airy model in the GUE takes the following form \cite{okounkov,Saad:2022kfe,Okuyama:2023pio}:
\begin{equation}
    \text{K}_\text{Airy}^{(2)}(S_0;y_1,y_2) = \frac{e^{S_0+\frac{8\beta^3}{3} e^{-2S_0}}}{4\sqrt{\pi} \, (2\beta)^{3/2}} \,\text{Erf} \left( e^{-S_0} \sqrt{2\beta(\beta^2+T^2)}\,\right) \,,\qquad \beta=\frac{y_1+y_2}{2} \,,\; T = \frac{y_1-y_2}{2i}\,.
\end{equation}
In the following, we will ignore the doubly exponential prefactor, which adds computational complexity and leaves the physics in most regimes unaffected; for instance, the doubly exponential term can be rigorously dropped in the $\uptau$-scaling limit $T\rightarrow \infty$ with $\uptau=Te^{-S_0}$ held fixed. We therefore focus on the $\uptau$-scaled form factor, defined as
\begin{equation}
\begin{split}
    {\cal K}_{\uptau\text{-Airy}}^{(2)}(\beta,\uptau) &= 
    \lim_{\substack {T,S_0 \rightarrow \infty \\ \uptau = T e^{-S_0} \text{ fix}}}  \, e^{-S_0}\, \text{K}_{\text{Airy}}^{(2)}\big(S_0;y_1=\beta+iT,y_2=\beta-iT\big) \\
    &=
    \frac{1}{4\sqrt{\pi} \, (2\beta)^{3/2}} \,\text{Erf} \left(  \sqrt{2\beta}\, \uptau\,\right) \,.
\end{split}
\label{eq:tauAiryGUE}
\end{equation}
While this quantity is finite and admits a convergent expansion in powers of $\uptau$, for notational consistency with the rest of this work, we define the following function of $y_{1,2}$ and $S_0$:
\begin{equation}
\begin{split}
    \text{K}_{\uptau\text{-Airy}}^{(2)}(S_0;y_1,y_2) &\equiv e^{S_0} \, {\cal K}_{\uptau\text{-Airy}}^{(2)}\left(\beta=\tfrac{y_1+y_2}{2},\uptau=e^{-S_0}\sqrt{y_1y_2}\right) 
    \\
    &=
 \sum_{g=0}^\infty a_g \, e^{-2gS_0}(y_1+y_2)^{g-1} \, (y_1y_2)^{g+\frac{1}{2}} \,,\qquad a_g = \frac{(-1)^g}{2\pi(2g+1)g!}\,.
\end{split}
\label{eq:tauAiryGUE2}
\end{equation}
This identification of $\beta$ and $\uptau$ is not unique in the $\uptau$-scaling limit. While the relation $\beta = \frac{y_1+y_2}{2}$ is exact, the identification of $\uptau$ receives corrections, which do however vanish in the $\uptau$-scaling limit. Indeed, the exact relation $y_{1,2}=\beta\pm i T$ implies
\begin{equation}
    e^{-S_0}\sqrt{y_1y_2} = \uptau + \frac{\beta^2}{2\uptau} \, e^{-2S_0} + {\cal O}\big( \uptau^{-3} e^{-4S_0} \big) \,.
\end{equation}
This is such that all corrections to the first term are manifestly subleading as $S_0\rightarrow \infty$ with $\beta$ and $\uptau$ held fixed.

%%%%%%%%%%%%%%%%%%%%%%%%%%%%%%%%%%%%%%%%%%%%%%%%
\subsubsection{Definition of the RMT${}_2$ lift}
%%%%%%%%%%%%%%%%%%%%%%%%%%%%%%%%%%%%%%%%%%%%%%%%

In the RMT${}_2$ uplift of the $\uptau$-scaled Airy model, ${\cal Z}^{(2)}_{\uptau\text{-Airy}}(\tau_1,\tau_2)$, we require that every spin sector independently reduces to the RMT spectral correlator \eqref{eq:tauAiryGUE} near extremality. However, we allow for a different spin-dependent weight in each sector. This means that the spin-$(j_1,j_2)$ sector near extremality should approach the following input:
\begin{equation}
\begin{split}
{\cal Z}^{(2)}_{\uptau\text{-Airy}|j_1,j_2}(y_1,y_2) &\;\stackrel{\substack{\text{near-}\\\text{extremal}}}{\longrightarrow}\; d_{j_1}^{-1} \, e^{-2\pi j_1(y_1+y_2)}\, \sqrt{y_1y_2}\, \text{K}^{(2)}_{\uptau\text{-Airy}}(S_0(j_1);y_1,y_2)   \, \delta_{j_1j_2}\,.
\end{split}
\label{eq:K2airy}
\end{equation}
This condition encapsulates the full universal sine kernel for the Airy spectral curve near extremality in all spin sectors of the modular-invariant spectrum.

The Mellin transform of the RMT input was already computed in \cite{Boruch:2025ilr} for the $m=0$ sector. The calculation is identical for all other sectors, so we find:
\beq
\label{eq:tauAiryf}
\boxed{\begin{aligned}
f^{(2)}_{\uptau\text{-Airy}}(S_0(m);\bar s_1,\bar s_2) &\equiv  {\cal M}_{y_1,y_2}\left[ \text{K}^{(2)}_{\uptau\text{-Airy}}(S_0(m);y_1,y_2)\right](-i\omega_1,-i\omega_2)
 \\
 &=\frac{e^{-2i \omega_+S_0(m)}}{12\pi (\omega_++i\epsilon)(\omega_+ - \frac{i}{2})} 
 \,\Gamma\left(\frac{1}{2}-\frac{i \omega_+}{2}+\frac{3i \omega_-}{2}\right)
 \Gamma\left(\frac{1}{2}-\frac{i \omega_+}{2}-\frac{3i \omega_-}{2}\right)
\end{aligned}
}
\eeq 
where $\epsilon \rightarrow 0$ is a regulator of the genus $g=0$ pole. This expression also directly follows from the more general discussion in Appendix \ref{app:mellinGeneralRho}, see the discussion around \eqref{eq:f2MellinGeneralFinal}. According to our general prescription, the Airy RMT${}_2$ is thus defined as 
\begin{equation}
\label{eq:ZtauAiryResult}
    {\cal Z}^{(2)}_{\uptau\text{-Airy}}(\tau_1,\tau_2) = \sum_{m \geq 0} d_m^{-1}  \int_{{\cal C}_\text{crit}^2} \frac{ds_1 ds_2}{(2\pi i)^2} \, \left[ P_{m}(\tau_1,s_1)P_{m}(\tau_2,s_2) \right]_\text{sym} \, f^{(2)}_{\uptau\text{-Airy}}(S_0(m);\bar s_1,\bar s_2) \,.
\end{equation}
We proceed to deconstruct this expression.

\paragraph{Genus zero pole.}
The integration in \eqref{eq:ZtauAiryResult}  along $s_{i} \equiv \frac{1}{2}+i\omega_{i} \in \frac{1}{2}+i\mathbb{R}$ can be written in terms of real integrals over $\omega_\pm = \frac{1}{3}(\omega_1\pm \omega_2)$, with
\begin{equation}
    \int_{{\cal C}_\text{crit}^2} \frac{ds_1 ds_2}{(2\pi i)^2} \, (\, \cdot \, ) = \frac{9}{2} \int_{\mathbb{R}^2} \frac{d\omega_+ d\omega_-}{(2\pi)^2} \, (\,\cdot\,) \,.
\end{equation}
Consider the genus zero pole of \eqref{eq:tauAiryf} at $\omega_+=-i\epsilon \rightarrow 0$. The contribution of this pole is most easily recognized by closing the $\omega_+$ contour in \eqref{eq:ZtauAiryResult}  first. Closing it in the lower half complex plane (as dictated by the exponential $e^{-2i\omega_+ S_0}$), this pole contributes as follows:
\begin{equation}
\begin{split}
    {\cal Z}^{(2)}_{\uptau\text{-Airy}}(\tau_1,\tau_2)\Big|_{\text{pole }\omega_+=-i\epsilon} 
     &= \sum_{m \geq 0} d_m^{-1}  \int_{{\cal C}_\text{crit}} \frac{ds}{2\pi i}  \frac{1}{2  \cosh(\pi \omega)} \, \big[P_m(\tau_1,\tfrac{1}{2}+i\omega) P_m(\tau_2,\tfrac{1}{2}-i\omega) \big]_\text{sym}
     \\
     &= {\cal Z}^{(0,2)}_{\text{RMT}_2}(\tau_1,\tau_2) \,,
    \end{split}
\end{equation}
where we rescaled $\omega_- = \frac{2}{3}\omega$ and rediscovered the universal $(g,n)=(0,2)$ expression \eqref{eq:CJpoincare}.

\paragraph{Evaluation of the contour integrals \eqref{eq:ZtauAiryResult}.}
Consider now the remaining poles in \eqref{eq:ZtauAiryResult}. We could again perform the $\omega_+$ integral first and the $\omega_-$ integral second. We leave this as an exercise, as it turns out to be more insightful to proceed in the opposite way. Take a sector with fixed $m>0$ of \eqref{eq:ZtauAiryResult} and perform the integral over $\omega_-$ first. The gamma functions in \eqref{eq:tauAiryf} have poles at
\begin{equation}
    \omega_- \text{ poles:} \qquad 
    \left\{ \begin{aligned} &\omega_-^{(k_1)} = \frac{\omega_+}{3} + \frac{(2k_1+1)i}{3} \qquad\;\;\, (k_1=0,1,2,\ldots)\\
    & \omega_-^{(k_2)} = -\frac{\omega_+}{3} - \frac{(2k_2+1)i}{3} \qquad (k_2=0,1,2,\ldots) \end{aligned}\right.
\end{equation}
For real $\omega_+$, the first series of poles lies in the upper half plane and the second series in the lower half plane.
Whether the $\omega_-$ contour needs to be closed in the upper or lower half plane is dictated by the relative size of $y_1$ and $y_2$ (recall that $P_m(\tau,s)$ is at large $y$ dominated by $P_{m|j=m}(\tau,s) \sim y^s e^{-2\pi m y}$). Without loss of generality, consider closure in the upper half plane, which picks up the first series of poles, resulting in\footnote{The $\omega_-$ contour also encounters simple poles of the Poincar\'e series. One can check that the subsequent $\omega_+$ integration contains residues that cancel these additional contributions. For simplicity, we leave this implicit and omit the Poincar\'e series' poles entirely.}
\begin{equation}
\label{eq:Z2Airyomegaplus}
\begin{split}
    {\cal Z}^{(2)}_{\uptau\text{-Airy}}(\tau_1,\tau_2) 
    &=  \sum_{m\geq 0} \frac{3}{2\pi d_m}\sum_{k\geq 0} \int_{\mathbb{R}} d\omega_+\; \left[ P_m\left(\tau_1,2i\omega_+-k\right)P_m\left(\tau_2,i\omega_++k+1\right) \right]_\text{sym}
    \\
    &\qquad\qquad\qquad  \times \frac{e^{-2i\omega_+ S_0(m)}}{6\pi (\omega_++i\epsilon)(2\omega_+-i)} \frac{(-1)^{k}\,\Gamma\left(1+k-i\omega_+ \right)}{k!}\,.
\end{split}
\end{equation}
Due to the factor $e^{-2i\omega_+ S_0(m)}$, it is necessary for convergence to close the remaining integral in the lower half plane. This picks up a simple genus-zero pole at $\omega_+=-i\epsilon \rightarrow 0$, and a series of poles due to the gamma function at $\omega_+^{(\ell)} = -(k+\ell+1)i$ for $\ell = 0,1,2,\ldots$. This gives:
\begin{equation}
\begin{split}
    &{\cal Z}^{(2)}_{\uptau\text{-Airy}}(\tau_1,\tau_2) 
    = {\cal Z}_{\text{RMT}_2}^{(0,2)}(\tau_1,\tau_2)
    \\
    &\quad + \sum_{m\geq 0} \frac{1}{2\pi d_m} \sum_{k,\ell\geq 0}  \left[ P_m\left(\tau_1,2+k+2\ell\right)P_m\left(\tau_2,2+2k+\ell\right) \right]_\text{sym} \frac{e^{-2(k+\ell+1) S_0(m)}\,(-1)^{k+\ell+1}}{ (k+\ell+1)(2k+2\ell+3)k!\ell!}
\end{split}
\end{equation}
where ${\cal Z}_{\text{RMT}_2}^{(0,2)}(\tau_1,\tau_2)$ is the universal (spectral curve-independent) genus-zero contribution due to the pole $\omega_+ = -i\epsilon \rightarrow 0$ (see above).
In the second line, we relabel $k+\ell+1 \equiv g$, which gives the final expression for the Airy RMT${}_2$ amplitude:
\begin{equation}
\boxed{
\begin{split}
    &{\cal Z}^{(2)}_{\uptau\text{-Airy}}(\tau_1,\tau_2) 
    = {\cal Z}_{\text{RMT}_2}^{(0,2)}(\tau_1,\tau_2)
    \\
    &\quad + \sum_{m\geq 0} d_m^{-1} \sum_{g\geq 1}a_g\, e^{-2g S_0(m)}\sum_{k= 0}^{g-1} \binom{g-1}{k} \left[ P_m\left(\tau_1,2g-k\right)P_m\left(\tau_2,g+k+1\right) \right]_\text{sym} 
\end{split}
}
\label{eq:GUEAiryFinal}
\end{equation}
with the same coefficients $a_g$ as in \eqref{eq:tauAiryGUE2}. We observe that each base spin sector $m$ admits its own genus expansion. For every genus $g$, there appears a finite number of products of Poincar\'e sums with integer-valued labels $s$.

%%%%%%%%%%%%%%%%%%%%%%%%%%%%%%%%%%%%%%%%%%%%%%%%
\subsubsection{Spectral decomposition}
\label{sec:GUEairySpectral}
%%%%%%%%%%%%%%%%%%%%%%%%%%%%%%%%%%%%%%%%%%%%%%%%

The expression \eqref{eq:GUEAiryFinal} determines the Airy RMT${}_2$ in the Poincare basis. Now we wish to study its spectral decomposition in terms of Eisenstein series and cusp forms. The spectral decomposition of the genus zero contribution ${\cal Z}^{(0,2)}_{\text{RMT}_2}(\tau_1,\tau_2)$, in the first line of \eqref{eq:GUEAiryFinal}, is worked out in Appendix~\ref{sec:CJnewderivation} and given in \eqref{eq:CJoverlaps}; as shown in \cite{DiUbaldo:2023qli}, it is equal to the two-boundary torus wormhole in AdS$_3$ pure gravity \cite{Cotler:2020ugk}. The rest of this section concerns the $g>0$ terms of \eqref{eq:GUEAiryFinal}.

\paragraph{Eisenstein ($m=0$) sector.}
The $m=0$ sector of Airy RMT${}_2$ was discussed in \cite{Boruch:2025ilr}. For this sector, it was found:
\beq
\label{eq:ZRMT2Eisensteins}
\begin{split}
&
{\cal Z}^{(g,2)}_{\uptau\text{-Airy}}(\tau_1,\tau_2|0,0)
=   a_g \,e^{-2gS_0} \sum_{k= 0}^{g-1} \binom{g-1}{k}\,\left[E_{2g-k}(\tau_1) E_{g+k+1}(\tau_2)\right]_\text{sym} \qquad (g>0)\,,
\end{split}
\eeq
which simply follows from \eqref{eq:GUEAiryFinal} by setting $m=0$ and using $P_{0}(\tau,s) = E_s(\tau)$. Note that all integer-index Eisenstein series $E_n$ in this result are consistent with the convergence condition $\text{Re}(n)>1$.

\paragraph{Spinning ($m>0$) sectors.}
To obtain the spectral overlaps for the genus $g$ Airy spectral correlator, we perform a double Kontorovich-Lebedev (KL) transform of the RMT input \eqref{eq:K2airy}, following \eqref{eq:KLgeneral}: 
\begin{equation}
\begin{split}
  &h_{\uptau\text{-Airy}|m_1,m_2}^{(g,2)}(\sGen_1,\sGen_2) \\
  & =4\delta_{m_1m_2}\int_0^\infty \frac{dy_1}{y_1}\, K_{\sGen_1-\frac{1}{2}}(2\pi m_1 y_1)
     \int_0^\infty \frac{dy_2}{y_2}\, K_{\sGen_2-\frac{1}{2}}(2\pi m_2 y_2)   e^{-2\pi (m_1 y_1+m_2 y_2)}\, \text{K}^{(g,2)}_{\uptau\text{-Airy}}(S_0(m_1);y_1,y_2)  \,.
\end{split}
\end{equation}
We can evaluate the overlaps by using the binomial expansion,
\es{}{
(y_1+y_2)^{g-1}= \sum_{k=0}^{g-1}\binom{g-1}{k}\, y_1^k \, y_2^{g-1-k} \,.
}
We then have a sum over products of KL transforms involving the convolution integral ${\cal B}_m(\omega,\nu)$ evaluated in \eqref{eq:convolutionKernel}:
\begin{align}
  &h_{\uptau\text{-Airy}|m_1,m_2}^{(g,2)}(\sGen_1,\sGen_2)\notag
  \\
  &\quad=4\delta_{m_1m_2} \, a_g\, e^{-2gS_0(m_1)} \sum_{k=0}^{g-1} \binom{g-1}{k} {\cal B}_{m_1}\left(\omega_1,-i(g+k+\tfrac{1}{2})\right) {\cal B}_{m_2}\left(\omega_2,-i(2g-k-\tfrac{1}{2})\right)  \notag\\
  &\quad =  \delta_{m_1m_2}\,\frac{4\pi a_g}{(4\pi m_1)^{3g}}\, e^{-2gS_0(m_1)} \sum_{k=0}^{g-1}\binom{g-1}{k} \frac{|\Gamma(g+\frac{1}{2}+k+i\w_1)|^2|\Gamma(2g-\frac{1}{2}-k+i\w_2)|^2}{\Gamma(g+1+k)\Gamma(2g-k)}
\end{align}
where $\sGen_i =\frac{1}{2}+ i \w_i$.
The expansion coefficients in the harmonic basis follow immediately, according to the general prescription \eqref{eq:spec11}:
\begin{equation}
\begin{split}
    \left(  {\cal Z}^{(g,2)}_{\uptau\text{-Airy}},\,\Psi_{\sGen_1}\Psi_{\sGen_2} \right)_{\mpos} &=  \sum_{m\geq 1} {\bar{\tt c}}_m^{(\sGen_1)}{\bar{\tt c}}_m^{(\sGen_2)} e^{-2gS_0(m)} \,\frac{2\pi a_g}{(4\pi m)^{3g}} \\
    &\qquad \times \sum_{k=0}^{g-1}\binom{g-1}{k} \frac{|\Gamma(g+\frac{1}{2}+k+i\w_1)|^2|\Gamma(2g-\frac{1}{2}-k+i\w_2)|^2}{\Gamma(g+1+k)\Gamma(2g-k)}\,.
\end{split}
\end{equation}
Note that the second line is independent of the spin $m$ and thus factorizes. That is, we can write (for $m>0$)
{\small
\begin{equation}
\boxed{
    \left(  {\cal Z}^{(g,2)}_{\uptau\text{-Airy}},\,\Psi_{\sGen_1}\Psi_{\sGen_2} \right)_{\mpos} =  \frac{\pi a_g}{2(4\pi)^{3g}}\,\widetilde{D}_{\Psi_{\bar\sGen_1}\times \Psi_{\bar\sGen_2}}(S_0;3g) \sum_{k=0}^{g-1}\binom{g-1}{k} \,  \, \frac{|\Gamma(g+k+\sGen_1)|^2|\Gamma(2g-1-k+\sGen_2)|^2}{\Gamma(g+1+k)\Gamma(2g-k)}
}
\label{eq:tauAiryFinalOverlap}
\end{equation}
}\normalsize
with the deformed square-product Dirichlet series
\begin{equation}
    \widetilde{D}_{\Psi_{\sGen_1}\times\Psi_{\sGen_2}}(S_0;t) \equiv 4\sum_{m\geq 1} \frac{{{\tt c}}_m^{(\sGen_1)}{{\tt c}}_m^{(\sGen_2)}}{m^t} \, e^{-\frac{2}{3} t\, S_0(m)} \,.
\end{equation}
For constant $S_0(m)\equiv S_0$ (and up to a factor $\zeta(2t)$), this is proportional to the standard second-power Rankin-Selberg L-function associated with Eisenstein series or Maass cusp forms. See Appendix \ref{app:Lfunctions} for some more details on this topic. Note that the deformation factor is the same as in the (a priori unrelated) L-functions appearing in the three-boundary wormhole of Section \ref{sec:multiboundary}.

The expression is quite a bit more complicated for a general genus; only for $g=0$ can it be written in terms of properties of the basis vectors. We see that the overlaps for the genus corrections to the spectral form factor are no longer diagonal in the spectral index, and in addition they contain overlaps between Eisenstein and cusp form basis vectors. We note that {\it the degree of non-diagonality increases with genus}, in a manner reminiscent of the periodic orbit expansion of RMT correlations for semiclassical chaotic systems  \cite{Sieber2001, Richter:2022sik, Saad:2022kfe}. It would be very interesting to make this analogy more precise.

\paragraph{Eisenstein ($m>0$) sector.}
The spectral overlaps for $m>0$ given in \eqref{eq:tauAiryFinalOverlap} are nontrivial even if the basis vectors are Eisenstein series. One might have thought these contributions to be unnecessary: the correct near-extremal limit in the scalar spectrum is already ascertained by the $m=0$ Eisenstein contribution (recall the remarks below \eqref{eq:RMT2defPmRep}). The $m>0$ Eisenstein contributions are an artifact of the construction using $\sl$ Poincar\'e sums.  Explicitly, their contribution is of the following form for fixed $m>0$:
\begin{align}
    &{\cal Z}_{\uptau\text{-Airy}}^{(g,2)}(\tau_1,\tau_2|m,m) \Big|_\text{Eisenstein} = e^{-2g S_0(m)}\, \frac{2\pi a_g}{(4\pi m)^{3g}} \\
    &\,\,\, \times \int_{{\cal C}_\text{crit}^2} \frac{ds_1 ds_2}{(4\pi i)^2} \, 
    \overline{{\tt a}}_m^{(s_1)} \overline{{\tt a}}_m^{(s_2)} \,\sum_{k=0}^{g-1}\binom{g-1}{k} \frac{|\Gamma(g+k+s_1)|^2|\Gamma(2g-k-1+s_2)|^2}{\Gamma(g+k+1)\Gamma(2g-k)}
    \, \left[E_{s_1}(\tau_1) \, E_{s_2}(\tau_2)\right]_\text{sym} .
    \nonumber
\end{align}
 We will discuss such closed, modular-invariant subsectors further in Section \ref{ssec:minimality}.

%%%%%%%%%%%%%%%%%%%%%%%%%%%%%%%%%%%%%%%%%%%%%%%%
\subsection{GUE Airy RMT${}_2$: plateau and modular corrections}
\label{sec:plateau}
%%%%%%%%%%%%%%%%%%%%%%%%%%%%%%%%%%%%%%%%%%%%%%%%

In this subsection our goal is to see the ramp-plateau transition in RMT$_2$, taking the modular corrections into account. In particular, for our construction of RMT$_2$ to be sensible, the modular corrections should be subleading to the original $\uptau$-scaled SFF, even {\it away} from the near-extremal limit.

\subsubsection{Recovering $\uptau$-scaled RMT in the Poincar\'e sum basis}
In the Poincar\'e sum basis, the $\uptau$-scaled Airy RMT${}_2$ is written in \eqref{eq:GUEAiryFinal}. In this basis, the late-time limit is easy to study: each Poincar\'e sum is evaluated at an integer value of $s$, but all values are within the strip of convergence of the sum. At late times, the expression is therefore term by term dominated by the simple asymptotics \eqref{eq:PmNE}. Thus, we find, in the spin $j$ sector,
\begin{align}
        &e^{-S_0(j)}\,{\cal Z}^{(2)}_{\uptau\text{-Airy}|j,j}(y_1,y_2) \nonumber \\
        & \; \stackrel{y_i\rightarrow \infty}{\longrightarrow} \;d_j^{-1}\sqrt{y_1y_2}\, e^{-2\pi j(y_1+y_2)} \left[\frac{\sqrt{y_1y_2}\,e^{-S_0(j)}}{2\pi(y_1+y_2)} +  \sum_{g\geq 1} a_g \, e^{-(2g+1)S_0(j)} \sum_{k=0}^{g-1} \binom{g-1}{k} \left[ y_1^{2g-k-\frac{1}{2}} y_2^{g+k+\frac{1}{2}}  \right]_\text{sym} \right] \nonumber \\
        &\;\; = d_j^{-1}\sqrt{y_1y_2}\, e^{-2\pi j(y_1+y_2)} \, \frac{1}{4\sqrt{\pi}(y_1+y_2)^{\frac{3}{2}}} \, \text{Erf} \left(e^{-S_0(j)} \sqrt{y_1y_2(y_1+y_2)}  \right) \nonumber
        \\
        &\;\; = d_j^{-1}\sqrt{y_1y_2}\, e^{-2\pi j(y_1+y_2)} \times {\cal K}^{(1)}_\text{RMT}(2\beta)\, \text{Erf}\left(\uptau \sqrt{2\beta}\right)\,,
    \label{eq:tauRMT}
\end{align}
where ${\cal K}^{(1)}_\text{RMT}(2\beta)$ is $e^{-S_0}$ times the RMT one-point correlator (i.e., partition function) at inverse temperature $2\beta$. The latter corresponds to the height of the asymptotic plateau (note that $\lim_{\uptau\rightarrow\infty}\text{Erf}(\uptau\sqrt{2\beta}) = 1$). The expression \eqref{eq:tauRMT} is indeed the correct $\uptau$-scaled RMT input in each spin sector of RMT${}_2$. This confirms the fundamental property \eqref{eq:axiom3}.

%%%%%%%%%%%%%%%%%%%%%%%%%%%%%%%%%%%%%%%%%%%%%%%%
\subsubsection{Modular corrections in the scalar sector}
%%%%%%%%%%%%%%%%%%%%%%%%%%%%%%%%%%%%%%%%%%%%%%%%
While the above analysis is enough to confirm the correct plateau asymptotics of ${\cal Z}^{(2)}_{\uptau\text{-Airy}}$, it is interesting to study the corresponding mechanism in the harmonic basis of Eisenstein series and cusp forms. In the following, we focus on the Eisenstein sector to illustrate a non-trivial resummation and the form of the {\it modular corrections} that render the RMT input modular-invariant. It would be interesting to extend this analysis to cusp forms. 

The $\uptau$-scaling limit consists in taking $y_1=\beta+iT, y_2=\beta-iT$ and scaling $T=e^{S_0}\uptau$ while keeping  $\beta=\frac{y_1+y_2}{2}$ and $\uptau$ fixed. Choosing the right variables is crucial to see the ramp-plateau transition in this limit. 
A well suited set of variables is given by \begin{equation}
e_1=y_1+y_2\,,\qquad e_2=y_1y_2 \,.
\end{equation}
These variables are the two {\it elementary symmetric polynomials} in two variables. They (and their generalization to $n$ variables) have been shown to form a ``natural basis" \cite{Eynard:2021zcj,eynard2023newformulaintersectionnumbers} for computing RMT correlators at large $N$ using topological recursion. The reason is that they make manifest the many non-trivial cancellations between Weil-Petersson volumes which are at the heart of the existence of the $\uptau$-scaling limit, as shown in \cite{Blommaert:2022lbh} (see also \cite{Weber:2022sov,DiUbaldo:2025cmg}). We will see that the scalar sector of RMT$_2$ can also be written in terms of symmetric polynomials, which allows us to identify clearly the $\uptau$-scaling limit as the highest degree monomial in $e_2=y_1y_2$ and fixed $e_1=y_1+y_2$.

The genus expansion for Airy RMT$_2$ in the scalar sector consists of a sum of terms of the form   $E_{2g-k} (y_1)E_{g+k+1}(y_2)$. Each can be expressed as (up to a proportionality constant)
\es{}{
E_{2g-k|0}(y_1) E_{g+k+1|0}(y_2)=\qty(y_1^{2g-k}+\varphi(2g-k)y_1^{1+k-2g})\qty(y_2^{g+k+1}+\varphi(g+k+1)y_2^{-g-k})
}
plus their $y_1\leftrightarrow y_2$ images. 
We have then for the spin ${\bf j = 0}$ sector of the Eisenstein (${\bf m=0}$) piece:
\begin{equation}
    \label{eq:scalareis}
\begin{split}
&{{\cal Z}_{\uptau\text{-Airy}|0,0}^{(g,2)}}(y_1,y_2|0,0)\\
& = \frac{a_g}{2\,e^{2gS_0}}
\sum_{k=0}^{g-1}\binom{g-1}{k}\qty(y_1^{2g-k}+\varphi(2g-k)y_1^{1+k-2g})\qty(y_2^{g+k+1}+\varphi(g+k+1)y_2^{-g-k})\\&+(y_1\leftrightarrow y_2)
\end{split}
\end{equation}
This is equal to $(y_1y_2)^{-(2g-1)}$ times a symmetric polynomial, which we can write in the basis of elementary symmetric polynomials $e_1=y_1+y_2,e_2=y_1y_2$. The scalar sector can then be written as
\es{}{
{{\cal Z}_{\uptau\text{-Airy}|0,0}^{(g,2)}}(y_1,y_2|0,0)=\frac{a_g\,(g-1)!}{2\,e^{2gS_0}}\,(y_1y_2)^{-(2g-1)}\sum_{\substack{d_1,d_2\geq 0 \\ d_1+d_2\leq 4g-1 }} W_{d_1,d_2}^{(g)}\,(y_1+y_2)^{d_1}(y_1y_2)^{d_2}
}
with some coefficients $W_{d_1,d_2}^{(g)}$.

We see a new feature of RMT$_2$ with respect to conventional RMT: the SFF at genus $g$ is no longer a finite degree symmetric polynomial but rather a finite degree symmetric Laurent polynomial.
We can determine that the total degree of the polynomial in this basis is $4g-1$: 
\es{}{
d_1+d_2\leq 4g-1
}
Taking into account the  overall factor of $e_2^{-(2g-1)}$, the total degree of ${{\cal Z}_{\uptau\text{-Airy}|0,0}^{(g,2)}}(y_1,y_2|0,0)$ is $2g$. 
In the $\uptau$-scaling limit the term with the highest power of $e_2$ is the leading one. This term determines the late time growth of ${{\cal Z}_{\uptau\text{-Airy}|0,0}^{(g,2)}}(y_1,y_2|0,0)$ and whether it displays the universal RMT ramp-plateau structure.  Note that the precise numerical coefficients in the sum \eqref{eq:scalareis} are important in rewriting it in terms of the symmetric polynomials and deducing the leading power of $e_2$.

\paragraph{Leading order in $\uptau$-scaling.} By explicitly computing the decomposition in elementary symmetric polynomials one can see that the term with the highest power in $e_2$ at genus $g$ is: 
\es{}{
{{\cal Z}_{\uptau\text{-Airy}|0,0}^{(g,2)}}(y_1,y_2|0,0) \supset a_g\, e^{-2gS_0} \, e_1^{g-1}e_2^{g+1} \qquad \Rightarrow \qquad W_{g-1,3g}^{(g)}=\frac{2}{(g-1)!}\,,}
which is exactly the term in \eqref{eq:scalareis} involving no $\varphi$-functions, i.e., proportional to $y_1^{2g-k} y_2^{g+k+1} + y_2^{2g-k} y_1^{g+k+1}$.

We can now take the $\uptau$-scaling limit of \eqref{eq:scalareis}: 
\es{}{
{{\cal Z}_{\uptau\text{-Airy}|0,0}^{(g,2)}}(y_1,y_2|0,0)= a_g\,e^{2S_0}\,(2\beta)^{g-1}\uptau^{2g+2} \left(1+\mathcal{O}(e^{-2S_0})\right) 
}
with the usual identifications $2\beta=y_1+y_2$ and $\uptau = \sqrt{y_1y_2}\, e^{-S_0} (1+ {\cal O}(e^{-2S_0}))$, and $g\geq1 $.
Indeed, we can now sum the genus expansion for the leading term in the $\uptau$-scaling limit and obtain the $\uptau$-scaled SFF for the standard Airy model:
\es{}{
(\uptau e^{S_0})^{-1}\sum_{g=1}^\infty{{\cal Z}_{\uptau\text{-Airy}|0,0}^{(g,2)}}(y_1,y_2|0,0)&=e^{S_0}\sum_{g=1}^{\infty}a_g \, (2\beta)^{g-1}\,\uptau^{2g+1}\left(1+\mathcal{O}(e^{-2S_0})\right) \\ 
&=\frac{e^{S_0}}{4\sqrt{\pi}(2\beta)^{3/2}}\,\erf\qty(\uptau\sqrt{2\beta})\left(1+\mathcal{O}(e^{-2S_0})\right)- \frac{\uptau\,e^{S_0}}{4\pi\beta}\\
&= \text{K}^{(2)}_{\uptau\text{-Airy}}(y_1,y_2)-\text{K}^{(0,2)}_{\text{RMT}}(y_1,y_2)\,,
}
where we removed the conventional factor $\sqrt{y_1y_2}$ common to all RMT${}_2$ amplitudes, and the linear ramp is supplied separately from $g=0$.

Thus we have shown that the $\uptau$-scaling limit of  Airy RMT$_2$ is equal to the $\uptau$-scaling limit of Airy RMT. In other words, {\it the modular corrections present in the Airy RMT$_2$ expression are subleading in the $\uptau$-scaling limit}.

\paragraph{Subleading orders in $\uptau$-scaling.}
We now move to subleading powers of $e_2$. The first subleading order would be ${\cal O}(e_2^g)$, corresponding to coefficients $W^{(g)}_{d_1,d_2=3g-1}$. However, we observe that the coefficients at this order always vanish:
\es{}{
W_{d_1,d_2=3g-1}^{(g)}=0 \qquad \text{for all } g \,.
}
The next subleading order is ${\cal O}(e_2^{g-1})$. Here, the relevant polynomials involve the coefficients $W^{(g)}_{d_1,d_2=3g-2}$ and take the following form:
\es{}{
\sum_{d_1=0}^{2}e_1^{d_1}W^{(g=1)}_{d_1,1}=-6\varphi(2)\,e_1, \qquad \sum_{d_1=0}^{3}e_1^{d_1}W^{(g=2)}_{d_1,4}=-4\varphi(3), \qquad W^{(g>2)}_{d_1,3g-2}=0\,.
}
This structure continues: at the next order, ${\cal O}(e_2^{g-2})$, we have $W^{(g)}_{d_1,d_2=3g-3}=0$ for $g>4$ (in all cases checked up to $g=30$). For instance, we find the following coefficients:
\es{}{
\sum_{d_1=0}^{3}e_1^{d_1}W^{(g=1)}_{d_1,0}&=2\varphi(2)\,e_1^3+2\varphi(2)^2\,, \qquad \sum_{d_1=0}^{4}e_1^{d_1}W^{(g=2)}_{d_1,3}=18\varphi(3)\,e_1^2-4\varphi(4)\,,\\
\sum_{d_1=0}^{5}e_1^{d_1}W^{(g=3)}_{d_1,6}&= 9\varphi(4)\,e_1\,,\qquad \sum_{d_1=0}^{6}e_1^{d_1}W^{(g=4)}_{d_1,9}= \frac{2}{3}\varphi(5)\,,\qquad W^{(g>4)}_{d_1,3g-3}=0\,.
}
This expansion gives the subleading terms away from the $\uptau$-scaling limit; some are already present in the original RMT, and some are due to modular corrections. 

More generally, we can show that
\begin{equation}
    W^{(g>2\ell-2)}_{d_1,d_2=3g-\ell}=0 \qquad (\ell\geq 1) \,.
    \label{eq:Wvanish}
\end{equation}
To see this, consider the singly reflected contribution in \eqref{eq:scalareis} (which has higher $e_2$ degree than the doubly reflected term):
\begin{equation}
{\cal Z}^{(g,2)}_{\uptau\text{-Airy}|0,0}(y_1,y_2|0,0) \supset \frac{a_g}{e^{2gS_0}} \, (y_1y_2)^{-(2g-1)} \times \left(y_1^{3g}+y_2^{3g}\right) \sum_{k=0}^{g-1} \binom{g-1}{k}\varphi(2g-k)\, e_2^k \,.
\label{eqLcount}
\end{equation}
Expressing $(y_1^{3g}+y_2^{3g})$ in terms of $e_1$ and $e_2$, the power of $e_2$ of this factor is at most $\lfloor \frac{3g}{2} \rfloor$. Therefore, the maximal $e_2$ degree of the symmetric polynomial part of \eqref{eqLcount} is $\lfloor \frac{3g}{2} \rfloor+(g-1)$. This implies \eqref{eq:Wvanish}.

%%%%%%%%%%%%%%%%%%%%%%%%%%%%%%%%%%%%%%%%%%%%%%%%
\subsection{GOE Airy RMT${}_2$}
\label{sec:GOEAiry}
%%%%%%%%%%%%%%%%%%%%%%%%%%%%%%%%%%%%%%%%%%%%%%%%

We will now extend our analysis of the Airy model to the GOE universality class. We focus on the new phenomena occurring in the GOE class, which is more generic in the sense of containing all potential terms in the SFF without any cancellations specific to the GUE. For simplicity, we work again in the $\uptau$-scaling limit.  

The $\uptau$-scaled GOE SFF for a general density of states $\rho(E) \equiv e^{S_0} \, \rho_0$ is given by the following piecewise Laplace transform of the high-energy `ramp' and low-energy `plateau' \cite{PhysRevE.72.046207,Blommaert:2022lbh,Saad:2022kfe,Weber:2024ieq,DiUbaldo:2025cmg}:
\begin{equation}\label{eq:GOESFF2}
\begin{split}
    {\cal K}^{(2)}_{\uptau\text{-RMT-GOE}}(\beta,\uptau) 
    & = \int_{E_*}^{\infty} dE\ e^{-2 \beta E} \left[\frac{\uptau}{\pi} - \frac{\uptau}{2\pi} \ \text{log} \left(\frac{\uptau}{\pi\rho_0(E)} + 1\right)\right] 
    \\
    &\quad + \int_0^{E_*}dE \, e^{-2\beta E} \left[ 2 \rho_0(E) - \frac{\uptau}{2\pi} \ \text{log} \left( \frac{\uptau +\pi \rho_0(E) }{\uptau-\pi\rho_0(E) } \right) \right]
    \end{split}
 \end{equation}
 where $\rho_0(E_*) = \frac{\uptau}{2\pi}$.
For the Airy density of states \eqref{eq:rhoAiry},
this results in the genus expansion \cite{Weber:2024ieq,DiUbaldo:2025cmg}
 \begin{equation}
 \label{eq:GOEairyResult}
 \begin{split}
 &{\cal K}^{(2)}_{\uptau\text{-Airy-GOE}}(\beta,\uptau) \\
 &\quad = 
  \frac{\uptau}{2\pi \beta} + \sum_{g=1,2,...}^\infty \left[ A_{g}(\rho_0^\text{Airy};\beta)+ B_{g}(\rho_0^\text{Airy};\beta) \, \log(\uptau) \right] \, \uptau^{2g+1} + \sum_{\tilde{g}=\frac{1}{2},\frac{3}{2},...}^\infty C_{\tilde{g}}(\rho_0^\text{Airy};\beta) \, \uptau^{2\tilde{g}+1}
\end{split}
 \end{equation}
 with 
\begin{equation}
\label{eq:AresultAiry}
\begin{split}
 A_{g}\big(\rho_0^\text{Airy};\beta\big) 
 &= -\frac{(-2\beta)^{g-1}}{\pi (2g+1) g!} + \frac{(-8\beta)^{g-1}}{\pi  g!} \left[\frac{1}{g}\;{}_2F_1(2g,2g,2g+1,-1) +  \psi(g+1)  - \log\left(8\beta\right)  \right] 
  \\
  B_{g}\big(\rho_0^\text{Airy};\beta\big) &= -\frac{2(-4)^{g-1}}{\pi g!} \,(2\beta)^{g-1}\,,
  \\
 C_{\tilde{g}}\big(\rho_0^\text{Airy};\beta\big) &= -\frac{4^{\tilde{g}-1}}{\pi \tilde{g}} \Gamma(1-\tilde{g}) \, (2\beta)^{\tilde{g}-1}\,,
 \end{split}
\end{equation}
 where $\psi(z) = \frac{d}{dz} \log \Gamma(z)$ is the digamma function. We recognize the first term in the $A_g$ coefficient as two times the GUE result \eqref{eq:tauAiryGUE}. Further, note the appearance of non-analytic $\log(\uptau)$ terms, as well as half-integer genus (crosscap) contributions leading to even powers of $\uptau$. These features are generically important, but absent in the GUE example discussed earlier. For a more detailed discussion of these features, see \cite{DiUbaldo:2025cmg}. Explicitly, the first few terms are:
\begin{equation}\label{eq:AiryGOESFF}
    \begin{split}
    {\cal K}^{(2)}_{\uptau\text{-Airy-GOE}}(\beta,\uptau) &= \frac{1}{2\pi \beta}\,\uptau -\frac{1}{\sqrt{2\pi\beta}}\, \uptau^2   - \frac{1}{\pi} \left(\frac{1}{3}+\gamma +\log(2\beta\uptau^2)\right) \uptau^3  +  \frac{8}{3\pi} \,(2\pi\beta)^{\frac{1}{2}} \, \uptau^4 \\
    &\quad + \frac{4\beta}{\pi} \left(-\frac{7}{60} + \gamma + \log(2\beta\uptau^2) \right)  \uptau^5  - \frac{64}{15\pi^2} \, (2\pi\beta)^{\frac{3}{2}}\,  \uptau^6  + \ldots 
    \end{split}
\end{equation}

%%%%%%%%%%%%%%%%%%%%%%%%%%%%%%%%%%%%%%%%%%%%%%%%
\subsubsection{Poincar\'e decomposition}
%%%%%%%%%%%%%%%%%%%%%%%%%%%%%%%%%%%%%%%%%%%%%%%%
Having described the GOE SFF for the standard Airy model in the genus expansion, let us now construct a modular-invariant RMT${}_2$ with the same GOE statistics. We will focus on the higher-genus corrections to the ramp.

Once again we assume the GOE result in each spin sector separately. That is, we seek a decomposition
\begin{equation}
    {\cal Z}^{(2)}_{\uptau\text{-Airy-GOE}}(\tau_1,\tau_2) = \sum_{m\geq 0} d_m^{-1}\int_{{\cal C}_\text{crit}^2} \frac{ds_1ds_2}{(2\pi i)^2} \, \left[ P_{m}(\tau_1,s_1)P_{m}(\tau_2,s_2) \right]_\text{sym} \, f^{(2)}_{\uptau\text{-Airy-GOE}}(S_0(m);\bar s_1,\bar s_2) 
\end{equation}
such that the near-extremal limit $y_{1,2}\rightarrow\infty$ produces (universal factors multiplying) the matrix model input with spin sector-dependent entropy.

Just like in the GUE Airy model, the spectral overlaps require the Mellin transform
\beq
\label{eq:tauAiryfGOE}
\begin{split}
f^{(2)}_{\uptau\text{-Airy-GOE}}(S_0(m);\bar s_1,\bar s_2) &\equiv  {\cal M}_{y_1,y_2}\left[ e^{S_0(m)} {\cal K}^{(2)}_{\uptau\text{-Airy-GOE}}\big(\beta=\tfrac{y_1+y_2}{2},\uptau=e^{-S_0(m)}\sqrt{y_1y_2}\big)\right](-i\omega_1,-i\omega_2)
\end{split}
\eeq 

Without detailed derivation, we simply state the result for the spectral overlaps reproducing \eqref{eq:GOEairyResult}. 
We find the following result (with $\bar s_i = \frac{1}{2}-i\omega_i$ and $\omega_\pm = \frac{1}{3}(\omega_1\pm \omega_2)$):
\begin{equation} \label{eq:GOEAiryMellin}
    \begin{split}
    f^{(2)}_{\uptau\text{-Airy-GOE}}(S_0;\bar{s}_1,\bar{s}_2) 
    &= e^{-2i\omega_+S_0} \, \Gamma\left(\frac{1}{2}-\frac{2i\omega_1}{3}+\frac{i\omega_2}{3}\right)\Gamma\left(\frac{1}{2}+\frac{i\omega_1}{3} - \frac{2i\omega_2}{3}\right) \mathfrak{f}^{(2)}_{\uptau\text{-Airy-GOE}} (\omega_+) \\
   \mathfrak{f}^{(2)}_{\uptau\text{-Airy-GOE}} (\omega_+) &=
   2\,\mathfrak{f}^{(2)}_{\uptau\text{-Airy-GUE}} (\omega_+)  -\frac{4^{i\omega_+}}{12\pi \omega_+^2} \bigg[ {}_2F_1(2i\omega_+,2i\omega_+,2i\omega_++1;-1)- \frac{2\pi\omega_+}{\sinh(2\pi \omega_+)} \bigg] 
   \end{split}
\end{equation}
where we recall from \eqref{eq:tauAiryf} the GUE expression $\mathfrak{f}^{(2)}_{\uptau\text{-Airy-GUE}} (\omega_+) = \left(12 \pi \omega_+ (\omega_+ - \frac{i}{2}) \right)^{-1}$. For ease of notation we have omitted a contour prescription (or, equivalently, an explicit regulator $\varepsilon$), but one can easily verify that the rule for performing the inverse Mellin transform (which involves deforming the contour $\omega_+ \in \mathbb{R}$) is as follows: the pole at $\omega_+=0$ should be considered as part of the LHP for the GUE piece; for the remaining GOE contribution, $\omega_+=0$ is considered as part of the UHP. This prescription ascertains the correct genus zero term. 

We have checked by inverting the Mellin transform that this expression yields the correct $\uptau$-scaled GOE Airy spectral form factor, see Appendix \ref{app:mellinGOEAiry}. The poles due to $\sinh(2\pi \omega_+)$ at $\omega_+ = -\frac{i}{2}(2g-1)$ for $g\in \mathbb{Z}_+$ are responsible for even powers $\uptau^{2g}$. The (double) poles at $\omega_+ = -ig$ yield odd powers $\uptau^{2g+1}$ as well as logarithmic terms.

%%%%%%%%%%%%%%%%%%%%%%%%%%%%%%%%%%%%%%%%%%%%%%%%
\subsubsection{Spectral decomposition}
%%%%%%%%%%%%%%%%%%%%%%%%%%%%%%%%%%%%%%%%%%%%%%%%
We can also compute the spectral decomposition in the basis of Eisenstein series and cusp forms. The base-spin-zero Eisenstein sector is directly determined by the Mellin amplitude \eqref{eq:GOEAiryMellin}. The cusp form sector requires the double Kontorovich-Lebedev transform of \eqref{eq:GOEAiryMellin}.
We analyze this integral transform in Appendix \ref{app:KL}. We find the following structure for the spectral overlaps (assuming at least one $\Psi_{\sGen_i}$ is a cusp form):
\begin{equation}
\begin{split}
\left({\cal Z}^{(2)}_{\uptau\text{-Airy-GOE}},\,\Psi_{\sGen_1}\Psi_{\sGen_2}\right)_{g>0}
&= \sum_{g\geq 1}\frac{\pi}{2} \left[  \left(\tilde{a}_g + \tilde{b}_g \, \partial_g \right) \left((4\pi)^{-3g} \,\widetilde{\mathfrak{H}}_{g-1,\frac{1}{2}+g}(\sGen_1,\sGen_2)\,\widetilde{D}_{\Psi_{\bar \sGen_1}\times\Psi_{\bar \sGen_2}}(S_0;3g)\right)  \right]
\\
&\quad + \sum_{\tilde g=\frac{1}{2},\frac{3}{2},\ldots} \frac{\pi}{2(4\pi)^{3\tilde g}} \, \tilde{c}_{\tilde g} \, \widetilde{D}_{\Psi_{\bar\sGen_1}\times\Psi_{\bar\sGen_2}}(S_0;3\tilde g) \, \widetilde{\mathfrak{H}}_{\tilde{g}-1,\frac{1}{2}+\tilde g}(\sGen_1,\sGen_2) \,,
\end{split}
\end{equation}
with the deformed Dirichlet series
\begin{equation}
    \begin{split}
      \widetilde{D}_{\Psi_{\sGen_1}\times\Psi_{\sGen_2}} (S_0;t)  &= 4\sum_{m\ge1} \frac{ {\tt c}^{(\sGen_1)}_m \,  {\tt c}^{(\sGen_2)}_m}{m^t} \, e^{-\frac{2}{3} t \, S_0(m)}  \,.
    \end{split}
\end{equation}
The coefficients $\{\tilde{a}_g,\tilde{b}_g,\tilde{c}_{\tilde{g}}\}$ are model-specific and can be found for the GOE Airy model in \eqref{eq:abctilde}. The function $\widetilde{\mathfrak{H}}_{\lambda,\mu}$ can also be found in Appendix \ref{app:KL} and it is universal for a large class of spectral curves in the GOE universality class.

%%%%%%%%%%%%%%%%%%%%%%%%%%%%%%%%%%%%%%%%%%%%%%%%
%%%%%%%%%%%%%%%%%%%%%%%%%%%%%%%%%%%%%%%%%%%%%%%%
\section{RMT${}_2$ universality and AdS$_3$ wormholes}
\label{sec:ramp}
%%%%%%%%%%%%%%%%%%%%%%%%%%%%%%%%%%%%%%%%%%%%%%%%
%%%%%%%%%%%%%%%%%%%%%%%%%%%%%%%%%%%%%%%%%%%%%%%%

In this section we analyze the cases $n=2,3$, building upon \cite{Haehl:2023tkr,DiUbaldo:2023qli,Haehl:2023xys,Haehl:2023mhf} and, for the $n=3$ case, \cite{Boruch:2025ilr}. Unlike the Airy case treated in detail in the previous section, the cases $n=2,3$ are universal across ensembles. We discuss their gravitational interpretation.

%%%%%%%%%%%%%%%%%%%%%%%%%%%%%%%%%%%%%%%%%%%%%%%%
\subsection{$n=2$: Linear ramp and the Cotler-Jensen wormhole}
%%%%%%%%%%%%%%%%%%%%%%%%%%%%%%%%%%%%%%%%%%%%%%%%

The simplest and most well-known example of RMT${}_2$ is the modular-invariant realization of the linear ramp in the spectral form factor \cite{Haehl:2023tkr,DiUbaldo:2023qli,Haehl:2023xys,Haehl:2023mhf,Boruch:2025ilr}, which matches a two-boundary wormhole calculation in pure AdS${}_3$ gravity \cite{Cotler:2020ugk,Cotler:2020hgz}. We review (and reconstruct) it here from the point of view of our general proposal.

Consider the two-point correlator in a double-scaled random matrix model, $\text{K}^{(2)}_{\text{RMT}}(S_0;y_1,y_2)=\left\langle\tr (e^{-y_1H}) \tr (e^{-y_2H}) \right\rangle_c$ and focus on its genus-zero contribution, i.e., the universal leading order approximation to the sine kernel:\footnote{The genus zero spectral form factor does not depend on $S_0$, so we suppress this label.}
\begin{equation}
\label{eq:rampGeneral}
\text{K}^{(0,2)}_{\text{RMT}}(y_1,y_2)= \frac{{\tt C}_\text{RMT}}{2\pi} \, \frac{\sqrt{y_1y_2}}{y_1+y_2} \,,\quad \text{where}~{{\tt C}_\text{RMT} := \frac{2}{\beta}}
\end{equation}
in the $\beta$-ensemble\footnote{Not to be confused with inverse temperature.} of random matrices (e.g. ${\tt C}_\text{RMT}\in\{1,2,\tfrac{1}{2}\}$ in the GUE/GOE/GSE universality classes). We wish to construct its RMT${}_2$ lift using the decomposition into Poincar\'e sums. 

Our general prescription instructs us to compute the Mellin transform of the linear ramp input:
\begin{equation}
    \begin{split}  
    f^{(0,2)}_{\text{RMT}|m_1,m_2}(\bar s_1,\bar s_2) &=  \int_0^\infty dy_1 dy_2 \; y_1^{-1-i\omega_1} y_2^{-1-i\omega_2} \left( d_{m_1}^{-1}\, \frac{{\tt C}_\text{RMT}}{2\pi}\, \frac{\sqrt{y_1 y_2}}{y_1+y_2}\,\delta_{m_1m_2}  \right)
    \\ 
    &=d_{m_1}^{-1} \; \frac{{\tt C}_\text{RMT}}{\cosh(\pi\omega_1)} \; \pi \delta(\omega_1+\omega_2) \,\delta_{m_1m_2} \,,\qquad (\bar s_i \equiv \tfrac{1}{2}-i\omega_i) \,.
\end{split}
\label{eq:f02}
\end{equation}
The RMT${}_2$ lift should then take the following form:
\begin{equation}
 {\cal Z}^{(0,2)}_{\text{RMT}_2}(\tau_1,\tau_2 | m_1,m_2)= \lim_{\epsilon\to0}\int_{{\cal C}_\text{crit}^2} \frac{ds_1 ds_2}{(2\pi i)^2} \;  f^{(0,2)}_{\text{RMT}|m_1,m_2}(\bar s_1,\bar s_2) \, P_{m_1}(\tau_1,s_1+\epsilon) P_{m_2}(\tau_2,s_2+\epsilon) 
\,,
\end{equation}
where $\epsilon$ consistently regulates the poles on the integration contour.\footnote{We leave the regulator implicit in the calculations below. The sum over $m$ in \eqref{eq:CJpoincare} converges absolutely for $\epsilon>1$ and then continues to $\epsilon\to 0$.}
The delta-function collapses one of the integrals and we get the result
\begin{equation}
\label{eq:CJpoincare}
\boxed{
    {\cal Z}_{\text{RMT}_2}^{(0,2)}(\tau_1,\tau_2) = \lim_{\epsilon\to0} \sum_{m\geq 0} d_m^{-1}\int_{{\cal C}_\text{crit}} \frac{ds}{2\pi i} \, \frac{{\tt C}_\text{RMT}}{2 \cosh(\pi \omega)} \, P_m(\tau_1,s+\epsilon) P_m(\tau_2,1-s+\epsilon) 
}
\end{equation}
This is a novel way of writing the AdS${}_3$ torus wormhole amplitude of \cite{Cotler:2020ugk}, which we discuss further below.
Although \eqref{eq:CJpoincare} naively involves a double Poincar\'e sum which diverges,\footnote{Even the single Poincar\'e sum diverges and is similarly regularized by using the spectral overlaps.} we can define it using spectral overlaps, which are valid beyond the region of convergence of the sum.

As required, \eqref{eq:CJpoincare} exhibits a linear ramp near extremality for every spin sector. Consider first $j_1,j_2>0$, where this immediately follows from the asymptotic expansion \eqref{eq:PmNE}:
\begin{equation}
\begin{split}
   &{\cal Z}^{(0,2)}_{\text{RMT}_2|j_1,j_2}(y_1,y_2) \\
   &\qquad\stackrel{\substack{\text{near-}\\\text{extremal}}}{\longrightarrow} \sum_{m\geq 0} d_m^{-1}\int_{{\cal C}_\text{crit}} \frac{ds}{2\pi i} \, \frac{{\tt C}_\text{RMT}}{2 \cosh(\pi \omega)} \, \left[ \delta_{mj_1} y_1^s \, e^{-2\pi m y_1} + \ldots\right]
   \left[ \delta_{mj_2} y_2^{1-s} \, e^{-2\pi m y_2} + \ldots\right] 
   \\
   &\qquad\;\;\; = \sum_{m\geq 0} d_{m}^{-1} \, \delta_{mj_1} \delta_{mj_2} \, e^{-2\pi m(y_1+y_2)} \, \sqrt{y_1y_2} \; \int_{\mathbb{R}} \frac{d\omega}{2\pi} \, \frac{{\tt C}_\text{RMT}}{2\cosh(\pi\omega)} \left( \frac{y_1}{y_2} \right)^{i\omega} + \ldots
   \\
   &\qquad\;\;\; = d_{j_1}^{-1}\, \delta_{j_1j_2} \, e^{-2\pi (j_1y_1+j_2y_2)} \, \sqrt{y_1y_2} \; \text{K}^{(0,2)}_\text{RMT}(y_1,y_2) + \ldots
\end{split}
\label{eq:ZppLimit0}
\end{equation}
which is the desired linear ramp (times the conventional factor $\sqrt{y_1y_2}$) with ground-state energy for the spin $j$ spectrum given by $E_0(j)= 2\pi j$. 

To recover the ramp in the scalar sector $j_1=j_2=0$, we need a more careful argument. The calculation \eqref{eq:ZppLimit0} still applies, but the second line receives an additional ``doubly reflected'' contribution, as shown for $P_0(\tau,s)$ in \eqref{eq:PmNE}. This means that this sector produces an undesired factor $2$ relative to \eqref{eq:ZppLimit0}:
\begin{equation}
\label{eq:plus2ramp}
    \begin{split}
        {\cal Z}^{(0,2)}_{\text{RMT}_2|0,0}(y_1,y_2|0,0)
        & \;\stackrel{\substack{\text{near-}\\\text{extremal}}}{\longrightarrow} \; 2\times \sqrt{y_1y_2}\,\text{K}_\text{RMT}^{(0,2)}(y_1,y_2)  + \ldots
    \end{split}
\end{equation}
This is rectified by a negative contribution from the $m>0$ sectors, which contribute as follows to the scalar amplitude (using the parity-even version of \eqref{eq:Pm0cusp}):
{\small
\begin{equation}
\label{eq:minusRamp}
    \begin{split}
        &\sum_{m>0}{\cal Z}^{(0,2)}_{\text{RMT}_2|0,0}(y_1,y_2|m,m)
        \\
        & \stackrel{\substack{\text{near-}\\\text{extremal}}}{\longrightarrow} \sum_{m>0} d_m^{-1} \int_{{\cal C}_\text{crit}}\frac{ds}{2\pi i} \frac{{\tt C}_\text{RMT}}{2\cosh(\pi\omega)} \, \left[y_1^{1-s}\,\frac{2\sqrt{\pi}\,\Gamma(s-\frac{1}{2}) \sigma_{1-2s}(m)}{\Gamma(s)\zeta(2s)} +\ldots\right] \left[y_2^{s}\,\frac{2\sqrt{\pi}\,\Gamma(\frac{1}{2}-s) \sigma_{2s-1}(m)}{\Gamma(1-s)\zeta(2-2s)}+\ldots \right]
        \\
        &\quad =\sqrt{y_1y_2}\int_{\mathbb{R}}\frac{d\omega }{2\pi} \,  \frac{{\tt C}_\text{RMT}}{2\cosh(\pi \omega)} \, \left( \frac{y_2}{y_1} \right)^{i\omega} \sum_{m>0} \frac{2\, \sigma_{2i\omega}(m) \sigma_{-2i\omega}(m)}{ |\zeta(2i\omega)|^2}
        \\
        &\quad = -\sqrt{y_1y_2} \int_{\mathbb{R}}\frac{d\omega}{2\pi}  \,  \frac{{\tt C}_\text{RMT}}{2\cosh(\pi \omega)} \, \left( \frac{y_2}{y_1} \right)^{i\omega} 
        \\
        &\quad = -\sqrt{y_1y_2}\,\text{K}_\text{RMT}^{(0,2)}(y_1,y_2)
    \end{split}
\end{equation}
}\normalsize
where we used the standard Dirichlet series for the divisor function with zeta-function regularization.\footnote{Specifically, we use the identity
\begin{equation}
    \sum_{m>0} \frac{ \sigma_{2i\omega}(m) \sigma_{-2i\omega}(m)}{m^u} = \frac{\zeta(u)^2\zeta(u+2i\omega)\zeta(u-2i\omega)}{\zeta(2u)} \qquad (\text{Re}(u)>1)\,,
\end{equation}
which analytically continues to $u=0$, using $\zeta(0)=-\frac{1}{2}$.}
Collecting contributions in the spin $j_1=j_2=0$ sector, we obtain two times the linear ramp from \eqref{eq:plus2ramp} plus the negative contribution \eqref{eq:minusRamp}; together these give the linear ramp with the correct coefficient:
\begin{equation}
    \sum_{m\geq 0}{\cal Z}^{(0,2)}_{\text{RMT}_2|0,0}(y_1,y_2|m,m)
    \;\stackrel{\substack{\text{near-}\\\text{extremal}}}{\longrightarrow} \; \sqrt{y_1y_2}\,\text{K}_\text{RMT}^{(0,2)}(y_1,y_2)\,.
\end{equation}

We derived \eqref{eq:CJpoincare} from first principles by imposing near-extremal asymptotics consistent with RMT universality, thus determining the Poincar\'e decomposition spin by spin.
As was already shown in \cite{DiUbaldo:2023qli,Haehl:2023xys}, such reasoning reproduces the pure gravity wormhole \cite{Cotler:2020ugk} as the minimal modular completion of the RMT input. To confirm our methods, we wish to re-derive the spectral decomposition of the universal RMT${}_2$ amplitude \eqref{eq:CJpoincare}:
\begin{equation}
\boxed{
\begin{aligned}
    \left( {\cal Z}^{(0,2)}_{\text{RMT}_2} \,,\, E_{\frac{1}{2}+i\omega_1} E_{\frac{1}{2}+i\omega_2} \right) &= \frac{{\tt C}_\text{RMT} }{2\cosh(\pi\omega_1)} \times 2\pi \left(\delta(\omega_1+\omega_2) + \varphi(1-s_2)\delta(\omega_1-\omega_2)\right) \,,\\
    \left( {\cal Z}^{(0,2)}_{\text{RMT}_2} \,,\, \phi_{n_1}\phi_{n_2} \right) &= \frac{{\tt C}_\text{RMT}}{2\cosh(\pi\omega_{n_1})} \times \delta_{n_1,n_2} \,,\\ 
    \left( {\cal Z}^{(0,2)}_{\text{RMT}_2} \,,\, E_{s_1}\,\phi_{n_2} \right) &= 0 \,,
\end{aligned}
}
\label{eq:CJoverlaps}
\end{equation}
This is, of course, the known `diagonal' spectral decomposition of the pure AdS${}_3$ gravity two-boundary torus wormhole \cite{Cotler:2020ugk}, which is the minimal modular completion of an independent linear ramp in each spin sector \cite{DiUbaldo:2023qli,Haehl:2023mhf}. The passage to this expression from the Poincar\'e representation \eqref{eq:CJpoincare} is performed in Appendix \ref{sec:CJnewderivation}.

\paragraph{On the {\tt maxRMT} principle.} The {\it spectral diagonality}  and {\it Hecke symmetry} of the AdS$_3$ two-boundary wormhole amplitude have been discussed before in the above references and were coined as the {\tt maxRMT} principle in \cite{DiUbaldo:2023qli}: these minimality properties respectively amount to $(i)$ the delta-function correlations in \eqref{eq:CJoverlaps}, and $(ii)$ the fact that Eisenstein and cusp form overlaps take the same functional form. We may now appreciate yet another manifestation of the simplicity of the universal $(g,n)=(0,2)$ RMT${}_2$ amplitude that matches the AdS${}_3$ gravity result: our more general construction shows that the spectral overlaps \eqref{eq:CJoverlaps} are {\it a priori} defined as Bessel transforms \eqref{eq:spec11Rep}, not Mellin transforms. In general there is no reason for these to be identical to the Mellin amplitude \eqref{eq:f02}. Yet, the universal result discussed above does exhibit this property. (See the preceding section for examples where this property does not hold.) To summarize: the spectral diagonality and Hecke symmetry manifest themselves in the property that
{\it the spectral overlaps \eqref{eq:CJoverlaps} are identical to the Poincar\'e overlaps \eqref{eq:CJpoincare}.}

%%%%%%%%%%%%%%%%%%%%%%%%%%%%%%%%%%%%%%%%%%%%%%%%
\subsection{$n=3$: RMT${}_2$ for the three-boundary wormhole}
\label{sec:multiboundary}
%%%%%%%%%%%%%%%%%%%%%%%%%%%%%%%%%%%%%%%%%%%%%%%%

In this section we discuss genus $g=0$ spectral correlators for $n= 3$ insertions of the partition function. The RMT input is independent of the spectral curve, and is given by \cite{Eynard:2015aea}\footnote{For notational simplicity, we use the normalization appropriate for the unitary class. In the orthogonal class, \eqref{03amplitude} gets multiplied by a factor $4$ \cite{Stanford:2019vob,Stanford:2023dtm}.}
\begin{equation}
    \label{03amplitude}
\text{K}^{(0,3)}_{\text{RMT}}(S_0;y_1,y_2,y_3)=\frac{e^{-S_0}}{(2\pi\gamma)^{3/2}}\,\sqrt{y_1y_2y_3} \,,
\end{equation}
where $\gamma$ is the unit of $y_i$, which will be convenient to keep track of. Since these amplitudes are universal in RMT (i.e., independent of the spectral curve), they provide valuable data points. In gravity, these topologies would correspond to three-boundary torus wormholes, which are yet to be computed independently.

\paragraph{RMT${}_2$ construction.} Let us now compute the Mellin transform to define the RMT${}_2$ lift. The expression \eqref{03amplitude} is factorized, therefore to obtain the spectral overlaps we can perform the required integral transforms in each variable separately. First, the Mellin space amplitude is given by \cite{Boruch:2025ilr}
\begin{equation}
\label{eq:f03}
    f^{(0,3)}_{\text{RMT}|m_1,m_2,m_3}(\bar s_1,\bar s_2,\bar s_3) = d_{m_1}^{-2}\,\delta_{m_1m_2}\delta_{m_2m_3}\,\frac{e^{-S_0(m_1)}}{(2\pi\gamma)^{3/2}} \, \prod_{i=1}^3 \left( \frac{2\epsilon}{\epsilon^2-(\frac{1}{2}-i\omega_i)^2} \right) \,,\qquad  \omega_i \in -\frac{i}{2} + \mathbb{R}\,,
\end{equation}
where $\epsilon >0$ is a regulator that is required for faithful Mellin transforms of monomials and powers such as $\sqrt{y_i}$, see \cite{Boruch:2025ilr} for more details. The integration contour of $\omega_i$ (or $s_i$) is shifted for Mellin transforms of polynomials.
The Mellin amplitude \eqref{eq:f03} defines the RMT${}_2$ amplitude in the Poincar\'e sum basis. The result factorizes in every base spin sector $m$:
\begin{equation}
\label{eq:n3RMT2def}
\begin{split}
    {\cal Z}^{(0,3)}_{\text{RMT}_2}(\tau_1,\tau_2,\tau_3) 
    &= \lim_{\epsilon\rightarrow 0^+} \sum_{m\geq 0} d_m^{-2}\,\frac{e^{-S_0(m)}}{(2\pi\gamma)^{3/2}}\,\prod_{i=1}^3 \left( \int_{1+i\mathbb{R}} \frac{ds_i}{2\pi i} \,\frac{2\epsilon}{\epsilon^2-(1- s_i)^2} \, P_m(\tau_i,s_i)\right)\,.
\end{split}
\end{equation}
Evaluating the integrals by residue theorem, we find 
\begin{equation}
\label{eq:n3RMT2def2}
\boxed{
    {\cal Z}^{(0,3)}_{\text{RMT}_2}(\tau_1,\tau_2,\tau_3) 
=  \frac{1}{(2\pi\gamma)^{3/2}} \left[ e^{-S_0(0)} \, \prod_{i=1}^3\widehat{E}_1(\tau_i)  + \frac{1}{4} \sum_{m> 0} \left( e^{-S_0(m)}\,\prod_{i=1}^3 P_m(\tau_i,1)\right)\right]
}
\end{equation}
where the first term ($m=0$) involves pole-subtracted Eisenstein series derived below (see \eqref{eq:EhatResult}).

In order to find the spectral overlaps, we follow the prescription \eqref{eq:spec11}. These are given in terms of the triple KL transform of $\text{K}^{(0,3)}_{\text{RMT}}$, which again factorizes. The relevant integral follows from \eqref{eq:convolutionKernel} and gives the following overlaps with three harmonic basis vectors $\Psi_{\sGen_i}\in\{E_{\frac{1}{2}+i\omega_i},\phi_{n_i},1\}$:
\begin{equation}
\boxed{
    \left(  {\cal Z}^{(0,3)}_{\text{RMT}_2},\,\Psi_{\sGen_1}\Psi_{\sGen_2} \Psi_{\sGen_3} \right) = 
    \,\sum_{m\geq 0} d_m \, {\bar{\tt c}}_m^{(\sGen_1)}{\bar{\tt c}}_m^{(\sGen_2)} {\bar{\tt c}}_m^{(\sGen_3)} \, \frac{e^{- S_0(m)}}{ (8\pi m\gamma)^{3/2}} \, \prod_{i=1}^3 \Gamma(\sGen_i)\, \Gamma(1-\sGen_i)
}
\label{eq:Z03result}
\end{equation}
We will discuss this expression in detail below. These spectral overlaps fully specify the three-boundary RMT${}_2$ amplitude. The correlations are not diagonal in any index and they contain mixed correlators between the Eisenstein series and cusp forms. In the following we study the different overlaps in more detail. Evidently, the case $m=0$ of \eqref{eq:Z03result} is ill-defined and requires a more careful treatment. 

The physical content of \eqref{eq:Z03result} can be summarized as follows (details below): the leading near-extremal RMT statistics is due to the triple-Eisenstein overlap (for spins $j_1=j_2=j_3=0$) and the triple-cusp form overlap (for spins $j_1=j_2=j_3>0$). The mixed overlaps are subleading near extremality. 
Note that even the $m>0$ sectors do contain triple-Eisenstein overlaps: while they follow naturally from our Poincar\'e sum construction, they are also subleading near extremality.

%%%%%%%%%%%%%%%%%%%%%%%%%%%%%%%%%%%%%%%%%%%%%%%%
\subsubsection{Scalar Eisenstein sector ($m=0$)}
\label{sec:EEEscalar}
%%%%%%%%%%%%%%%%%%%%%%%%%%%%%%%%%%%%%%%%%%%%%%%%

Consider the universal RMT${}_2$ three-point function \eqref{eq:n3RMT2def} for $m= 0$. This case was analyzed in \cite{Boruch:2025ilr} and we refer to it as the {\it scalar Eisenstein sector}, keeping in mind that this nomenclature refers to the base spin $m=0$, not the physical spins $(j_1,j_2,j_3)$, which are all supported by this modular-invariant piece. The integral in \eqref{eq:n3RMT2def} can be performed for $m=0$:
\begin{equation}
    \begin{split}
        \lim_{\epsilon\rightarrow 0} \int_{1+i\mathbb{R}} \frac{ds_i}{2\pi i} \,\frac{2\epsilon}{\epsilon^2-(1-s_i)^2} \, P_{m=0}(\tau_i,s_i) 
        &= 
        \lim_{\epsilon\rightarrow 0} \int_{\mathbb{R}} \frac{d\omega_i}{2\pi } \,\frac{2\epsilon}{\epsilon^2+\omega_i^2} \, E_{1+i\omega_i}(\tau_i)
        \\
        &= \widehat{E}_1(\tau_i) 
        + \lim_{\epsilon\rightarrow 0} \int_{\mathbb{R}} \frac{d\omega_i}{2\pi } \,\frac{2\epsilon}{\epsilon^2+\omega_i^2} \, \frac{3}{\pi i \omega_i}
        \\
        &= \widehat{E}_1(\tau_i)  \qquad \text{(p.v. prescription)}
    \end{split}
\end{equation}
where we separated the Eisenstein series into its pole plus finite part,
\begin{equation}\label{e1hatdef}
    \widehat{E}_1(\tau) \equiv \lim_{s\rightarrow 1} \left( E_s - \frac{3}{\pi (s-1)} \right) \,,
\end{equation}
and used a principal value prescription to remove a moduli-independent divergence.
In conclusion, the $m=0$ sector of the RMT${}_2$ three-point correlator is
\begin{equation}
\label{eq:EhatResult}
    {\cal Z}^{(0,3)}_{\text{RMT}_2}(\tau_1,\tau_2,\tau_3 | 0,0,0) = \frac{e^{-S_0(0)}}{(2\pi \gamma)^{3/2}} \,  \widehat{E}_1(\tau_1) \widehat{E}_1(\tau_2) \widehat{E}_1(\tau_3)\,.
\end{equation}
In the scalar spectrum (i.e., the spins of all Eisenstein series $j_1=j_2=j_3=0$), this contribution is large and it reproduces the RMT input. Indeed, the spin-zero component of the regularized Eisenstein series satisfies $\widehat{E}_{1|j=0}(y) = y + {\cal O}(\log y)$ such that
\begin{equation}
\label{eq:n3limit}
    {\cal Z}^{(0,3)}_{\text{RMT}_2|\{j_i=0\}}(y_1,y_2,y_3 | 0,0,0) \; \stackrel{\substack{\text{near-}\\ \text{extremal}}}{\longrightarrow} \; \frac{e^{-S_0(0)}}{(2\pi \gamma)^{3/2}} \, y_1y_2y_3 = \sqrt{y_1y_2y_3} \, \text{K}^{(0,3)}_\text{RMT}(S_0(0);y_1,y_2,y_3) \,.
\end{equation}
On the other hand, if any spin $j_i \neq 0$, the $m_i=0$ piece \eqref{eq:EhatResult} is subleading compared to the RMT expectation; RMT in spinning sectors is captured by other components of ${\cal Z}^{(0,3)}_{\text{RMT}_2}(\tau_1,\tau_2,\tau_3)$, as we will see momentarily.

%%%%%%%%%%%%%%%%%%%%%%%%%%%%%%%%%%%%%%%%%%%%%%%%
\subsubsection{Triple Eisenstein overlaps ($m>0$)}
\label{sec:EEEmneq0}
%%%%%%%%%%%%%%%%%%%%%%%%%%%%%%%%%%%%%%%%%%%%%%%%

Consider the case where all basis vectors in \eqref{eq:Z03result} are Eisenstein series. We note that the result factorizes and can thus be written as follows:
\es{}{
\label{eq:EEE}
\left( {\cal Z}_{\text{RMT}_2}^{(0,3)},E_{s_1}E_{s_2}E_{s_3}\right)= \frac{1}{4(8\pi \gamma)^{3/2}} \,\widetilde{D}_{E_{1-s_1}\times E_{1-s_2} \times E_{1-s_3}}\left(S_0;\frac{3}{2}\right)  
\qty(\prod_{i=1}^{3}\Gamma(s_i)\Gamma(1-s_i))  + [m\!=\!0 \text{ term}]
}
with the {\it deformed Eisenstein L-function}
\begin{equation}
    \widetilde{D}_{E_{s_1}\times E_{s_2} \times E_{s_3}}(S_0;t) 
    \equiv 8\sum_{m> 0} \frac{{\tt a}_{m}^{(s_1)}{\tt a}_{m}^{(s_2)}{\tt a}_{m}^{(s_3)}}{m^t}\, e^{-\frac{2}{3}t\,S_0(m)}\,.
\end{equation}
The $S_0(m)$-dependent exponential is the deformation; without it, this would be the standard triple product Dirichlet series. See Appendix \ref{app:Lfunctions} for more details.

The $m=0$ term was discussed separately above (see Section \ref{sec:EEEscalar}) and was seen to dominate in the scalar spectrum ($j_1=j_2=j_3=0$). Let us therefore consider the $m>0$ terms in \eqref{eq:EEE}. We expect the following hierarchy: for $j_1=j_2=j_3=0$, the scalar Eisenstein piece \eqref{eq:n3limit} dominates and it is ${\cal O}(y_1y_2y_3)$. For $j_1=j_2=j_3=m>0$, the contribution \eqref{eq:EEE} is small compared to the scalar contribution, but still ${\cal O}(\sqrt{y_1y_2y_3})$.
We will now confirm this. 

The triple Eisenstein overlap clearly factorizes into three identical factors, each with an integration kernel $\Gamma(s_i)\Gamma(1-s_i)$ that is reminiscent of the two-boundary wormhole. To evaluate the integrals over Eisenstein series, we require the following auxiliary integral, which computes the imprint of the spin $m>0$ Eisenstein seed onto the physical spin $j_i>0$ sector:
\begin{equation}
\int_{\mathbb{R}}\frac{d\omega_i}{4\pi} \, E_{\frac{1}{2}+i\omega_i|j_i}(y_i)
\, \frac{{\tt a}_m^{(1-s_i)}}{(8\pi m \gamma)^{1/2}} \, e^{-\frac{1}{3}S_0(m)} \, \Gamma(s_i)\Gamma(1-s_i)
\equiv  \frac{e^{-\frac{1}{3}S_0(m)}}{(8\pi m \gamma)^{1/2}}\times  \frac{1}{4\pi}\,e^{-2\pi j_i y_i} \, I_{m|j_i}(y_i)\,.
\end{equation}
The full $n=3$ wormhole triple Eisenstein piece consists of three such factors (diagonal in $m$, but with arbitrary physical spins $j_1,j_2,j_3$).
For simplicity, consider $j_1=j_2=j_3\equiv j$ and $m_1=m_2=m_3 \equiv m$ (other cases are similar). Then we get the following Eisenstein contribution:
\begin{equation}
\label{eq:ZjjjmmmEisenstein}
{\cal Z}^{(0,3)}_{\text{RMT}_2|j,j,j}(\tau_1,\tau_2,\tau_3 | m,m,m)
\supset \frac{2\,e^{-S_0(m)}}{(8\pi m \gamma)^{3/2}} \times \frac{1}{(4\pi)^3} \, e^{-2\pi j (y_1+y_2+y_3)} \, I_{m|j}(y_1) \, I_{m|j}(y_2)  \, I_{m|j}(y_3) \,,
\end{equation}
where we scaled out conventional constants and the near-extremal ground state factor for the spin $j$ spectrum.
In Appendix \ref{sec:appEEE}, we study $I_{m|j}(y)$ in more detail and confirm the following behavior:
\begin{itemize}
    \item $jm$ a perfect square (in particular $j=m$): the imprint of the base spin $m$ on the physical spin $j$ sector is largest. Evaluating the integral $I_{m|j}(y)$, we find that the result is well fit by $I_{m|j}(y) \sim \sqrt{y}$. 
    Removing the usual extra factor $\sqrt{y}$ of all RMT${}_2$ amplitudes, we are left with a constant. This is subleading to the universal RMT expectation \eqref{03amplitude}.
    \item $jm$ not a perfect square: in this case the imprint is parametrically smaller, tending to a constant: $I_{m|j}(y) \sim \text{const}$. After removing the conventional factor $\sqrt{y}$, this becomes a decaying function of $y$.
\end{itemize}
In summary, this implies for the triple Eisenstein overlap of the RMT${}_2$ amplitude:
\begin{equation}
     \eqref{eq:ZjjjmmmEisenstein} \; \stackrel{\substack{\text{near-}\\ \text{extremal}}}{\longrightarrow} \, \left\{
     \begin{aligned}
    & \text{const} \times e^{-S_0(m)} \,e^{-2\pi j (y_1+y_2+y_3)}\, \sqrt{y_1y_2y_3}
     \qquad (jm \text{ perfect square}) \\
   &  \text{const} \times e^{-S_0(m)} \,e^{-2\pi j (y_1+y_2+y_3)}\qquad\qquad\quad\;\;\, (\text{otherwise}) 
     \end{aligned}
     \right.
    \label{eq:EEEfinal}
\end{equation}
Even though the aligned case ($j=m$) is enhanced, it is still subleading compared to the RMT expectation, cf.\ \eqref{eq:n3limit}. This shows that, near the cusp, the Eisenstein piece of ${\cal Z}^{(0,3)}_{\text{RMT}_2}(\tau_1,\tau_2,\tau_3)$ is only dominant in the scalar sector $j_1=j_2=j_3=0$, where it reduces to the RMT prediction. To uncover the leading RMT scaling in the spinning spectrum, we turn to the Maass cusp forms.

%%%%%%%%%%%%%%%%%%%%%%%%%%%%%%%%%%%%%%%%%%%%%%%%
\subsubsection{Triple cusp form overlap}
\label{sec:phiphiphi}
%%%%%%%%%%%%%%%%%%%%%%%%%%%%%%%%%%%%%%%%%%%%%%%%

Suppose that the three basis elements in \eqref{eq:Z03result} are cusp forms:
\es{eq:phiphiphi0}{
\left(  {\cal Z}^{(0,3)}_{\text{RMT}_2},\,\phi_{n_1}\phi_{n_2}\phi_{n_3} \right) =
    \frac{1}{4(8\pi \gamma)^{3/2}} \, \widetilde{D}_{\phi_{n_1}\times\phi_{n_2}\times\phi_{n_3}}\left(S_0;\frac{3}{2}\right) \, \left( \prod_{i=1}^3 \frac{\pi}{\cosh(\pi \omega_{n_i})} \right)\,,
}
with the {\it deformed triple-product Dirichlet series}
\begin{equation}
    \widetilde{D}_{\phi_{n_1}\times\phi_{n_2}\times\phi_{n_3}}(S_0;t) \equiv 8\sum_{m>0} \frac{{\tt b}_m^{(n_1)}{\tt b}_m^{(n_2)}{\tt b}_m^{(n_3)}}{m^t} \, e^{-\frac{2}{3} t \, S_0(m)} \,.
\end{equation}
See again Appendix \ref{app:Lfunctions} for more details on the Dirichlet series and L-functions relevant in this context.

In Appendix \ref{app:cuspcuspcusp}, we present a numerical analysis of the sum over the triple product of cusp forms with the overlaps \eqref{eq:phiphiphi0}.  Our numerical findings confirm that this piece contributes to the spin-$\{j_1,j_2,j_3\}$ near-extremal spectrum as follows:
\begin{equation}
\begin{split}
    &{\cal Z}^{(0,3)}_{\text{RMT}_2|j_1,j_2,j_3}(y_1,y_2,y_3|m,m,m)\\
    &\quad \supset\frac{e^{- S_0(m) -2\pi(j_1y_1+j_2y_2+j_3y_3)}}{(2\pi\gamma)^{3/2}} \times  \frac{1}{4}\left( y_1 \, \delta_{mj_1} \right)\left( y_2 \, \delta_{mj_2} \right)\left( y_3 \, \delta_{mj_3} \right) + \ldots \qquad (y_i \rightarrow \infty) 
    \\ 
    &\quad \equiv \frac{1}{4} \,e^{ -2\pi(j_1y_1+j_2y_2+j_3y_3)}\,    \sqrt{y_1y_2y_3} \, \text{K}_{\text{RMT}}^{(0,3)}(S_0(m);y_1,y_2,y_3) \,\delta_{mj_1}\delta_{mj_2}\delta_{mj_3} + \ldots   \qquad (y_i\rightarrow\infty)
\end{split}
\end{equation}
This is consistent with our basic axiom \eqref{eq:NErequirement}, thus confirming that the triple-cusp form overlap is responsible for near-extremal RMT universality in the spinning spectrum.

%%%%%%%%%%%%%%%%%%%%%%%%%%%%%%%%%%%%%%%%%%%%%%%%
\subsubsection{Mixed overlaps}
%%%%%%%%%%%%%%%%%%%%%%%%%%%%%%%%%%%%%%%%%%%%%%%%

Suppose that there is at least one cusp form among the three elements, then $m\geq 0$ sum will start from $m=1$, since the cusp forms have no scalar component: 
{\small
\begin{equation}
\label{eq:phiphiphi}
    \begin{split}
        \left(  {\cal Z}^{(0,3)}_{\text{RMT}_2},\,\phi_{n_1}\phi_{n_2}\, E_{s_3} \right) &=2
    \left(\sum_{m> 0} {\tt b}_m^{(n_1)} {\tt b}_m^{(n_2)} \,{{\tt a}}_m^{(\frac{1}{2}-i\omega_3)} \, \frac{e^{- S_0(m)}}{ (8\pi m\gamma)^{3/2}} \right) \frac{\pi^3}{\cosh(\pi \omega_{n_1})\cosh(\pi \omega_{n_2})\cosh(\pi \omega_3)} 
    \\
    \left(  {\cal Z}^{(0,3)}_{\text{RMT}_2},\,\phi_{n_1}E_{s_2} E_{s_3} \right) &=2
    \left(\sum_{m> 0} {{\tt b}}_m^{(n_1)}\,{{\tt a}}_m^{(\frac{1}{2}-i\omega_2)} {{\tt a}}_m^{(\frac{1}{2}-i\omega_3)} \, \frac{e^{- S_0(m)}}{ (8\pi m\gamma)^{3/2}} \right)  \frac{\pi^3}{\cosh(\pi \omega_{n_1})\cosh(\pi \omega_2)\cosh(\pi \omega_3)} 
    \end{split}
\end{equation}
}\normalsize
One can again define deformed mixed Eisenstein-cusp form L-functions to write the universal prefactor in a suggestive form.
Due to the factorization for every fixed $m$, the contributions of these overlaps to ${\cal Z}^{(0,3)}_{\text{RMT}_2}$ can be computed numerically using the results from above. It follows immediately that the cusp forms contribute at leading order on their respective torus, while the Eisenstein series are subleading. For example, for $mj_3$ a perfect square, the first line of \eqref{eq:phiphiphi} contributes as 
\begin{equation}
\begin{split}
{\cal Z}^{(0,3)}_{\text{RMT}_2|j_1,j_2,j_3}(y_1,y_2,y_3) & \supset \sum_{n_1,n_2} \phi_{n_1|j_1}(y_1)\phi_{n_2|j_2}(y_2)\int_{{\cal C}_\text{crit}}\frac{ds_3}{4\pi i} \,  E_{s_3|j_3}(y_3) \left(  {\cal Z}^{(0,3)}_{\text{RMT}_2},\,\phi_{n_1}\phi_{n_2}\, E_{s_3} \right) 
\\
&= \sum_{m>0}  \frac{e^{-S_0(m)-2\pi(j_1y_1+j_2y_2+j_3y_3)}}{2(2\pi \gamma)^{3/2}} \, \left( y_1 \,\delta_{mj_1} \right)\left( y_2 \,\delta_{mj_2} \right) \left( c(m,j_3) \, \sqrt{y_3}  \right) + \text{subl.}
\end{split}
\end{equation}
where $c(m,j_3)$ is a numerical constant that depends on spins, similar to \eqref{eq:EEEfinal}.

%%%%%%%%%%%%%%%%%%%%%%%%%%%%%%%%%%%%%%%%%%%%%%%%
\subsection{Minimality and gravity} \label{ssec:minimality}
%%%%%%%%%%%%%%%%%%%%%%%%%%%%%%%%%%%%%%%%%%%%%%%%

A natural gravitational ansatz for the three-boundary AdS$_3$ pure gravity wormhole amplitude would be given by three sums over modular images of the seed \eqref{03amplitude} appropriately weighted with spins:\footnote{A closely related matrix model ansatz for multiboundary torus wormholes, completed via independent sums over modular images, was presented in \cite{deBoer:2025oge}. Our construction provides an explicit Poincar\'e series realization of this type of completion, including the regularization of the scalar sector and the spectral decomposition.}
\begin{equation}
\begin{split}
    {\cal Z}_{\text{grav}}(\tau_1,\tau_2,\tau_3) &\stackrel{?}{=} \sum_{\substack{\gamma_1,\gamma_2,\gamma_3 \in \\ \Gamma_\infty\setminus\text{SL}(2,\mathbb{Z})}} \, {\cal Z}^{(0,3)}_\text{seed}\left( \gamma_1\tau_1,\gamma_2\tau_2,\gamma_3\tau_3\right)
    \,,
\end{split}
\end{equation}
with the spin-diagonal (and parity-even on each torus) seed
\begin{equation}
\begin{split}
 {\cal Z}^{(0,3)}_\text{seed}(\tau_1,\tau_2,\tau_3)=\sum_{m_i\geq 0} d_{m_1}\delta_{m_1m_2}\delta_{m_2m_3} \, \sqrt{y_1y_2y_3} \,\text{K}^{(0,3)}_\text{RMT}(S_0(m_1);y_1,y_2,y_3)\,\prod_{k=1}^3 \cos(2\pi m_k x_k)\, e^{- 2\pi m_k y_k}
\end{split}
\end{equation}
By exchanging the SL$(2,\mathbb{Z})$ sums and the sum over spins $m_i$, one obtains Poincar\'e series, which must be regularized at $m=0$. The RMT${}_2$ construction provides a natural regularization scheme and leads to \eqref{eq:n3RMT2def2}, a pole-subtracted avatar of this ansatz.

More generally, while our axioms (Section \ref{sec:axioms}) reflect an aspiration towards minimality, we already noted that the explicit Poincar\'e sum construction of RMT$_2$ (Section \ref{sec:construction}) is only one of infinitely many solutions to the axioms. One may view it as a particularly natural solution with a distinctly gravitational flavor, since the SL$(2,\mathbb Z)$ sums implement the bulk mapping class group. Nevertheless, the axioms alone do not enforce this choice: a simple gravitational ansatz of the form 
\begin{equation}
\label{eq:naive_grav_ansatz}
\begin{split}
    {\cal Z}_{\text{grav}}(\tau_1,\dots,\tau_n) &\stackrel{?}{=} \sum_{\substack{\gamma_1,\dots,\gamma_n \in \\ \Gamma_\infty\setminus\text{SL}(2,\mathbb{Z})}} \, {\cal Z}_\text{seed}\left( \gamma_1\tau_1,\dots,\gamma_n\tau_n\right)
    \,,
\end{split} 
\end{equation}
may contain a subsector that is closed under modular transformations and vanishes in the near-extremal limit, and therefore does not follow simply from requirements of modularity and near-extremal random matrix statistics. For example, in the case of the three-boundary wormhole, the Eisenstein overlaps associated with base spins $m>0$ (such as \eqref{eq:EEE}) as well as the mixed overlaps (such as \eqref{eq:phiphiphi}) form such closed subsectors; these terms could be dropped and all axioms would still hold. 

This discussion demonstrates the sense in which the notion of `minimality' is, so far as we currently understand, not inherent to a definition of RMT${}_2$. The crucial question for holography is how the AdS$_3$ wormhole amplitudes fix the freedom in the modular completion of the near-extremal ansatz: do they realize such closed subsectors, and if they do, what is their physical meaning? Due to the natural appearance of $\text{SL}(2,\mathbb{Z})$ sums in gravity, it is tempting to conjecture an alignment of gravitational wormholes with the Poincar\'e frame, but this is speculative -- and underscores the need for a genuine, bottom-up AdS$_3$ calculation of $n>2$ wormhole amplitudes.

%%%%%%%%%%%%%%%%%%%%%%%%%%%%%%%%%%%%%%%%%%%%%%%%
\begin{center}
\begin{tikzpicture}
    \draw[line width=0.4pt]
        (-2,0) -- (-0.5,0);
    \draw[line width=1.2pt]
        (-0.5,0) -- (0.5,0);
    \draw[line width=0.4pt]
        (0.5,0) -- (2,0);
\end{tikzpicture}
\end{center}
%%%%%%%%%%%%%%%%%%%%%%%%%%%%%%%%%%%%%%%%%%%%%%%%
\subsection*{Acknowledgments}
We thank Andreas Blommaert, Tom Hartman, Luca V. Iliesiu, Daniel Jafferis, Kristan Jensen, Greg Moore, Sridip Pal, Boris Post, Steve Shenker, Douglas Stanford, Elisa Tabor, Gustavo Joaquin Turiaci and Edward Witten for useful discussions. JB is supported by the Leinweber Foundation and the Leinweber Institute for Theoretical Physics at Stanford University. GDU is supported by the
Japan Science and Technology Agency (JST) as part
of Adopting Sustainable Partnerships for Innovative Research Ecosystem (ASPIRE), Grant No.\ JPMJAP2318. GDU is partly supported by the Leinweber Institute
for Theoretical Physics at UC Berkeley, by the Department of Energy, Office of Science, Office of High Energy Physics through the award DE-SC0025522, and by the
Department of Energy through QuantISED award DE-SC0019380. FMH is supported by UK Research and Innovation (UKRI) under the UK government's Horizon Europe Funding Guarantee EP/X030334/1. This research was supported in part by grant NSF PHY-2309135 to the Kavli Institute for Theoretical Physics (KITP). MR is supported by a Discovery grant from NSERC. This work was supported in part by ERC Starting Grant 853507. This work was made possible by Institut Pascal at Universit\'e Paris-Saclay with the support of the program ``Investissements d'avenir'' ANR-11-IDEX-0003-01.\\
AI tools have been used in the final stages of the project for inspiration, minor editing tasks, and adversarial review.

%%%%%%%%%%%%%%%%%%%%%%%%%%%%%%%%%%%%%%%%%%%%%%%%
\appendix

%%%%%%%%%%%%%%%%%%%%%%%%%%%%%%%%%%%%%%%%%%%%%%%%
%%%%%%%%%%%%%%%%%%%%%%%%%%%%%%%%%%%%%%%%%%%%%%%%
\section{RMT in Mellin space}
\label{app:RMTmellin}
%%%%%%%%%%%%%%%%%%%%%%%%%%%%%%%%%%%%%%%%%%%%%%%%
%%%%%%%%%%%%%%%%%%%%%%%%%%%%%%%%%%%%%%%%%%%%%%%%

In this appendix we explain some of the techniques involved in computing Mellin transforms of RMT spectral correlators. RMT in Mellin space is interesting in its own right, as it provides a novel perspective on the topological expansion. For simplicity, we focus on the GUE universality class, but we do consider general spectral curves for this case. See Section \ref{sec:GOEAiry} for one example of a GOE matrix model; a more general formalism would be interesting to develop.

%%%%%%%%%%%%%%%%%%%%%%%%%%%%%%%%%%%%%%%%%%%%%%%%
\subsection{$\uptau$-scaling and random matrix universality for general spectral curve}
%%%%%%%%%%%%%%%%%%%%%%%%%%%%%%%%%%%%%%%%%%%%%%%%

Before discussing specific examples, we will study RMT in Mellin space for a general class of models. We assume a spectral curve with general expansion and a square root edge:
\beq
\label{eq:rhoGeneral}
\begin{split}
\rho_0(E) &= \frac{1}{2\pi} \, \sqrt{E}  \, \sum_{k\geq 0} r_{k} \, E^k\,,\qquad r_0=1\,.
\end{split}
\eeq
We will explore the $\uptau$-scaling limit of the spectral form factor associated with spectral curves of this form.

\paragraph{$\uptau$-scaling limit.} 
Consider the spectral form factor (SFF), i.e., $\text{K}_\text{RMT}^{(2)}(y_1,y_2)$ analytically continued to the following quantity:
\beq
 \text{K}^{(2)}_\text{RMT}(\beta,T) \equiv \text{K}^{(2)}_\text{RMT}(y_1=\beta+iT\,,\, y_2 = \beta-iT ) \,.
\eeq
It is convenient to discuss a simplifying limit, which clarifies the behavior of the spectral correlators at very late times and focuses on universal quantities. Following \cite{Saad:2022kfe} (see also \cite{PhysRevE.72.046207,Okuyama:2020ncd,Blommaert:2022lbh,DiUbaldo:2025cmg}) we recall the $\uptau$-scaling limit defined in \eqref{tauscdef}:
\begin{equation}
    T, \, e^{S_0} \rightarrow \infty\,,  \qquad \uptau = T\, e^{-S_0} \text{ fixed.}
\end{equation}
In this limit we define
\begin{align}\label{genusexpansionSFF}
{\cal K}_{\uptau\text{-RMT}}^{(2)}(\beta,\uptau) 
% \equiv e^{-S_0} \, \text{K}_{\uptau\text{-RMT}}^{(2)}(\beta,\uptau)
\equiv \lim_{\substack{{T,\,e^{S_0} \rightarrow \infty} \\ \uptau = Te^{-S_0} \text{ fix} }}
 \; e^{-S_0} \, \text{K}^{(2)}_\text{RMT}(\beta,T) \,,
\end{align}
A crucial observation is that the right-hand side is finite and it admits a {\it topological expansion} in powers of $\uptau$ with a nonzero radius of convergence. That is, we can expand in genus:
\begin{align}\label{genusexpansionSFF2}
{\cal K}_{\uptau\text{-RMT}}^{(2)}(\beta,\uptau)
 = \sum_{g = 0}^{\infty}  {\cal K}^{(2,g)}_{\uptau\text{-RMT}}(\beta,\uptau) \,,
\end{align}
where $\text{K}^{(2,g)}_{\uptau\text{-RMT}} \sim \uptau^{2g+1}$ encodes the genus $g$ contribution, weighted by $e^{-(2g+1)S_0}$ (hidden inside $\uptau$).
This scaling thus allows us to explicitly study the transition from the ``ramp'' to the late-time ``plateau''.

Random matrix correlators in the $\uptau$-scaling limit are universally determined by the spectral curve $\rho_0(E)$. The universal form is most easily expressed in microcanonical variables, where it takes the well-known piecewise ramp/plateau structure \cite{Liu:2018hlr}. This is related to the canonical expression by a Laplace transform:
\begin{equation}
   {\cal K}^{(2)}_{\uptau\text{-RMT}}(\beta,\uptau) 
   = \int_{0}^{\infty} dE\, e^{-2 \beta E} \, \text{min} \left\{\frac{\uptau}{2\pi}\,,\,\rho_0(E) \right\}  \,.
\end{equation}
It was shown in \cite{Saad:2022kfe,Blommaert:2022lbh} that the coefficients in the $\uptau$-scaled SFF can be obtained in terms of low-energy residues. The result is:
\beq
\begin{split}
{\cal K}_{\uptau\text{-RMT}}^{(2)}(\beta,\uptau) &= \frac{\uptau}{4\pi \beta} - \frac{1}{2\pi} \sum_{g \geq 1}  \frac{c_{2g}(\rho_0;\beta) }{g (2g+1)} \, \uptau^{2g+1}\,,\\
  c_{2g}(\rho_0;\beta) &\equiv \frac{1}{(2\pi)^{2g}} \oint_0 \frac{dE}{2\pi i}\; \frac{e^{-2\beta E}}{(\rho_0(E))^{2g}} \,,
\end{split}
\label{eq:GUEgeneralK}
\eeq
where the contour integral is counterclockwise around $E=0$. In particular, the coefficients $c_{2g}$ are only sensitive to the low-energy edge of the spectrum.

\paragraph{Expansion in powers of $\uptau$ and $\beta$.}
Due to the general structure of \eqref{eq:GUEgeneralK}, it can be useful to define coefficients for inverse integer powers of $\rho_0$:
\beq 
\label{eq:rhoInvExpansion}
\begin{split}
  \frac{1}{\left(\rho_0(E)\right)^{2g}} &= (2\pi)^{2g} \, E^{-g} \, \sum_{k\geq 0} \alpha_k(g) \, E^k \,.
\end{split}
\eeq 
The relation between the expansions \eqref{eq:rhoGeneral} and \eqref{eq:rhoInvExpansion} is
\beq
\label{eq:alphaRel}
 \alpha_k(g) = \sum_{\substack{m_1,m_2,m_3,\ldots \geq 0 \\ \sum_n n m_n = k}} \frac{\Gamma(2g+\sum_n m_n)}{\Gamma(2g)} \, \prod_n \frac{(-r_n)^{m_n}}{m_n!}  \,.
\eeq
For spectral curves of the form \eqref{eq:rhoGeneral}, the integrand in the coefficient $c_{2g}$ has a pole of order $g$ at $E=0$. The residues therefore take the following general form:
\begin{equation}
c_{2g}(\rho_0;\beta) = \sum_{k=0}^{g-1} \frac{\alpha_k(g)}{\Gamma(g-k)}\, (-2\beta)^{g-k-1} \,.
\end{equation}
In summary, we have the general structure
\beq
\begin{split}
{\cal K}_{\uptau\text{-RMT}}^{(2)}(\beta,\uptau) &= \frac{\uptau}{4\pi \beta} - \frac{1}{2\pi} \sum_{g \geq 1}  \uptau^{2g+1} \sum_{k=0}^{g-1} \frac{\alpha_k(g)}{g(2g+1)\Gamma(g-k)}\, (-2\beta)^{g-k-1}  \,.
\end{split}
\label{eq:GUEgeneralK2}
\eeq
For concreteness, in the Airy model one has $r_0=1$ and $r_{k\geq 1}=0$; equivalently, $\alpha_0=1$ and $\alpha_{k\geq 1}=0$, so the powers of $\uptau$ and $\beta$ in \eqref{eq:GUEgeneralK2} are always paired in the simplest form $\uptau^{2g+1} \beta^{g-1}$. See \cite{DiUbaldo:2025cmg} for more examples in the same spirit.

%%%%%%%%%%%%%%%%%%%%%%%%%%%%%%%%%%%%%%%%%%%%%%%%
\subsection{Mellin transform for general spectral curves}
\label{app:mellinGeneralRho}
%%%%%%%%%%%%%%%%%%%%%%%%%%%%%%%%%%%%%%%%%%%%%%%%

To formulate ($\uptau$-scaled) RMT in Mellin space, we compute the Mellin transform of the general result \eqref{eq:GUEgeneralK}. This can be performed abstractly:
\begin{equation}
\begin{split}
 f^{(2)}_{\uptau\text{-RMT}}(\bar s_1,\bar s_2) 
 &\equiv {\cal M}_{y_1,y_2} \left[ e^{S_0}\,{\cal K}^{(2)}_{\uptau\text{-RMT}} \left( \beta=\frac{y_1+y_2}{2} \,,\, \uptau = e^{-S_0}\sqrt{y_1y_2} \right) \right](-i\omega_1,-i\omega_2) \\
 &= \int_0^\infty \frac{dy_1}{y_1^{1+i\omega_1}}\int_0^\infty \frac{dy_2}{y_2^{1+i\omega_2}} \, e^{S_0}\,{\cal K}^{(2)}_{\uptau\text{-RMT}} \left( \beta=\frac{y_1+y_2}{2} \,,\, \uptau = e^{-S_0}\sqrt{y_1y_2} \right)
\\
& = \frac{\pi \delta(\omega_1+\omega_2)}{\cosh(\pi\omega_1)}  - \sum_{g\geq 1} \frac{e^{-2gS_0}}{2\pi g(2g+1)}  \Gamma\left(\frac{1}{2}+g -i\omega_1 \right) \Gamma\left( \frac{1}{2}+g-i\omega_2 \right) \mathfrak{c}_{2g}(\rho_0;\omega_+)
\end{split}
\label{eq:f2MellinGeneral}
\end{equation}
where the first term corresponds to the universal linear ramp ($g=0$) and the model-specific coefficients that only depend on the sum $\omega_+=\frac{1}{3}(\omega_1+\omega_2)$ are given by
\begin{equation}
 \mathfrak{c}_{2g}(\rho_0;\omega_+) = \frac{1}{(2\pi)^{2g}} \oint_0\frac{dE}{2\pi i} \, \frac{E^{3i\omega_+-2g-1}}{(\rho_0(E))^{2g}} \,.
\end{equation}
Naively, the contour integral now appears problematic because $E$ appears with a non-integer power. However, when inverting the Mellin transform (or evaluating RMT${}_2$ amplitudes), the $\omega_+$ integral localizes on poles at $i\omega_+=g-\frac{k}{3}$. The relevant power of $E$ in the contour integral is then in fact integer after all and the integral yields a simple residue. More generally, note that
\begin{equation}
 \oint_0 \frac{dE}{2\pi i} \, E^{\alpha-1} = \delta_{\alpha,0} \qquad (\alpha\in\mathbb{Z}) \,.
 \label{eq:deltaId}
\end{equation}
We can formally work with $\delta_{\alpha,0}$ for non-integer values of $\alpha$. This simply stands for a coefficient which is unity if $\alpha=0$ and zero for any other $\alpha$.\footnote{This formal manipulation can be made more precise by performing the contour integration in \eqref{eq:deltaId} explicitly over a circle with radius $\varepsilon$. The result is then:
\begin{equation}
\oint_{|E|=\varepsilon} \frac{dE}{2\pi i} \, E^{\alpha-1} = \frac{\varepsilon^\alpha}{2\pi i \alpha} \left( e^{2\pi i \alpha}-1\right)\,.
\label{eq:deltaDefCircle}
\end{equation}
For $\alpha \geq 0$ or $\alpha\in\mathbb{Z}$, this analytic continuation indeed produces $\delta_{\alpha,0}$ in the limit $\varepsilon\rightarrow 0$. For other values of $\alpha$ it is used formally.}
Using the expansion \eqref{eq:rhoInvExpansion}, we then find formally:
\begin{equation}
 \mathfrak{c}_{2g}(\rho_0;\omega_+) 
  = \sum_{k\geq 0} \alpha_k(g) \, \delta_{3i\omega_+,3g-k} \,.
\end{equation}

To summarize, the universal $\uptau$-scaled RMT Mellin space form factor for an arbitrary spectral curve takes the form
\begin{equation}
\begin{split}
 f^{(2)}_{\uptau\text{-RMT}}(\bar s_1,\bar s_2) 
 &= \frac{\pi \delta(\omega_1+\omega_2)}{\cosh(\pi\omega_1)}  - \sum_{g\geq 1} \frac{\Gamma\left(\frac{1}{2}+g -i\omega_1 \right) \Gamma\left( \frac{1}{2}+g-i\omega_2 \right)}{2\pi g(2g+1) \,e^{2gS_0}}   \sum_{k\geq 0} \alpha_k(g) \, \delta_{3i\omega_+,3g-k} \\
 &= \frac{\pi \delta(\omega_1+\omega_2)}{\cosh(\pi\omega_1)}  - \sum_{k\geq 0}  \frac{\Gamma\left(\frac{1}{2}+\frac{k}{3}-\frac{2i\omega_1}{3}+\frac{i\omega_2}{3} \right) \Gamma\left( \frac{1}{2}+\frac{k}{3}+\frac{i\omega_1}{3}-\frac{2i\omega_2}{3} \right)}{6\pi (i\omega_++\tfrac{k}{3})(2i\omega_+ + \tfrac{2k}{3}+1) \, e^{(2i\omega_++\frac{2k}{3})S_0}} \,\alpha_k(i\omega_++\tfrac{k}{3}) \,,
\end{split}
\end{equation}
where we formally performed the sum over $g$.\footnote{Performing the sum over $g$ using the Kronecker-delta requires the unusual replacement $\delta_{3i\omega_+,3g-k} = \frac{1}{3} \delta_{i\omega_+,g-\frac{k}{3}}$. This transformation is similar to the transformation of a Dirac-delta function and it follows from the regularized definition \eqref{eq:deltaDefCircle}.} The topological expansion is thus completely reorganized into the analytic structure of the gamma functions and $\alpha_k$ coefficients.

As a final step, we can conveniently absorb the separate $g=0$ term into the sum over gamma functions: let us introduce a small regulator $\epsilon$ to indicate a convention where the simple pole at $\omega_+ = \frac{k}{3}i$ lies in the lower half plane for $k=0$. We claim that this accounts for the genus-zero pole, leading to our final formula:
\begin{equation}
\boxed{
 f^{(2)}_{\uptau\text{-RMT}}(\bar s_1,\bar s_2) 
 = - \sum_{k\geq 0}  \frac{\Gamma\left(\frac{1}{2}+\frac{k}{3}-\frac{i\omega_+}{2}+\frac{3i\omega_-}{2}\right) \Gamma\left( \frac{1}{2}+\frac{k}{3}-\frac{i\omega_+}{2}-\frac{3i\omega_-}{2}\right)}{6\pi (i\omega_++\tfrac{k}{3}-\epsilon)(2i\omega_+ + \tfrac{2k}{3}+1)\,e^{(2i\omega_++\frac{2k}{3})S_0}} \,\alpha_k(i\omega_++\tfrac{k}{3})\,.
}
\label{eq:f2MellinGeneralFinal}
\end{equation}
This is the non-perturbative Mellin space RMT correlator for arbitrary spectral curve (encoded in the coefficients $\alpha_k$ via \eqref{eq:rhoInvExpansion}). For instance, in the Airy model, the sum over $k$ only has a single term, $\alpha_0^\text{Airy}=1$ with $\alpha_{k\geq 1}^\text{Airy}=0$, which reproduces \eqref{eq:tauAiryf}.\footnote{For general spectral curves, \eqref{eq:f2MellinGeneralFinal} is formal and may require resummation.}

\paragraph{Exercise: inverse Mellin transform.}
As an instructive cross-check of \eqref{eq:f2MellinGeneralFinal}, we confirm that the inverse Mellin transform works as advertised. That is, we compute:
\begin{equation}
\begin{split}
& {\cal M}^{-1}_{-i\omega_1,-i\omega_2} \left[ f^{(2)}_{\uptau\text{-RMT}}(\bar s_1,\bar s_2) \right](y_1,y_2) \\
 & = -  \sum_{k\geq 0} \int_{\mathbb{R}} \frac{ 9d\omega_+ d\omega_-}{2(2\pi)^2}\, ( y_1 y_2 )^{\frac{3i \omega_+}{2}} \left( \frac{y_1}{y_2} \right)^{\frac{3i \omega_-}{2}}\,\frac{\Gamma\left(\frac{1}{2}+\frac{k}{3}-\frac{i\omega_+}{2}+\frac{3i\omega_-}{2}\right) \Gamma\left( \frac{1}{2}+\frac{k}{3}-\frac{i\omega_+}{2}-\frac{3i\omega_-}{2}\right)\alpha_k(i\omega_++\tfrac{k}{3})}{6\pi (i\omega_++\tfrac{k}{3}-\epsilon)(2i\omega_+ + \tfrac{2k}{3}+1)\,e^{(2i\omega_++\frac{2k}{3})S_0}} 
 \end{split}
\end{equation}
The $\omega_-$ integral can be performed in complete generality: we deform the $\omega_-$ contour into the upper half plane (if $\frac{y_1}{y_2}>1$) or lower half plane (if $\frac{y_1}{y_2}<1$) and collect poles of the gamma functions. This gives:
\begin{equation}
\begin{split}
& {\cal M}^{-1}_{-i\omega_1,-i\omega_2} \left[ f^{(2)}_{\uptau\text{-RMT}}(\bar s_1,\bar s_2) \right](y_1,y_2) \\
  &= -\frac{1}{4\pi^2}\sum_{k\geq 0} \int_{\mathbb{R}} d\omega_+\, (y_1y_2)^{i\omega_++\frac{2k+3}{6}}\left(y_1+y_2\right)^{i\omega_+-\frac{2k}{3}-1} \, \frac{\Gamma\left( 1 +\frac{2k}{3}-i\omega_+ \right)\, \alpha_k(i\omega_++\tfrac{k}{3})}{ (i\omega_++\tfrac{k}{3}-\epsilon)(2i\omega_+ + \tfrac{2k}{3}+1)\,e^{(2i\omega_++\frac{2k}{3})S_0}} .
\end{split}
\end{equation}
To perform the remaining integral, we deform the contour into the lower half plane. There are two types of poles: 
\begin{enumerate}
\item A simple pole at $\omega_+ = (\frac{k}{3}-\epsilon) i$, which is associated with the genus-zero linear ramp. This pole only lies in the lower half plane for $k=0$. In this case, $\alpha_{k}(0) = \delta_{k,0}$ is the only non-vanishing coefficient (for any spectral curve).
\item A series of gamma function  poles at $\omega_+ = -\frac{i}{3}(2k+3n)$ for $n=1,2,3,\ldots$. As we will see momentarily, these poles are associated with the topological expansion. We observe that $\alpha_k(i\omega_++\frac{k}{3}) \rightarrow \alpha_k(n+k)$ does get evaluated on an integer argument, as it must. This confirms the consistency of our manipulations with the analytically continued Kronecker-delta.
\end{enumerate}
The contribution of these residues is as follows:
\begin{equation}
\begin{split}
& {\cal M}^{-1}_{-i\omega_1,-i\omega_2} \left[ f^{(2)}_{\uptau\text{-RMT}}(\bar s_1,\bar s_2) \right](y_1,y_2) \\
  & = \frac{1}{2\pi} \frac{\sqrt{y_1y_2}}{y_1+y_2} -\frac{1}{2\pi}\sum_{n\geq 1}\sum_{k\geq 0}(y_1y_2)^{n+k+\frac{1}{2}} (y_1+y_2)^{n-1}\,\frac{(-1)^{n-1} \,\alpha_k(n+k)}{(n+k)(2n+2k+1)(n-1)!\, e^{2(n+k)S_0}}   \\
  &= \frac{1}{2\pi} \frac{\sqrt{y_1y_2}}{y_1+y_2} -\frac{1}{2\pi}\sum_{g\geq 1}\sum_{k= 0}^{g-1}(y_1y_2)^{g+\frac{1}{2}} (y_1+y_2)^{g-k-1}\,\frac{(-1)^{g-k-1} \alpha_k(g)}{g(2g+1)(g-k-1)!\,e^{2gS_0}} 
  \\
  &= e^{S_0} \, {\cal K}^{(2)}_{\uptau\text{-RMT}}\left(\frac{y_1+y_2}{2},\,e^{-S_0}\sqrt{y_1y_2}\right) \,,
\end{split}
\end{equation}
where we relabeled $\sum_{n\geq 1}\sum_{k\geq 0} f(n,k) = \sum_{g\geq 1} \sum_{k=0}^{g-1} f(g-k,k)$. We have thus reproduced the starting point \eqref{eq:GUEgeneralK2}.

%%%%%%%%%%%%%%%%%%%%%%%%%%%%%%%%%%%%%%%%%%%%%%%%
\subsection{A different method: analytic continuation in genus}
\label{app:analyticContGenus}
%%%%%%%%%%%%%%%%%%%%%%%%%%%%%%%%%%%%%%%%%%%%%%%%

There are other methods to compute Mellin transforms of the form \eqref{eq:f2MellinGeneral}. We briefly present one, which is based on an analytic continuation in genus.

We begin by writing the general topological expansion \eqref{eq:GUEgeneralK} in terms of the model-specific coefficients $\alpha_k(g)$, defined in \eqref{eq:rhoInvExpansion}, as a Sommerfeld-Watson-like contour integral around non-negative integers:
\begin{equation}
\begin{split}
 {\cal K}^{(2)}_{\uptau\text{-RMT}} \left( \beta,\uptau \right) = \sum_{k\geq 0} \oint_{\mathbb{Z}_+} \frac{dG}{2\pi i}\frac{1}{\sin(\pi G)} \frac{ (-1)^k\alpha_k(G)}{2G(2G+1)\Gamma(G-k)} \, \uptau^{2G+1} \,(2\beta)^{G-k-1}\,.
\end{split}
\label{eq:KanalyticG}
\end{equation}
The sum over $k$ automatically truncates when the pole of $\sin(\pi G)$ is lifted by the factor $\Gamma(G-k)$. It is easy to confirm that the contour integral also reproduces the $g=0$ term correctly.

\begin{figure}
    \centering
    \includegraphics[width=\textwidth]{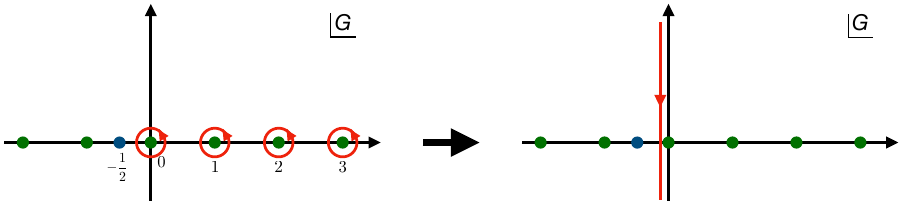}
    \caption{Contour deformation performed in the analytic continuation of the genus expansion.}
    \label{fig:contourAiryg}
\end{figure}

We now return to the Mellin transform \eqref{eq:f2MellinGeneral}:
\begin{equation}
\begin{split}
 &f^{(2)}_{\uptau\text{-RMT}}(\bar s_1,\bar s_2) \equiv {\cal M}_{y_1,y_2} \left[ e^{S_0} \, {\cal K}^{(2)}_{\uptau\text{-RMT}} \left( \beta=\frac{y_1+y_2}{2} \,,\, \uptau = e^{-S_0}\sqrt{y_1y_2} \right) \right](-i\omega_1,-i\omega_2) \,.
\end{split}
\label{eq:f2MellinGeneralG}
\end{equation}
We wish to perform the Mellin transform inside the $G$-integral, using the identity
\begin{equation}
\label{eq:MellinIdentity}
\begin{split}
    &{\cal M}_{y_1,y_2} \left[(y_1y_2)^{G+\frac{1}{2}} \,(y_1+y_2)^{G-k-1} \right](-i\omega_1,-i\omega_2) 
    \\
    &\qquad = \frac{2\pi}{3} \,\frac{ \Gamma\left(\frac{1}{2}+G-i\omega_1 \right) \Gamma\left( \frac{1}{2} -2 G+k + i\omega_1 \right)}{\Gamma(1-G+k)} \, \delta\left(\omega_++i G- \tfrac{ik}{3}\right) \,.
\end{split}
\end{equation}
However, this identity is only valid for $-\frac12<\text{Re}(G)<\frac{2k+1}{4}$. In order to use it inside \eqref{eq:KanalyticG}, we first need to deform the $G$ contour  into a downward vertical contour just to the left of the origin for $k=0$ (and appropriately shifted for $k>0$):
\begin{equation}
\begin{split}
 {\cal K}^{(2)}_{\uptau\text{-RMT}} \left( \beta,\uptau \right) = \sum_{k\geq 0} \int_{\frac{k}{3}-\epsilon-i\mathbb{R}} \frac{dG}{2\pi i}\frac{1}{\sin(\pi G)} \frac{ (-1)^k\alpha_k(G)}{2G(2G+1)\Gamma(G-k)} \, \uptau^{2G+1} \,(2\beta)^{G-k-1}\,.
\end{split}
\end{equation}
This is illustrated in Fig.\ \ref{fig:contourAiryg}.
The use of the identity \eqref{eq:MellinIdentity} is now permitted inside the $G$-integral, and we get:
\begin{equation}
\begin{split}
 &f^{(2)}_{\uptau\text{-RMT}}(\bar s_1,\bar s_2) \\
 &\quad = \sum_{k\geq 0} \int_{\frac{k}{3}-\epsilon-i\mathbb{R}} \frac{dG}{2\pi i}  \frac{ \alpha_k(G) \,e^{-2GS_0}}{3G(2G+1)}  \,\Gamma\left(\frac{1}{2}+G-i\omega_1 \right) \Gamma\left( \frac{1}{2} -2 G+k + i\omega_1 \right) \, \delta\left(\omega_++i G- \tfrac{ik}{3}\right)
 \\
 &\quad = - \sum_{k\geq 0} \frac{ \alpha_k(i\omega_++\tfrac{k}{3}) \, e^{-2\left(\frac{k}{3}+i\omega_+\right)S_0}}{6\pi(i\omega_++\tfrac{k}{3})(2i\omega_++\tfrac{2k}{3}+1)}  \,\Gamma\left(\frac{1}{2}+\frac{k}{3}-\frac{i\omega_+}{2}+\frac{3i\omega_-}{2}\right) \Gamma\left(\frac{1}{2}+\frac{k}{3}-\frac{i\omega_+}{2}-\frac{3i\omega_-}{2}\right) .
\end{split}
\label{eq:f2MellinGeneralG2}
\end{equation}
This agrees with our earlier result \eqref{eq:f2MellinGeneralFinal}.
The complex Dirac delta-function in the intermediate expression is a standard tool in Mellin transforms; it was possible to employ it consistently by deforming the $G$ contour such that the support of the delta-function lies on the integration contour.

\paragraph{The plateau pole.} We briefly comment on the pole at $G=-\frac{1}{2}$ of the integrand \eqref{eq:KanalyticG} (shown in blue in Fig.\ \ref{fig:contourAiryg}). This pole is not picked up by the natural contour deformation that develops the genus expansion. It would be picked up by deforming the vertical $G$ contour to the {\it left}, which introduces a non-vanishing arc contribution at infinity. We ignore this arc, but only comment on the role of the pole at $G=-\frac{1}{2}$. Its residue is as follows:
\begin{equation}
\label{eq:plateauPole}
    {\cal K}^{(2)}_{\uptau\text{-RMT}}(\beta,\uptau) \bigg|_{G=-\frac{1}{2} \text{ pole}} = \frac{1}{2\pi}\sum_{k\geq 0} \Gamma\left(\frac{3}{2}+k\right) \, \alpha_k(-\tfrac{1}{2})\,(2\beta)^{-\frac{3}{2}-k} \,.
\end{equation}
At the special value $g=-\frac{1}{2}$, the coefficients $\alpha_k$ are the same as the coefficients $r_k$ appearing directly in the spectral curve. Indeed, comparing \eqref{eq:rhoGeneral} and \eqref{eq:rhoInvExpansion} for $g=-\frac{1}{2}$, we conclude
\begin{equation}
\label{eq:alphar}
    \alpha_k(-\tfrac{1}{2}) = r_k \,.
\end{equation}
We claim that the $G=-\frac{1}{2}$ pole \eqref{eq:plateauPole} encodes the asymptotic ``plateau'' ($T\rightarrow \infty$ limit) of the matrix model. On general grounds, the plateau for a spectral curve in canonical variables has the height
\begin{equation}
\begin{split}
 \lim_{\uptau\rightarrow\infty} {\cal K}^{(2)}_{\uptau\text{-RMT}}(\beta,\uptau) = 
\lim_{S_0\rightarrow\infty}\, e^{-S_0}\,\text{K}^{(1)}_\text{RMT}(2\beta)
    &= \int_0^\infty dE \, e^{-2\beta E} \, \rho_0(E)
    \\
    &= \frac{1}{2\pi} \sum_{k\geq 0}r_k \int_0^\infty dE \, e^{-2\beta E} \, E^{k+\frac{1}{2}} 
    \\
    &= \sum_{k\geq 0} \frac{\Gamma\left(\frac{3}{2}+k\right)}{2\pi} \, r_k\,(2\beta)^{-\frac{3}{2}-k} \,.
\end{split}
\end{equation}
Using the identification \eqref{eq:alphar}, we conclude that the $G=-\frac{1}{2}$ pole encodes precisely the plateau value:
\begin{equation}
    {\cal K}^{(2)}_{\uptau\text{-RMT}}(\beta,\uptau) \bigg|_{G=-\frac{1}{2} \text{ pole}} = e^{-S_0}\,\text{K}^{(1)}_\text{RMT}(2\beta) \,.
\end{equation}
To summarize, the analytic structure of the Mellin space RMT correlator not only encodes the genus expansion via its analytic structure, but it also non-perturbatively resums the plateau into an isolated simple pole.

%%%%%%%%%%%%%%%%%%%%%%%%%%%%%%%%%%%%%%%%%%%%%%%%
\subsection{Kontorovich-Lebedev transform for general spectral curve}
\label{app:KLtransform}
%%%%%%%%%%%%%%%%%%%%%%%%%%%%%%%%%%%%%%%%%%%%%%%%

The Kontorovich-Lebedev transform generalizes the Mellin transform and plays an analogous role in our construction for base spins $m>0$. 

\paragraph{KL transform for $\uptau$-scaled spectral form factor.} 
We can work it out for a general spectral curve $\rho_0(E)$ and the universal $\uptau$-scaled spectral two-point correlator \eqref{eq:GUEgeneralK2}. The relevant KL transform is
\begin{equation}
\begin{split}
h^{(2)}_{\uptau\text{-RMT}|m,m}(\sGen_1,\sGen_2) = 4 \int_0^\infty \frac{dy_1}{y_1}\frac{dy_2}{y_2} \, e^{-2\pi m (y_1+y_2)} K_{i\omega_1}(2\pi m y_1) K_{i\omega_2}(2\pi m y_2) \, \text{K}^{(2)}_{\uptau\text{-RMT}}(S_0(m);y_1,y_2)\,,
\end{split}
\end{equation}
where we already set $m_1=m_2\equiv m$ and we assume $m>0$ (the case $m=0$ is captured by the Mellin transform discussed separately in Appendix \ref{app:RMTmellin}). Plugging in \eqref{eq:GUEgeneralK2}, we see that the KL transform is determined by integrals of the form (for $g>0$)
\begin{equation}
 \begin{split}
   &\int_0^\infty \frac{dy_1}{y_1}\frac{dy_2}{y_2} \, e^{-2\pi m (y_1+y_2)} K_{i\omega_1}(2\pi m y_1) K_{i\omega_2}(2\pi m y_2) (y_1y_2)^{g+\frac{1}{2}} (y_1+y_2)^{g-k-1} \\
   &\quad = \sum_{n=0}^{g-k-1} \binom{g-k-1}{n}\, {\cal B}_m\big(\omega_1,-i(g+n+\tfrac{1}{2})\big)\, {\cal B}_m\big(\omega_2,-i(2g-n-k-\tfrac{1}{2})\big) \\
   &\quad = \frac{\pi}{(4\pi m)^{3g-k}} \sum_{n=0}^{g-k-1} \binom{g-k-1}{n} \, \frac{|\Gamma(g+n+\tfrac{1}{2}+i\omega_1)|^2| \Gamma(2g-n-k-\tfrac{1}{2}-i\omega_2)|^2}{\Gamma(g+n+1)\Gamma(2g-n-k)} \,,
 \end{split}
\end{equation}
where we used \eqref{eq:convolutionKernel}. Putting these pieces together, we find the following KL transform:
\begin{equation}
\boxed{
\begin{aligned}
&h^{(2)}_{\uptau\text{-RMT}|m,m}(\sGen_1,\sGen_2) \\
&\quad  = [g=0] + \sum_{g\geq 1} \sum_{k=0}^{g-1}  \frac{2(-1)^{g-k}\,e^{-2gS_0(m)} \, \alpha_k(g)}{g(2g+1)(4\pi m)^{3g-k}}  \sum_{n=0}^{g-k-1} \frac{|\Gamma(g+n+\sGen_1)|^2 |\Gamma(2g-n-k-\sGen_2)|^2}{\Gamma(g-k-n)\Gamma(2g-n-k) n! (g+n)!}\,.
\end{aligned}
}
\label{eq:generalKL}
\end{equation}
where the genus $g=0$ term is universal and as in Appendix~\ref{sec:CJnewderivation}. This expression should be understood as a formal (or asymptotic) genus expansion, which is sensible term by term. Viewed as an infinite series, it requires a resummation prescription to yield a finite result.

%%%%%%%%%%%%%%%%%%%%%%%%%%%%%%%%%%%%%%%%%%%%%%%%
\subsection{An example beyond $\uptau$-scaling: full GUE Airy model}
\label{app:fullAiry}
%%%%%%%%%%%%%%%%%%%%%%%%%%%%%%%%%%%%%%%%%%%%%%%%

In this subsection we give one example of an RMT in Mellin space, which is not $\uptau$-scaled and indeed more general than \eqref{eq:GUEgeneralK2} or \eqref{eq:f2MellinGeneralFinal}.

The example consists of the full GUE Airy form factor of Section~\ref{sec:GUEairyRMT2}, retaining the full $S_0$-dependence. That is, using $2\beta=y_1+y_2$ and $\beta^2+T^2=y_1y_2$, the RMT input is
\begin{equation}
\label{eq:fullAiryInput}
\text{K}_{\text{Airy}}^{(2)}(S_0;y_1,y_2)
= \frac{e^{S_0+\frac13e^{-2S_0}(y_1+y_2)^3}}{4\sqrt{\pi}(y_1+y_2)^{3/2}}\, \text{Erf}\!\left(e^{-S_0}\sqrt{y_1y_2(y_1+y_2)}\right)\,.
\end{equation}
This example requires an additional prescription: at fixed positive $\frac{y_1}{y_2}$, the error function approaches one as $y_1+y_2\to\infty$, while the exponential grows faster than any power. The Mellin integral \eqref{eq:mellinBasic} therefore has no convergence strip for the full input.\footnote{Relatedly, the moderate growth assumption in axiom 1 of the RMT$_2$ construction holds term by term in the $e^{-S_0}$ expansion, but not for the full resummed expression \eqref{eq:fullAiryInput}.} In the following, we will define an analytically regularized Mellin amplitude and verify its inverse by a corresponding contour continuation.

\paragraph{Regulated Mellin transform.}
Introduce an auxiliary damping parameter $\lambda$ and define the regularized Mellin amplitude as
\begin{equation}
\label{eq:fullAiryReg}
\begin{split}
f_{\text{Airy},\lambda}^{(2)}(S_0;\bar s_1,\bar s_2)
\equiv \int_0^\infty\frac{dy_1}{y_1^{1+i\omega_1}} \frac{dy_2}{y_2^{1+i\omega_2}}\, e^{-(\lambda+\frac13)e^{-2S_0}(y_1+y_2)^3}\, \text{K}_{\text{Airy}}^{(2)}(S_0;y_1,y_2).
\end{split}
\end{equation}
For $\lambda>0$ the integrand is damped at large $y_i$. The value $\lambda=0$ gives the $\uptau$-scaled Airy expression \eqref{eq:tauAiryGUE2}, while the exact model corresponds to the analytic continuation $\lambda\to -\frac13$. We first evaluate the Mellin transform \eqref{eq:fullAiryReg} in its convergence domain and then continue in $\lambda$.

Using an integral representation for the error function, the regulated form factor can be written as
\begin{equation}
\label{eq:fullAiryErfRep}
\begin{split}
&e^{-(\lambda+\frac13)e^{-2S_0}(y_1+y_2)^3}\,
\text{K}_{\text{Airy}}^{(2)}(S_0;y_1,y_2)
\\
&\quad=
\frac{\sqrt{y_1y_2}}{2\pi(y_1+y_2)}
\int_0^1dt\,
\exp\!\left[-e^{-2S_0}(y_1+y_2)^3
\left(\lambda+\frac{y_1y_2}{(y_1+y_2)^2}\,t^2\right)\right]\,.
\end{split}
\end{equation}
Set $r=y_1+y_2$ and $x=y_1/(y_1+y_2)$, so that $dy_1dy_2=r\,dr\,dx$. The Mellin transform becomes
\begin{equation}
\label{eq:fullAiryRadial}
\begin{split}
f_{\text{Airy},\lambda}^{(2)}
&=\frac1{2\pi}
\int_0^1dx\,x^{-\frac12-i\omega_1}(1-x)^{-\frac12-i\omega_2}
\int_0^1dt\int_0^\infty dr\,r^{-1-3i\omega_+}
e^{-e^{-2S_0}r^3[\lambda+x(1-x)t^2]}
\\
&=\frac{e^{-2i\omega_+S_0}\Gamma(-i\omega_+)}{6\pi}
\int_0^1dx\,x^{-\frac12-i\omega_1}(1-x)^{-\frac12-i\omega_2}
\int_0^1dt\,[\lambda+x(1-x)t^2]^{i\omega_+}.
\end{split}
\end{equation}
For $\lambda>0$, these manipulations are justified by absolute convergence
when $\text{Re}(-i\omega_+)>0$ and $\text{Re}\!\left(\frac12-i\omega_i\right)>0$.
In particular, we may take $\omega_+\in\mathbb R+i\epsilon$ and $\omega_-\in\mathbb R$, with $\epsilon>0$.

For $\lambda>\frac14$, we expand the last factor in \eqref{eq:fullAiryRadial}:
\begin{equation}
[\lambda+x(1-x)t^2]^{i\omega_+} =\lambda^{i\omega_+} \sum_{n=0}^{\infty}\frac{(-i\omega_+)_n}{n!} \left(-\frac1\lambda\right)^n[x(1-x)]^nt^{2n}\,.
\end{equation}
The integrals are now elementary and give:
\begin{equation}
\label{eq:fullAirySeries}
\begin{split}
f_{\text{Airy},\lambda}^{(2)}
&=\frac{e^{-2i\omega_+S_0}\lambda^{i\omega_+}}{6\pi} \sum_{n=0}^{\infty} \frac{(-1)^n\Gamma(n-i\omega_+)}{\lambda^n(2n+1)n!}
\, \frac{\Gamma(n+\frac12-i\omega_1)\Gamma(n+\frac12-i\omega_2)}{\Gamma(2n+1-3i\omega_+)}
\\
&=
\frac{e^{-2i\omega_+S_0}\lambda^{i\omega_+}}{6\pi} \frac{\Gamma(-i\omega_+)\Gamma(\frac12-i\omega_1)\Gamma(\frac12-i\omega_2)}{\Gamma(1-3i\omega_+)}\;
{}_4F_3\!\left(
\begin{matrix}
-i\omega_+,\ \frac12,\ \frac12-i\omega_1,\ \frac12-i\omega_2\\
\frac32,\ \frac12-\frac{3i\omega_+}{2},\
1-\frac{3i\omega_+}{2}
\end{matrix}
;-\frac1{4\lambda}\right)\,.
\end{split}
\end{equation}

\paragraph{Continuation to the full Airy model.}
Having found the closed-form Mellin transform \eqref{eq:fullAirySeries}, we analytically continue from $\lambda>0$ through the upper or lower half-plane and define
\begin{equation}
\label{eq:fullAiryLateral}
f_{\text{Airy},\pm}^{(2)}
\equiv\lim_{\delta\to0^+}
f_{\text{Airy},-\frac13\pm i\delta}^{(2)}
\,.
\end{equation}
The paths may be chosen with $|\lambda|>\frac14$, so that the hypergeometric series stays inside its disk of convergence. Noting $(-\frac13\pm i0)^{i\omega_+}
=3^{-i\omega_+}e^{\mp\pi\omega_+}$, the resulting amplitudes are
\begin{equation}
\label{eq:fullAiryMellin}
\boxed{
\begin{aligned}
f_{\text{Airy},\pm}^{(2)}(S_0;\bar s_1,\bar s_2)
&=
\frac{e^{-2i\omega_+S_0}3^{-i\omega_+}e^{\mp\pi\omega_+}}{6\pi}
\frac{\Gamma(-i\omega_+)
\Gamma(\frac12-i\omega_1)\Gamma(\frac12-i\omega_2)}
{\Gamma(1-3i\omega_+)}
\\
&\qquad\times
{}_4F_3\!\left(
\begin{matrix}
-i\omega_+,\ \frac12,\ \frac12-i\omega_1,\ \frac12-i\omega_2\\
\frac32,\ \frac12-\frac{3i\omega_+}{2},\
1-\frac{3i\omega_+}{2}
\end{matrix}
;\frac34\right).
\end{aligned}}
\end{equation}
The two expressions differ by the factor $e^{\mp\pi\omega_+}$. Their inverse Mellin transforms will nevertheless agree when the contours are continued consistently, thus showing that the choice of deformation was arbitrary. On the critical line, the genus-zero pole is approached as $\omega_+\to\omega_++i0$ at fixed $\omega_-$, in accordance with \eqref{eq:tauAiryf}.

As a check, we can instead take $\lambda\to0^+$ in
\eqref{eq:fullAiryRadial}. This limit is justified for
$0<\text{Re}(-i\omega_+)<\frac12$ and
$\text{Re}(\frac12-i\omega_i+i\omega_+)>0$. For this case, the integrals give:
\begin{equation}
\label{eq:fullAiryTauCheck}
\begin{split}
f_{\uptau\text{-Airy}}^{(2)}
&=
\frac{e^{-2i\omega_+S_0}\Gamma(-i\omega_+)}
{6\pi(1+2i\omega_+)\Gamma(1-i\omega_+)}\,
\prod_{i=1}^2\Gamma\!\left(\frac12-i\omega_i+i\omega_+\right)
\\
&=
\frac{e^{-2i\omega_+S_0}}{12\pi\omega_+(\omega_+-\frac i2)}\,
\Gamma\!\left(\frac12-\frac{i\omega_+}{2}+\frac{3i\omega_-}{2}\right)
\Gamma\!\left(\frac12-\frac{i\omega_+}{2}-\frac{3i\omega_-}{2}\right)\,.
\end{split}
\end{equation}
This indeed reproduces the $\uptau$-scaled result \eqref{eq:tauAiryf}, including its prescription for the genus-zero pole for a correctly regulated contour.

\paragraph{Inverse Mellin transform by contour deformation.}
To verify \eqref{eq:fullAiryMellin}, it is instructive to compute the inverse Mellin transform. To justify the inversion, we begin with $\lambda>\frac14$ and $\omega_+\in\mathbb R+i\epsilon$, $\omega_-\in\mathbb R$. The inverse transform is
\begin{equation}
\label{eq:fullAiryInverseDef}
{\cal M}^{-1}\!\left[f_{\text{Airy},\lambda}^{(2)}\right](y_1,y_2)
=\frac92\int\frac{d\omega_+d\omega_-}{(2\pi)^2}\,
(y_1y_2)^{\frac{3i\omega_+}{2}}
\left(\frac{y_1}{y_2}\right)^{\frac{3i\omega_-}{2}}
f_{\text{Airy},\lambda}^{(2)}\,.
\end{equation}
We insert the series \eqref{eq:fullAirySeries} and perform the $\omega_-$ integral first. For $y_1>y_2$, we close the contour in the upper half-plane (the other case is analogous). The poles of $\Gamma(n+\frac12-\frac{3i\omega_+}{2}+\frac{3i\omega_-}{2})$
are at
\begin{equation}
 \text{poles:}\qquad   \omega_-=\omega_++\frac{2i}{3}\left(n+\frac12+k\right),
\qquad k=0,1,2,\ldots 
\end{equation}
Summing their residues gives
\begin{equation}
\label{eq:fullAiryOmegaMinus}
\begin{split}
&\int_{\mathbb R}\frac{d\omega_-}{2\pi}
\left(\frac{y_1}{y_2}\right)^{\frac{3i\omega_-}{2}}
\Gamma\!\left(n+\frac12-\frac{3i\omega_+}{2}
+\frac{3i\omega_-}{2}\right)
\Gamma\!\left(n+\frac12-\frac{3i\omega_+}{2}
-\frac{3i\omega_-}{2}\right)
\\
&\quad=
\frac23\,\Gamma(2n+1-3i\omega_+)\,
\frac{(y_1y_2)^{n+\frac12-\frac{3i\omega_+}{2}}}
{(y_1+y_2)^{2n+1-3i\omega_+}}\,.
\end{split}
\end{equation}
The gamma function in \eqref{eq:fullAiryOmegaMinus} cancels the
denominator in \eqref{eq:fullAirySeries}. Setting $G=i\omega_+$, we find
\begin{equation}
\label{eq:fullAiryInverseScale}
\begin{split}
{\cal M}^{-1}\!\left[f_{\text{Airy},\lambda}^{(2)}\right]
&=
\frac{\sqrt{y_1y_2}}{2\pi(y_1+y_2)}
\sum_{n=0}^{\infty}
\frac{\left(-\frac{y_1y_2}{\lambda(y_1+y_2)^2}\right)^n}{(2n+1)n!}
\int_{-\epsilon+i\mathbb R}\frac{dG}{2\pi i}\,
\Gamma(n-G)
\left[\lambda e^{-2S_0}(y_1+y_2)^3\right]^G.
\end{split}
\end{equation}
For $\lambda>0$, we deform the upward vertical $G$ contour into a contour that runs from $+\infty-i0^+$ to the left of $G=0$ and returns to $+\infty+i0^+$, surrounding the nonnegative integers clockwise. The integrand decays factorially along the horizontal ends of the contour. Consequently, this loop representation remains convergent when $\lambda$ is continued to $-1/3\pm i0$. The original vertical contour does not: the continuation produces $e^{\pm i\pi G}$, which overwhelms the Gamma function's exponential decay in one of its tails. The contour deformation is therefore made in the convergent region before continuing $\lambda$.

The poles of $\Gamma(n-G)$ occur at $G=n+k$, giving a sum over residues that simplifies to the following expression:
\begin{equation}
\label{eq:fullAiryResidues}
\begin{split}
&\int_{\cal C}\frac{dG}{2\pi i}\,
\Gamma(n-G)\left[\lambda \,e^{-2S_0}(y_1+y_2)^3\right]^G
=
\left[\lambda\, e^{-2S_0}(y_1+y_2)^3\right]^n
e^{-\lambda \,e^{-2S_0}(y_1+y_2)^3}\,.
\end{split}
\end{equation}
At $\lambda=-1/3\pm i0$, the two choices of phase agree on every residue, since $e^{+i\pi(n+k)}=e^{-i\pi(n+k)}=(-1)^{n+k}$. Substituting into \eqref{eq:fullAiryInverseScale}, the Mellin inverse with the given contour continuation is
\begin{equation}
\label{eq:fullAiryDoubleSeries}
\begin{split}
{\cal M}^{-1}_{\cal C}\!\left[f_{\text{Airy},\pm}^{(2)}\right]
&=
\frac{\sqrt{y_1y_2}}{2\pi(y_1+y_2)} \sum_{n,k\geq0}\frac{(-1)^n} {(2n+1)n!k!}\, \big[y_1y_2(y_1+y_2)\,e^{-2S_0}\big]^n \left[\frac{e^{-2S_0}(y_1+y_2)^3}{3}\right]^k
\\
&=
\frac{\sqrt{y_1y_2}}{2\pi(y_1+y_2)} \,e^{\frac13e^{-2S_0}(y_1+y_2)^3} \int_0^1dt\,e^{-y_1y_2(y_1+y_2)t^2\,e^{-2S_0}}
\\
&=
\frac{e^{S_0+\frac13e^{-2S_0}(y_1+y_2)^3}}
{4\sqrt{\pi}(y_1+y_2)^{3/2}}\,
\text{Erf}\!\left(e^{-S_0}\sqrt{y_1y_2(y_1+y_2)}\right)
\\
&=\text{K}_{\text{Airy}}^{(2)}(S_0;y_1,y_2)\,.
\end{split}
\end{equation}
The double sum in the first line is absolutely convergent for finite positive $y_i$; its term with $n+k=g$ is of genus $g$. We have thus recovered the full Airy model RMT correlator. Either of the analytically continued contour prescriptions works equally. To summarize the procedure, we simply restate that the Mellin transform was defined by continuation from $\lambda>0$, and its inverse was continued after its vertical contour had been deformed to $\mathcal C$.

%%%%%%%%%%%%%%%%%%%%%%%%%%%%%%%%%%%%%%%%%%%%%%%%
%%%%%%%%%%%%%%%%%%%%%%%%%%%%%%%%%%%%%%%%%%%%%%%%
\section{Some properties of Poincar\'e sums}
\label{app:poincare}
%%%%%%%%%%%%%%%%%%%%%%%%%%%%%%%%%%%%%%%%%%%%%%%%
%%%%%%%%%%%%%%%%%%%%%%%%%%%%%%%%%%%%%%%%%%%%%%%%

In this appendix, we give some more details about the Poincar\'e sums $P_m(\tau,s)$ defined in \eqref{eq:PmDef0}.

\paragraph{Fourier decomposition and cuspidal behavior.}
Recall the definition and Fourier decomposition of the basic spin-$m$ Poincar\'e sum:
\be 
\label{eq:PmFourierApp}
\begin{split}
{\cal P}_{m} (\tau ,s) &\equiv \sum_{\gamma \in \Gamma_\infty\setminus \text{SL}(2,\mathbb{Z})} \text{Im}(\gamma \tau)^s \, e^{2\pi i m(\gamma \tau)}
= 
\sum_{j \in \mathbb{Z}} e^{2\pi i j x} \left( 
y^s \,  \delta_{jm} \, e^{-2\pi m y}
+y^{1-s} \,\mathcal{G}_{m|j}(y,s)  \right)  
\end{split}
\ee
The first (defining) expression converges absolutely for $\text{Re}(s)>1$. Since we are mostly interested in integrals over the critical line $s\in {\cal C}_\text{crit} = \frac{1}{2}+i\mathbb{R}$, the Fourier representation is more useful.

The spin-off-diagonal term ${\cal G}_{m|j}$ is given in terms of Kloosterman sums $S(m,j;c)$,\footnote{Recall $S(m,j;c) = \sum_\ell e^{2\pi i (m \ell + j \bar\ell)/c}$ with $1\leq \ell \leq c$ such that $\text{gcd}(\ell,c)=1$ and $\ell \bar \ell = 1 \,\text{mod} \, c$.} which can be written as follows:
\be
\begin{split}
\mathcal{G}_{m|j}(y,s) &=  \sum_{c=1}^\infty c^{-2s} \, S(m,j;c) \int_{\mathbb{R}}\frac{d\alpha}{(\alpha^2+1)^s} \, e^{-2\pi i j\alpha y  - \frac{2\pi i m}{c^2 y (\alpha+i)}} 
\\
&=\sum_{c=1}^\infty c^{-2s} \, S(m,j;c) \sum_{k=0}^\infty  \frac{1}{k!}\, \left(-\frac{2\pi i m}{c^2 y}\right)^k \int_{\mathbb{R}}d\alpha  \,(\alpha+i)^{-s-k} (\alpha-i)^{-s} \, e^{-2\pi i j \alpha y}\\
&= \pi^s (|j|y)^{s-1} \sum_{k=0}^{\infty} \frac{(-2\pi m)^k}{k!} \left(\frac{\pi |j|}{y}\right)^{\frac{k}{2}}
\frac{W_{\frac{k}{2}\text{sgn}(j),\,
   s+\frac{k-1}{2}}\!\left(4\pi |j|y\right)}{\Gamma\!\left(
s+\frac{k}{2}\left(1+\text{sgn}(j)\right)
\right)}\sum_{c=1}^{\infty}\frac{S(m,j;c)}{c^{2s+2k}}\qquad (j\neq 0) \,,
\end{split}
\label{eq:KloosterG}
\ee
where $W_{\kappa,\mu}(z)$ is a Whittaker function.
For $j>0$ this admits the following large $y$ asymptotic expansion:
\be
\begin{split}
\mathcal{G}_{m|j>0}(y,s) 
&\sim e^{-2\pi jy} \sum_{p=0}^\infty \frac{ \pi\,\Gamma(p+s)}{ \Gamma(s)p! \, 2^{p+s-1}}  \left( \sqrt{\frac{j}{m}} \,y \right)^{s-p-1} \,  \sum_{c=1}^\infty  \frac{ S(m,j;c)}{c^{s+p+1}}\,  J_{s-p-1}\left(\frac{4\pi}{c}\sqrt{jm}\right) \,.
\end{split}
\label{eq:KloosterG2}
\ee

\begin{figure}
\begin{center}
\includegraphics[width=.7\textwidth]{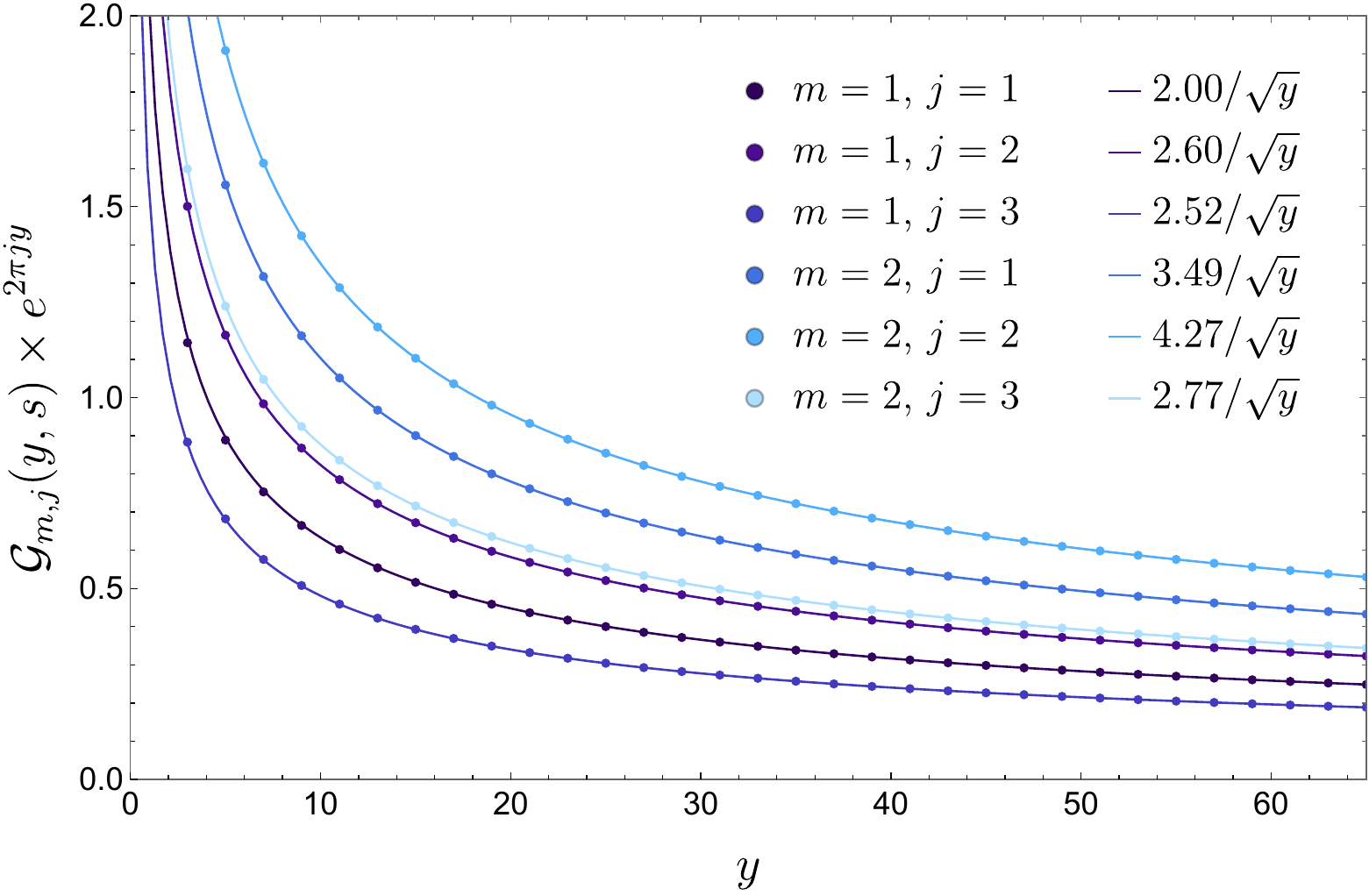}
\end{center}
\caption{Numerical evaluation of the remainder term ${\cal G}_{m|j}(y,s)$ for $s=\frac{1}{2}$ and different values of spins. In all cases (including the diagonal part of the remainder term, ${\cal G}_{j|j}$) the result is very well fit by $1/\sqrt{y}$ for large $y$. This confirms that these contributions are subleading compared to the first term in \eqref{eq:PmFourierApp}.}
\label{fig:kloosterman}
\end{figure}

The remainder term ${\cal G}_{m|j\neq 0}$ is subleading at the cusp $y \rightarrow\infty$.
This is illustrated numerically in Fig.\ \ref{fig:kloosterman}.
All data is consistent with ${\cal O}(y^{-1/2})$ scaling (after removal of the universal Boltzmann factor $e^{-2\pi |j| y}$). This compares to ${\cal O}(y^0)$ scaling of the $j=m$ diagonal term in the decomposition \eqref{eq:PmFourierApp} (stripping off the $y^{1-s}$ and $y^s$ prefactors, respectively). That is, in the spin-$j$ sector, the contribution of ${\cal P}_{m=j}$ is dominant relative to all ${\cal P}_{m\neq j}$.

The large $y$ asymptotics of the spin $j=0$ component is particularly simple and in fact enhanced. It can be computed as follows:
\be
\begin{split}
\mathcal{G}_{m|0}(y,s) &=  \sum_{c=1}^\infty c^{-2s} \, S(m,0;c) \int_{\mathbb{R}}\frac{d\alpha}{(\alpha^2+1)^s} \, e^{ - \frac{2\pi i m}{c^2 y (\alpha+i)}} 
\\
&=  \sum_{c=1}^\infty c^{-2s} \, S(m,0;c) \frac{\sqrt{\pi}\,\Gamma(s-\frac{1}{2})}{\Gamma(s)} + {\cal O}(y^{-1})
\\
&= \frac{\sqrt{\pi}\,\Gamma(s-\frac{1}{2}) \sigma_{1-2s}(m)}{\Gamma(s)\zeta(2s)}+ {\cal O}(y^{-1})
\end{split}
\label{eq:KloosterGj0}
\ee
where we used a large $y$ approximation at fixed $m$ to set $\text{exp}( - \frac{2\pi i m}{c^2 y (\alpha+i)}) = 1 + {\cal O}(\frac{m}{c^2 y})$, and we used that $S(m,0;c)$ is a Ramanujan sum that evaluates to the expression shown.
This implies the following large $y$ asymptotics for the Poincar\'e sums with $m>0$:
\begin{equation}
\label{eq:Pm0cusp}
    {\cal P}_{m|0}(y,s) = y^{1-s} \left[ \frac{\sqrt{\pi}\,\Gamma(s-\frac{1}{2}) \sigma_{1-2s}(m)}{\Gamma(s)\zeta(2s)}+ {\cal O}(y^{-1}) \right] \qquad (m>0) \,.
\end{equation}
Note that for $s\in{\cal C}_\text{crit}$ and fixed $m>0$, this component is of the same order as the Eisenstein series $E_{s|0}(y)$.

In the main text, we also use the parity-even remainder term, defined as
\begin{equation}
    G_{m|j}(y,s) = \frac{d_m}{2} \, \left( {\cal G}_{m|j}(y,s) + {\cal G}_{m|-j}(y,s) \right) \,.
\end{equation}

\paragraph{Overlaps with the harmonic basis.} We review the derivation of the important relation \eqref{eq:goldfeldOverlaps}, which provides the spectral decomposition of Poincar\'e series. To compute the overlap, we use the unfolding trick (assuming $m>0$):
\begin{equation}
\label{eq:goldfeldOverlapsApp}
\begin{split}
\left( {\cal P}_m (\,\cdot\, ,s) , \Psi_\sGen \right) 
&=
\sum_{j\geq 0} \bar{\tt c}_j^{(\sGen)} \int_{\cal F} \frac{dx dy}{y^2} \, \left( \sum_{\gamma\in\Gamma_\infty\setminus \text{SL}(2,\mathbb{Z})} \text{Im}(\gamma\tau)^s \, e^{2\pi i m \cdot \gamma\tau} \right) d_j\cos(2\pi j x) \sqrt{y} K_{\sGen-\frac{1}{2}}(2\pi j y ) 
\\
&= \sum_{j\geq 0} \bar{\tt c}_j^{(\sGen)} \int_0^\infty dy \int_{-\frac{1}{2}}^{\frac{1}{2}}dx \, y^{s-\frac{3}{2}}\,K_{\sGen-\frac{1}{2}}(2\pi j y ) \, e^{-2\pi m y} \, d_j\cos(2\pi jx) \, e^{2\pi i m x}  \\
&= \bar{\tt c}_m^{(\sGen)} \int_0^\infty dy \, y^{s-\frac{3}{2}}\,K_{\sGen-\frac{1}{2}}(2\pi m y)\,e^{-2\pi m y}
\\
&=\frac{\bar{\tt c}^{(\sGen)}_m\,\sqrt{\pi}}{(4\pi m)^{s-\frac{1}{2}}}
\frac{\Gamma\left(s-\sGen\right) \Gamma\left(s+\sGen-1\right)}{\Gamma(s)}\,.
\end{split}
\end{equation}

For the parity-conjugated Poincar\'e sum, we find the same result for overlaps with Eisenstein series and parity-even cusp forms. 
Therefore, the parity-even Poincar\'e sums satisfy the same relation (up to the conventional factor 2):
\begin{equation}
\label{eq:goldfeldOverlapsParity}
\begin{split}
\left( P_m(\,\cdot\, ,s) , \Psi_\sGen \right) 
&=\frac{2\bar{\tt c}^{(\sGen)}_m\,\sqrt{\pi}}{(4\pi m)^{s-\frac{1}{2}}}
\frac{\Gamma\left(s-\sGen\right) \Gamma\left(s+\sGen-1\right)}{\Gamma(s)} \qquad (m>0)\,.
\end{split}
\end{equation}

%%%%%%%%%%%%%%%%%%%%%%%%%%%%%%%%%%%%%%%%%%%%%%%%
%%%%%%%%%%%%%%%%%%%%%%%%%%%%%%%%%%%%%%%%%%%%%%%%
\section{Harmonic and Poincar\'e decompositions as WDW wavefunctions}
\label{app:wdw}
%%%%%%%%%%%%%%%%%%%%%%%%%%%%%%%%%%%%%%%%%%%%%%%%
%%%%%%%%%%%%%%%%%%%%%%%%%%%%%%%%%%%%%%%%%%%%%%%%

The harmonic basis and Poincar\'e frame admit physical interpretations as two different quantizations of the Wheeler-DeWitt (WDW) equation of 3d gravity on $T^2\times \mathbb{R}$. Useful references can be found in \cite{Carlip_2005,Marolf:2000iq,Freidel:2008sh,Moncrief:1989dx,DeWitt:1967yk,Held:2025mai}.

We derive the WDW equation starting from the homogeneous ansatz for the Euclidean metric on $T^2\times \mathbb{R}$ \cite{Carlip_2005}: 
\begin{equation}
\label{eq:summaryWdWmetric}
 ds^2=N(\rho)^2d\rho^2+\frac{A(\rho)}{y(\rho)}\left|d\sigma_1+\tau(\rho)d\sigma_2\right|^2\,,
 \qquad \tau=x+iy\,,\qquad \sigma_i\sim\sigma_i+1\,.
\end{equation}
Here $N$ is the lapse, $A$ is a scale factor and $\tau(\rho)$ is the modulus of the torus.  Substituting \eqref{eq:summaryWdWmetric} into the Einstein-Hilbert action, including the Gibbons-Hawking term, gives the reduced action
\begin{equation}
\label{eq:summaryWdWaction}
 S_E=\frac{1}{\kappa}\int d\rho\left[-\frac{\dot A^{\,2}}{2NA}
 +\frac{A}{2Ny^2}\left(\dot x^2+\dot y^2\right)+2\Lambda NA\right]\,,
 \qquad \kappa\equiv16\pi G\,,
\end{equation}
where $\Lambda$ is the cosmological constant. The lapse $N$ has no kinetic term and enforces the Hamiltonian constraint.  Introducing the variable $q\equiv\log A$ and canonically conjugate momenta gives the constraint:
\begin{equation}
\label{eq:summaryWdWconstraint}
 {\cal C}=-p_q^2+y^2\left(p_x^2+p_y^2\right)-\frac{4\Lambda}{\kappa^2}e^{2q}=0\,.
\end{equation}
Canonical quantization promotes this constraint to the WDW equation on wavefunctions $\Phi(\tau,q)$
\begin{equation}
\label{eq:summaryWdWeq}
 \widehat{\cal C}_{\rm WDW}\Phi(\tau,q)=0\,,\qquad
 \left[\partial_q^2+\Delta_{\mathbb H}-\frac{4\Lambda}{\kappa^2}e^{2q}\right]\Phi(\tau,q)=0\,,
 \qquad
 \Delta_{\mathbb H}\equiv-y^2\left(\partial_x^2+\partial_y^2\right)\,.
\end{equation}
We see that the torus part of the WDW operator $\widehat{\cal C}_{\rm WDW}$ is precisely the hyperbolic Laplacian on the upper half-plane.  We can solve the equation by separating the conformal factor mode $q$ and the torus mode $\tau$:
\begin{equation}
\label{eq:summaryWdWseparation}
 \Phi(q,\tau)=\chi_s(q)\psi_s(\tau)\,,
\end{equation}
which results in
\begin{equation}
 \Delta_{\mathbb H}\psi_s(\tau)=s(1-s)\psi_s(\tau)\,, \qquad \qquad \left(\partial_q^2+s(1-s)-\frac{4\Lambda}{\kappa^2}e^{2q}\right)\chi_s(q)=0\,.
\end{equation}
The torus WDW wavefunctions $\psi_s(\tau)$ locally satisfy the Laplace equation on the upper-half plane, defining kinematical states which are  not subject to global constraints \cite{Held:2025mai,Marolf:2000iq} that define the physical states. Physical states must satisfy diffeomorphism invariance. For 3d gravity on a (spatial) Riemann surface $\Sigma$, diffeomorphism invariance includes invariance under large diffeomorphisms given by the Mapping Class Group of the surface $\text{MCG}(\Sigma)$, which for the torus is $\sl$. Thus we need to impose the constraint that physical states solve not only the WDW equation but are in addition invariant under $\sl$:
\es{}{
\psi_s(\tau)=\psi_s(\gamma \tau).
}
There are two natural ways to implement the constraint, as familiar from the general quantization of constrained systems \cite{constrainedq}.

\begin{enumerate}
    \item  \emph{Constrain first and quantize second}. We  identify $\tau\sim\gamma\tau$ and quantize directly on the fundamental domain ${\cal F}$.  Then the Hilbert space of normalizable states is 
\begin{equation}
\label{eq:summaryWdWHilbert}
 {\cal H}_{\rm phys}=L^2({\cal F})\,,
\end{equation}
and diagonalizing the Laplacian on the fundamental domain $\Delta_{\mathcal{F}}$ gives a complete $\delta$-normalized basis which is simply the harmonic basis: the continuum of real analytic Eisenstein series $E_s$, and the discrete spectrum of  Maass cusp forms $\phi_n$: 
\es{}{
\Delta_{\mathcal{F}}\Psi_s(\tau)=s(1-s)\Psi_s(\tau), \qquad \Psi_s(\tau)\in \{E_s(\tau),\phi_n(\tau),1\}
}

This gives a gravitational interpretation of the CFT trace formula. Consider an arbitrary state $\ket{\Psi}\in{\cal H}_{\rm phys}$ in the Hilbert space of 3d gravity on $T^2 \times \mathbb{R}$, with wavefunction $\braket{\tau}{\Psi}=\Psi_{\rm WDW}(\tau)$ satisfying the WDW equation. We can decompose this state into eigenfunctions of the WDW equation which results in the same spectral decomposition as the CFT trace formula: 
\es{}{
\Psi_{\rm WDW}(\tau)= \int_{\mathbb{R}} \frac{d\w}{4\pi} (\Psi_{\rm WDW},E_{\frac{1}{2}+i\w }) E_{\frac{1}{2}+i\w}(\tau) +\sum_{n=1}^{\infty} (\Psi_{\rm WDW},\phi_{n})\phi_n(\tau) + \frac{3}{\pi}\,(\Psi_{\rm WDW},1)\,.
}
The space of normalizable WDW wavefunctions $\Psi_{\rm WDW}(\tau)$ then corresponds to the space of partition functions ${\cal Z}(\tau)$.

\item \emph{Quantize first and constrain second}.  We solve the WDW equation locally by diagonalizing the Laplacian on $\mathbb H$. The eigenfunctions are:
\begin{equation}
\label{eq:summaryWdWkinematical}
 \psi_{s}(\tau)=\sum_{m\in \mathbb{Z}}c_{m,s}\,e^{2\pi i m x }\psi_{m,s}(y),  \qquad
 \psi_{m,s}(y)=\sqrt{y}\,
 K_{s-\frac12}(2\pi |m|y)\,, \;\; \psi_{0,s}(y)=y^s\,.
\end{equation}
For integer $m$ they are invariant under the translation subgroup $\Gamma_\infty\subset \sl$. We can now impose the constraint on the WDW wavefunctions by group averaging \cite{Marolf:2000iq, Held:2025mai}. Group averaging consists in summing (or integrating) over diffeomorphisms, leading to an invariant physical state. This is a  projection onto the Hilbert space of wavefunctions that satisfy the constraint.\footnote{See \cite{Held:2025mai} for a detailed discussion.} In the case of a discrete group $\Gamma=\sl$, the projector is given by a sum over the group elements: 
\es{}{
\mathcal{P}_{\Gamma}[\psi(\tau)]=\sum_{\gamma \in \Gamma_{\infty}\setminus\Gamma}\psi(\gamma \tau).
}
The sum is understood by regularization when necessary. The resulting WDW wavefunctions are Poincar\'{e} sums of the local solutions $\psi_s(\tau)$:
\es{}{
\hat{P}_{m,s}(\tau)\equiv\mathcal{P}_{\Gamma}[e^{2\pi i mx}\psi_{m,s}(y)]=\sum_{\gamma \in \Gamma_{\infty}\setminus\text{SL}(2,\mathbb{Z})} e^{2\pi im \Re(\gamma \tau)} \psi_{m,s}(\Im(\gamma\tau))
}
The $m=0$ case is proportional to the Eisenstein series $E_s(\tau)$. For $m\neq 0$ this leads to Poincar\'{e} sums of Bessel functions \cite{Goldfeld1979, Goldfeld2020TraceFormulaeSL2R,Angelantonj:2011br,Angelantonj:2012gw}. Group averaging does produce a set of WDW wavefunctions that satisfy the constraints; however, in general, they are non-orthogonal and it is not guaranteed that they form either a complete set of functions or a resolution of the identity. The $P_m(\,\cdot\,,s)$ with $\text{Re}(s)>1$ span a dense subspace of $L^2({\cal F})$ but are overcomplete. Moreover, they do not form a resolution of the identity, hence mathematically they are a so-called ``frame'' \cite{frames}.
In our work, we  use a different but closely related Poincar\'{e} series as a basis, which leads to simpler Mellin transforms rather than Bessel transforms. 
As presented previously our $P_m$ is obtained by group averaging the simpler Mellin seed
\begin{equation}
\label{eq:summaryMellinSeed}
 f_{m,s}(\tau)=y^s e^{2\pi i m\tau}=y^s e^{-2\pi my}e^{2\pi i mx}\,,
\end{equation}
(and its parity image), rather than the Bessel seed in \eqref{eq:summaryWdWkinematical}.  For $m=0$, the two constructions coincide, since $f_{0,s}=y^s$ and $P_0=E_s$.  For $m>0$, they do not: the Mellin seed is not a Laplacian eigenfunction,
\begin{equation}
\label{eq:summaryMellinNotEigen}
 \left(\Delta_{\mathbb H}-s(1-s)\right)f_{m,s}
 =4\pi m s\,y\,f_{m,s}\qquad (m>0)\,.
\end{equation}
Consequently, for $m>0$ the Poincar\'e sums used below are not individual WDW eigenstates but wavepackets of harmonic eigenstates,
\begin{equation}
\label{eq:summaryPmWavepacket}
 P_m(\tau,s)=\sumint_\sGen
 \left(P_m(\,\cdot\,,s),\Psi_\sGen\right)\Psi_\sGen(\tau)\,,
\end{equation}
with coefficients given in \eqref{eq:goldfeldOverlaps}. 
\end{enumerate}

To summarize, the harmonic basis is the modular-invariant WDW eigenbasis obtained by imposing the constraint of modular invariance before quantization. The $P_m$ frame is instead an overcomplete family of group-averaged wavepackets: it is closely related to, but distinct from, the Niebur-Poincar\'e series \cite{Goldfeld1979} produced by quantizing first and imposing the large-diffeomorphism constraint afterwards.

%%%%%%%%%%%%%%%%%%%%%%%%%%%%%%%%%%%%%%%%%%%%%%%%
%%%%%%%%%%%%%%%%%%%%%%%%%%%%%%%%%%%%%%%%%%%%%%%%
\section{Parity-odd sector}
\label{app:parity}
%%%%%%%%%%%%%%%%%%%%%%%%%%%%%%%%%%%%%%%%%%%%%%%%
%%%%%%%%%%%%%%%%%%%%%%%%%%%%%%%%%%%%%%%%%%%%%%%%

In the bulk of this work, we assume a parity-invariant spectrum. That is, the CFT's torus partition function is assumed to be invariant under $x \mapsto -x$. In terms of the spectrum, for every state with energy $E$ and spin $j$, there exists a corresponding parity-conjugate state with energy $E$ and spin $-j$. We resolve this redundancy by always assuming $j\geq 0$ and $m\geq 0$.

In CFTs that do not satisfy this assumption, the inclusion of parity-odd Maass cusp forms and Poincar\'e sums in our discussion would be required. In this appendix we review a few of their properties.

In the definition of $\Psi_\sGen$ in \eqref{eq:PsiDef2}, we restricted to parity-even basis functions. If we included parity-odd Maass cusp forms, we would extend the definition as follows:
\be 
\label{eq:PsiDef2mod}
 \begin{split}
   \Psi_\sGen^+ \equiv E_{\frac{1}{2}+i\omega} \, : \qquad\;\;\, {\tt c}_j^{(\sGen)} &\equiv {\tt c}_{-j}^{(\sGen)} =  {\tt a}_j^{(s)}   \equiv \frac{2\sigma_{2i\omega}(j)}{|j|^{i\omega} \Lambda(-i\omega)}  \,,\\
    \Psi_\sGen^+ \equiv \phi_n^{+} \, : \qquad\;\;\, {\tt c}_j^{(\sGen)} &\equiv  {\tt c}_{-j}^{(\sGen)} =    {\tt b}_j^{(n,+)} \qquad (j>0)\,,\\
    \Psi_\sGen^- \equiv \phi_n^{-}\, : \qquad\;\;\, {\tt c}_j^{(\sGen)} &\equiv  -{\tt c}_{-j}^{(\sGen)} =  -i\,{\tt b}_j^{(n,-)} \qquad (j>0)\,,
 \end{split}
\ee
where ${\tt b}_j^{(n,\pm)}$ are the Fourier coefficients of even ($+$) and odd ($-$) Maass cusp forms. It is then useful to introduce the following notation for even and odd Fourier coefficients:
\begin{equation}
\label{eq:cEvenOdd}
   {\tt c}^{(\sGen,\pm)}_j \equiv  {\tt c}^{(\sGen)}_j\pm {\tt c}^{(\sGen)}_{-j} \,, 
\end{equation}
The cosine coefficients are ${\tt c}_j^{(\sGen,+)}$, whereas the sine coefficients are $i{\tt c}_j^{(\sGen,-)}$. For an odd cusp form, the latter equal $2{\tt b}_j^{(n,-)}$.

Similarly, the Poincar\'e sum $P_m(\tau,s)\equiv P_m^+(\tau,s)$, defined in \eqref{eq:PmDef0} is parity-even. More generally, we can define two distinct combinations, 
\begin{equation}
\label{eq:PpmDef}
    P^{\pm}_m(\tau,s) \equiv \frac{d_m}{2}\,\left( {\cal P}_m(\tau,s) \pm {\cal P}_m(-\bar{\tau},s) \right) \,.
\end{equation}
To confirm this definition, note that parity acts as follows on the Poincar\'e sums:\footnote{Alternatively, one could consider complex conjugate Poincar\'e sums: $\overline{{\cal P}_m(\tau,s)}={\cal P}_{m}(-\bar\tau,\bar s)$.}
\es{}{
\text{parity:} \qquad {\cal P}_m(\tau,s) \; \mapsto \;  {\cal P}_m(-\bar \tau,s)&={\cal P}_m(-\Re(\tau),\Im(\tau),s)\,.
}
Note that $\Re(\gamma (-\bar \tau))$ and $\Im(\gamma (-\bar \tau))$ are actually complicated functions of $\Re(\gamma \tau), \Im(\gamma\tau)$. However, they are related to a different element of $\sl$ which is $\gamma (-\bar \tau)= -\overline{ \gamma^*\tau}$ where $\gamma^* = \left(\begin{smallmatrix} a & -b \\ -c & d \end{smallmatrix}\right)$. Since this is another element of the sum, we can just relabel it. So parity acts as: 
\es{}{
\text{parity:} \qquad {\cal P}_m(\tau,s) \; \mapsto \; 
{\cal P}_m(-\bar \tau,s)=\sum_\gamma \Im(\gamma\tau)^s e^{-2\pi i m \overline{\gamma \tau}}=\sum_\gamma \Im(\gamma\tau)^se^{-2\pi i m \Re(\gamma \tau)-2\pi m \Im(\gamma \tau)}\,.
}
The parity conjugate Poincar\'e sums and hence the combinations $P_m^\pm(\tau,s)$ therefore decay at the cusp for $m>0$. 

The overlaps of Poincar\'e series with harmonic basis vectors become:
\begin{equation}
\label{eq:goldfeldOverlapsParity2}
\begin{split}
\left( P_m^\pm(\,\cdot\, ,s) , \Psi_\sGen^\pm \right) 
&=\frac{\bar{\tt c}^{(\sGen,\pm)}_m\,\sqrt{\pi}}{(4\pi m)^{s-\frac{1}{2}}}
\frac{\Gamma\left(s-\sGen\right) \Gamma\left(s+\sGen-1\right)}{\Gamma(s)}\,,\qquad
\left( P_m^\pm(\,\cdot\, ,s) , \Psi_\sGen^\mp \right)=0\,.
\end{split}
\end{equation}

%%%%%%%%%%%%%%%%%%%%%%%%%%%%%%%%%%%%%%%%%%%%%%%%
%%%%%%%%%%%%%%%%%%%%%%%%%%%%%%%%%%%%%%%%%%%%%%%%
\section{Zero modes}
\label{app:zeromodes}
%%%%%%%%%%%%%%%%%%%%%%%%%%%%%%%%%%%%%%%%%%%%%%%%
%%%%%%%%%%%%%%%%%%%%%%%%%%%%%%%%%%%%%%%%%%%%%%%%

Zero modes in the spectral decomposition are overlaps of an RMT${}_2$ amplitude with the (modular-invariant) constant function. These are fixed by the requirement that the critical line integrals over Eisenstein series must admit a consistent contour deformation: recall that $E_{s=\frac{1}{2}+i\omega}$ has a pole at $s=1$. Deforming the critical line contours towards infinity should lead to a cancellation of all spurious residues picked up in the process. We illustrate this for $n=1$ and $n=2$; higher $n$ are analogous.

\paragraph{${\bf n=1}$:} As an example, consider a one-boundary amplitude \cite{Collier:2022emf}, 
\begin{equation}
\label{eq:n1ex}
    {\cal Z}^{(1)}_{\text{RMT}_2}(\tau) = \int_{{\cal C}_\text{crit}} \frac{ds}{2\pi i} \,f^{(1)}_\text{RMT}(\bar s) \, E_s(\tau)  \;+\; \text{\tt [zero mode]} \;+\; \text{\tt [cusp forms]} .
\end{equation}
In this spectral decomposition, the zero mode is the overlap with the constant function, i.e., the `modular average' of ${\cal Z}^{(1)}_{\text{RMT}_2}$:
\begin{equation}
\begin{split}
    \text{\tt [zero mode]} &\equiv 
    \frac{1}{\text{vol}({\cal F})} \int_{\cal F} \frac{dx dy}{y^2} \, {\cal Z}^{(1)}_{\text{RMT}_2}(\tau) \\
    &= \frac{1}{\text{vol}({\cal F})} \,\lim_{s\rightarrow 1}\int_{\cal F} \frac{dx dy}{y^2} \, {\cal Z}^{(1)}_{\text{RMT}_2}(\tau) \, E_{1-s}(\tau)
    \\
    &= \frac{1}{\text{vol}({\cal F})} \,\lim_{s\rightarrow 1} \, \left( {\cal Z}^{(1)}_{\text{RMT}_2} \,,\, E_s \right)
    \\
    &= \frac{1}{\text{vol}({\cal F})} \, f_\text{RMT}^{(1)}(\bar s=0) \,.
\end{split}
\end{equation}
This has the following property: we may deform the $s$ contour in \eqref{eq:n1ex} towards $\text{Re}(s) \rightarrow \infty$. This deformation picks up a residue at $s=1$, which is
\begin{equation}
    -\oint_{s=1} \frac{ds}{2\pi i} \, f_\text{RMT}^{(1)}(\bar s \equiv 1-s) E_s(\tau) = -f_\text{RMT}^{(1)}(0) \, \text{Res}_{s=1} E_s(\tau) = -\frac{1}{\text{vol}({\cal F})} \, f_\text{RMT}^{(1)}(0)  \,,
\end{equation}
thus canceling the zero mode.

\paragraph{${\bf n=2}$:} 
Consider now the case of two boundaries. The full Eisenstein sector of RMT${}_2$  for $n=2$ takes the following form:
\begin{equation}
\label{eq:RMT2liftFull}
\begin{split}
{\cal Z}_{\text{RMT}_2}^{(2)}(\tau_1,\tau_2) 
&\equiv \int_{{\cal C}_\text{crit}} \frac{ds_1}{2\pi i}  \int_{{\cal C}_\text{crit}}  \frac{ds_2}{2\pi i}\,  {f}^{(2)}_{\text{RMT}}(\bar s_1,\bar s_2) \;\frac{1}{2}\left( E_{s_1}(\tau_1) E_{s_2}(\tau_2)+E_{s_1}(\tau_2) E_{s_2}(\tau_1) \right)
\\
&\quad + \frac{2}{\text{vol}({\cal F})} \int_{{\cal C}_\text{crit}} \frac{ds_1}{2\pi i} \, {f}^{(2)}_{\text{RMT}}(\bar s_1,\bar s_2=0) \,\frac{1}{2}\left( E_{s_1}(\tau_1) + E_{s_1}(\tau_2)\right)
\\
&\quad +  \frac{1}{\text{vol}({\cal F})^2}  \,f^{(2)}_{\text{RMT}}(\bar s_1=0,\bar s_2=0)
\;\;+\;\; \text{\tt [cusp forms]}
\end{split}
\end{equation}
where the second and third lines would be abbreviated as `$+\text{\tt [zero modes]}$' in the main text.

We begin with the constant term in the last line of \eqref{eq:RMT2liftFull}.
We recognize this term as a double `modular average':
\begin{equation}
\label{eq:ZRMT2const}
\begin{split}
\left( {\cal Z}_{\text{RMT}_2}^{(2)} \right)
&\equiv \frac{1}{\text{vol}({\cal F})^2} \int_{{\cal F}} \frac{dx_1 dy_1}{y_1^2} \int_{{\cal F}} \frac{dx_2 dy_2}{y_2^2} \,{\cal Z}_{\text{RMT}_2}^{(2)}(\tau_1,\tau_2) \\
&= \frac{1}{\text{vol}({\cal F})^2}\, \lim_{s_1 \rightarrow 1} \, \lim_{s_2 \rightarrow 1} \,\int_{{\cal F}} \frac{dx_1 dy_1}{y_1^2} \int_{{\cal F}} \frac{dx_2 dy_2}{y_2^2} \,{\cal Z}_{\text{RMT}_2}^{(2)}(\tau_1,\tau_2) E_{1-s_1}(\tau_1) E_{1-s_2}
(\tau_2) \\
&=  \frac{1}{\text{vol}({\cal F})^2}\, \lim_{s_1 \rightarrow 1} \, \lim_{s_2 \rightarrow 1} \, \left(\left( {\cal Z}_{\text{RMT}_2}^{(2)} (\,\star\,, \, \boldsymbol{\cdot} \,),\, E_{s_1}(\,\star\,)\right),E_{s_2}(\, \boldsymbol{\cdot} \,) \right) 
\\
 &= \frac{1}{\text{vol}({\cal F})^2}\,   f^{(2)}_{\text{RMT}}(\bar s_1=0,\bar s_2=0)\,.
\end{split}
\end{equation}

A similar computation shows that the overlap with one Eisenstein series and one constant can be interpreted as a `half-sided modular average':
\begin{equation}
\label{eq:ZRMT2halfconst}
\begin{split}
\left( {\cal Z}_{\text{RMT}_2}^{(2)}(\tau_1,\,\cdot\,) \right)
 &\equiv \frac{1}{\text{vol}({\cal F})} \int_{{\cal F}} \frac{dx_2 dy_2}{y_2^2}  \,{\cal Z}_{\text{RMT}_2}^{(2)}(\tau_1,\tau_2) 
 \\
&= \frac{1}{\text{vol}({\cal F})}\, \lim_{s_2\rightarrow 1} \int_{{\cal F}} \frac{dx_2 dy_2}{y_2^2}  \,{\cal Z}_{\text{RMT}_2}^{(2)}(\tau_1,\tau_2) \, E_{1-s_2}(\tau_2)\\
&= \frac{1}{\text{vol}({\cal F})}\, \lim_{s_2\rightarrow 1}  \left( {\cal Z}_{\text{RMT}_2}^{(2)}(\tau_1,\,\boldsymbol{\cdot}\,) \,,\, E_{s_2}(\,\boldsymbol{\cdot} \,) \right) \\
&= \frac{1}{\text{vol}({\cal F})}\,
 \int_{\frac{1}{2}+i\mathbb{R}}\frac{ds_1}{2\pi i}\; f^{(2)}_{\text{RMT}}(\bar s_1,\bar s_2=0) \, E_{s_1}(\tau_1) \\
 &\quad + \frac{1}{\text{vol}({\cal F})^2}\,   f^{(2)}_{\text{RMT}}(\bar s_1=0,\bar s_2=0)
 \,.
\end{split}
\end{equation}

Finally, we note that \eqref{eq:RMT2liftFull} can again be written in more compact form by deforming the integration contours. The second and third lines are precisely such as to cancel poles that are due to Eisenstein series rather than due to the actual spectral overlap. This becomes manifest if we move the contours for $s_1$ and $s_2$ to the right, just past the poles at $s_1=1$ and $s_2=1$. We then find:
\begin{equation}
\label{eq:RMT2liftFull2}
\begin{split}
{\cal Z}_{\text{RMT}_2}^{(2)}(\tau_1,\tau_2) 
&\equiv \int_{1+\epsilon+i\mathbb{R}} \frac{ds_1}{2\pi i}  \int_{1+\epsilon+i\mathbb{R}}  \frac{ds_2}{2\pi i}\,  {f}^{(2)}_{\text{RMT}}(\bar s_1,\bar s_2) \;\frac{1}{2}\left( E_{s_1}(\tau_1) E_{s_2}(\tau_2)+E_{s_1}(\tau_2) E_{s_2}(\tau_1) \right)\\
&\quad + \text{ residues due to } f^{(2)}_\text{RMT} 
\;\; + \;\;  \text{\tt [cusp forms]}\,,
\end{split}
\end{equation}
where the second line indicates that there may be other (physical) residues picked up due to poles of the spectral overlap.

%%%%%%%%%%%%%%%%%%%%%%%%%%%%%%%%%%%%%%%%%%%%%%%%
%%%%%%%%%%%%%%%%%%%%%%%%%%%%%%%%%%%%%%%%%%%%%%%%
\section{Dirichlet series and L-functions}
\label{app:Lfunctions}
%%%%%%%%%%%%%%%%%%%%%%%%%%%%%%%%%%%%%%%%%%%%%%%%
%%%%%%%%%%%%%%%%%%%%%%%%%%%%%%%%%%%%%%%%%%%%%%%%

In this appendix we define and analyze some L-functions relevant to our analysis. Most of the stated results are well-known, see, for example, \cite{Shimura,Sankaranarayanan,NewtonLfunc,Blomer,Iwaniec2021AnalyticNT}.

\paragraph{L-functions and Dirichlet series.}
We begin with the L-function associated with a completed Eisenstein series $E_s^*(\tau) = \Lambda(s) E_s(\tau)$:
\begin{equation}
    D_{E_s^*}(t) \equiv \frac{\Lambda(s)}{2} \sum_{m\geq 1} \frac{{\tt a}_m^{(s)}}{m^t} = \sum_{m\geq 1} \frac{m^{-i\omega}\sigma_{2i\omega}(m)}{m^{t}} \qquad (s=\tfrac{1}{2}+i\omega\in{\cal C}_\text{crit},\;\text{Re}(t)>1)\,.
\end{equation}
This is the standard notion, which satisfies a nice functional equation. In our conventions, it is useful to also consider the corresponding function for non-completed Eisenstein series $E_s(\tau)$, and define:
\begin{equation}
    D_{E_s}(t) \equiv 2 \sum_{m\geq 1} \frac{{\tt a}_m^{(s)}}{m^t} \qquad (s=\tfrac{1}{2}+i\omega\in{\cal C}_\text{crit},\;\text{Re}(t)>1)\,.
\end{equation}
The meromorphic continuation of this series is given by
\begin{equation}
    D_{E_s^*}(t) = \zeta(t+s-\tfrac{1}{2})\zeta(t-s+\tfrac{1}{2}) \,,
\end{equation}
and similarly for $D_{E_s}(t)$.
This can also be written in a standard Euler representation:
\begin{equation}
    D_{E_s^*}(t) = \prod_{p \text{ prime}} \;\frac{1}{\left( 1- \alpha_p\, p^{-t} \right) \left( 1 - \alpha_p^{-1}\, p^{-t}\right)} \equiv L_{E_s^*}(t)\,,\qquad \alpha_p=p^{s-\frac{1}{2}}  \,,
\end{equation}
where $\alpha_p$ are called Satake parameters.

Similarly, we define a Dirichlet series for the $n$-th Maass cusp form
\begin{equation}
    D_{\phi_n}(t) = \sum_{m\geq 1} \frac{{\tt b}_m^{(n)}}{m^t} \qquad (\text{Re}(t)>1)\,.
\end{equation}
This can also be written as an Euler product L-function:
\begin{equation}
    D_{\phi_n}(t) = {\tt b}_1^{(n)} \, L_{\phi_n}(t)  \,,\qquad 
    L_{\phi_n}(t) = \prod_{p \text{ prime}} \;\frac{1}{1-a_p^{(n)} p^{-t} + p^{-2t}}\,,\qquad a_j^{(n)} \equiv \frac{{\tt b}_j^{(n)}}{{\tt b}_1^{(n)}}\,.
\end{equation}
Written in terms of Satake parameters, these are seen to satisfy the non-trivial relation $\alpha_p + \alpha_p^{-1} = a^{(n)}_p$,  such that
\begin{equation}
    1 - a^{(n)}_p\, p^{-t}+p^{-2t} = \left( 1- \alpha_p \, p^{-t} \right) \left( 1- \alpha_p^{-1} \, p^{-t} \right) \,.
\end{equation}

\paragraph{Second power L-functions.}
We can consider the Rankin-Selberg L-function for the product of two cusp forms:
\begin{equation}
   L_{\phi_{n_1} \times \phi_{n_2}}(t) 
   =  \prod_{p\text{ prime}} \,\prod_{\epsilon_i=\pm} \frac{1}{\left(1-\alpha_{n_1,p}^{\epsilon_1}\,\alpha_{n_2,p}^{\epsilon_2}\, p^{-t}\right)}
   \,,
\label{eq:Lphin1phin2}
\end{equation}
where $\alpha_{n_i,p}+\alpha_{n_i,p}^{-1}=a_p^{(n_i)}$ for both $i=1,2$.
This is related to the Dirichlet series
\begin{equation}
    D_{\phi_{n_1} \times \phi_{n_2}}(t) = \sum_{m\geq 1} \frac{{\tt b}^{(n_1)}_{m}\,{\tt b}^{(n_2)}_{m}}{m^t} \qquad (\text{Re}(t)>1) 
\end{equation}
by a local correction factor, 
\begin{equation}
    D_{\phi_{n_1} \times \phi_{n_2}}(t) = {\tt b}_1^{(n_1)}{\tt b}_1^{(n_2)}\,L_{\phi_{n_1} \times \phi_{n_2}}(t) \, \prod_{p\text{ primes}} \left( 1 - p^{-2t} \right)  = {\tt b}_1^{(n_1)}{\tt b}_1^{(n_2)}\frac{L_{\phi_{n_1} \times \phi_{n_2}}(t)}{\zeta(2t)} \, \,.
\end{equation}

A special case is the symmetric-square L-function,
\begin{equation}
   L_{\text{sym}^2\phi_n}(t) 
   \equiv  \prod_{p\text{ prime}}  \frac{1}{\left(1-\alpha_p^{2}\, p^{-t}\right)\left(1- p^{-t}\right)\left(1-\alpha_p^{-2}\, p^{-t}\right)} 
   = \frac{\zeta(2t)}{{\tt b}_1^{(n)}} \sum_{m\geq 1} \frac{{\tt b}^{(n)}_{m^2}}{m^t}
   \,,
   \label{eq:Lphinphin}
\end{equation}
which differs from $L_{\phi_n\times\phi_n}(t)$ by a factor $\zeta(t)$.

\paragraph{$\ell$-th power L-functions.} 
The above construction generalizes to arbitrary powers. For $\ell$ cusp forms, we define 
\begin{equation}
\begin{split}
    L_{\phi_{n_1}\times\ldots\times \phi_{n_\ell}}(t) &\equiv \prod_{p \text{ prime}} \; \prod_{\epsilon_i=\pm} \, \frac{1}{\left( 1 - \alpha_{n_1,p}^{\epsilon_1} \cdots \alpha_{n_\ell,p}^{\epsilon_\ell}\, p^{-t}\right)} \,,\\
    D_{\phi_{n_1}\times\ldots\times \phi_{n_\ell}}(t) &\equiv \sum_{m\geq 1} \frac{{\tt b}^{(n_1)}_m \cdots {\tt b}^{(n_\ell)}_m}{m^t} \qquad (\text{Re}(t)>1) \,.
\end{split}
\end{equation}
For $\ell\geq 3$, the correction factors relating the $\ell$-product L-function and the Dirichlet series are non-trivial:
\begin{equation}
   D_{\phi_{n_1} \times\ldots \times \phi_{n_\ell}}(t) 
   = \left( \prod_{i=1}^\ell {\tt b}_1^{(n_i)} \right) L_{\phi_{n_1} \times\ldots \times \phi_{n_\ell}}(t) \prod_{p\text{ primes}} \; M_p(p^{-t})
   \,,
\end{equation}
where $M_p$ is a polynomial that depends non-trivially on the Satake parameters for every $p$.

\paragraph{Deformed Dirichlet series.}
In our construction of RMT${}_2$, we encounter $\ell$-th power Dirichlet series with a deformation parameter. In general, we find it useful to define for the deformation parameter $S_0(m)\geq 0$ a generalized Dirichlet series that captures any combination of Eisenstein series and cusp forms:
\begin{equation}
\boxed{\widetilde{D}_{\Psi_{\sGen_1}\times\ldots\times \Psi_{\sGen_\ell}}(S_0;t) \equiv 2^\ell \sum_{m\geq 1} \frac{{\tt c}^{(\sGen_1)}_m \cdots {\tt c}^{(\sGen_\ell)}_m}{m^t} \, e^{-\frac{2}{3} t\,S_0(m)} \qquad (\text{Re}(t)>1)}
\end{equation}

%%%%%%%%%%%%%%%%%%%%%%%%%%%%%%%%%%%%%%%%%%%%%%%%
%%%%%%%%%%%%%%%%%%%%%%%%%%%%%%%%%%%%%%%%%%%%%%%%
\section{Other ways to write RMT${}_2$ correlators}\label{app:alt}
%%%%%%%%%%%%%%%%%%%%%%%%%%%%%%%%%%%%%%%%%%%%%%%%
%%%%%%%%%%%%%%%%%%%%%%%%%%%%%%%%%%%%%%%%%%%%%%%%

Our definition of the RMT${}_2$ lift can be written in a variety of alternative forms, which can be useful in certain contexts. We give two examples of this below.

\paragraph{RMT${}_2$ as a Poincar\'e sum of RMT.} Consider the definition of RMT${}_2$ in \eqref{eq:RMT2defPmRep}. We can evaluate it by formally exchanging the contour integrals with the Poincar\'e sums (we don't indicate the zero modes here):
{\small
\begin{align}
   &{\cal Z}^{(n)}_{\text{RMT}_2} (\btau) \\
   & = 
    \sum_{m\geq 0} d_m \int_{{\cal C}_\text{crit}^n} \left( \prod_{i=1}^n \frac{ds_i}{2\pi i} \, \sum_{\gamma_i\in \Gamma_{\infty}\setminus\text{SL}(2,\mathbb{Z})} \text{Im}(\gamma_i \tau_i)^{s_i} \, \cos\big(2\pi m \text{Re}(\gamma_i\tau_i)\big) e^{-2\pi m \text{Im}(\gamma_i\tau_i)} \right) \, f^{(n)}_\text{RMT}(S_0(m);\bar{\bf s} )  \nonumber \\
    & = \sum_{m\geq 0} d_m\sum_{\substack{\gamma_1,\ldots,\gamma_n \in \\\Gamma_{\infty}\setminus\text{SL}(2,\mathbb{Z}) }}   \left( \prod_{i=1}^n  \cos\big(2\pi m \text{Re}(\gamma_i\tau_i)\big) e^{-2\pi m \text{Im}(\gamma_i\tau_i)} \right)
    \int_{{\cal C}_\text{crit}^n} \left( \prod_{i=1}^n \frac{ds_i}{2\pi i}\, \text{Im}(\gamma_i \tau_i)^{s_i} \right) f^{(n)}_\text{RMT}(S_0(m);\bar{\bf s} ) \nonumber\\
    & =\sum_{m\geq 0} d_m\sum_{\substack{\gamma_1,\ldots,\gamma_n \in \\ \Gamma_{\infty}\setminus\text{SL}(2,\mathbb{Z}) }}   \left( \prod_{i=1}^n   \cos\big(2\pi m \text{Re}(\gamma_i\tau_i)\big) e^{-2\pi m \text{Im}(\gamma_i\tau_i)} \sqrt{\text{Im}(\gamma_i\tau_i)}\right) \text{K}^{(n)}_\text{RMT}\big(S_0(m); \text{Im}(\gamma_1\tau_1),\ldots,\text{Im}(\gamma_n\tau_n) \big) \nonumber
\end{align}
}\normalsize
where we recognized the $s_i$ integrals as inverse Mellin transforms in the variables $y_i' \equiv \text{Im}(\gamma_i \tau_i)$. This shows that the RMT${}_2$ amplitude can be written as (multiple) Poincar\'e sums over the RMT seed, dressed with appropriate spin-dependent factors and the conventional $\sqrt{y_i'}$ normalization.

\paragraph{RMT${}_2$ as a matrix integral.}
We can write the RMT${}_2$ correlators as explicit integrals over the underlying RMT random matrix ensemble,
\begin{equation}
    \langle \, \cdot \, \rangle \equiv \frac{1}{\text{Z}_\text{RMT}}\int dH \, ( \, \cdot \, ) \, e^{-N \tr V(H)} \,.
\end{equation}
To see this, recall that the RMT form factor can be written as follows (for concreteness, we consider GUE):
\es{}{
\text{K}_\text{RMT}^{(n)}({\bf y})
&=\left\langle \tr(e^{-y_1 H})\cdots\tr(e^{-y_n H})\right\rangle _{\rm c}\\
&=
\frac{1}{\text{Z}_\text{RMT}}\int dH\, \tr(e^{-y_1 H})\cdots \tr(e^{-y_n H}) \, e^{-N\tr V(H)}\\
&=\frac{1}{\text{Z}_\text{RMT}}  \int_{\mathbb{R}^N}d^N x\, \left(\sum_{i_1,\ldots,i_n =1}^N e^{-y_1x_{i_1}-\ldots-y_n x_{i_n}} \right)\, e^{-N\sum_{i}V(x_i)+\sum_{i<j}\log(x_i-x_j)^2}\\
&\equiv \frac{1}{\text{Z}_\text{RMT}}  \,\sum_{i_1,\ldots,i_n =1}^N\int_{\mathbb{R}^N}d^N x\,  e^{-y_1x_{i_1}-\ldots-y_n x_{i_n}} \, e^{-N\,V_\text{eff}({\bf x})} \,.
}
We have left implicit the fact that we are only considering the connected part of the correlator. 
We can now perform the Mellin transform by exchanging it with the matrix integral. We use that:\footnote{This equation holds for $\Re(x)>0$ and $\Im(\omega)>0$. The extension to negative $x$ is formal.}
\es{}{
\mathcal{M}_{y}[e^{-y x}](-i\w)= \Gamma(-i\w)\,x^{i\w}\,.
}
We then obtain the Mellin amplitude:
\es{eq:overmatinte}{
f^{(n)}_\text{RMT}(\bar{\bf s})
&= \Gamma(-i\w_1)\cdots\Gamma(-i\w_n)\, \frac{1}{\text{Z}_\text{RMT}}\sum_{i_1,\ldots,i_n}\int_{\mathbb{R}^N}d^N x \, e^{-NV_\text{eff}({\bf x})} \, x_{i_1}^{i\w_1}\cdots x_{i_n}^{i\w_n}\\
&=
\Gamma(-i\w_1)\cdots\Gamma(-i\w_n)\,\frac{1}{\text{Z}_\text{RMT}}\int dH \, e^{-N \tr V(H)}\, \tr(H^{i\w_1})\cdots \tr(H^{i\w_n}) \\
&=
\Gamma(-i\w_1)\cdots\Gamma(-i\w_n)\int dx_1 \cdots dx_n\, \la \rho(x_1) \cdots \rho(x_n)\ra_{c} \;x_1^{i\w_1} \cdots x_n^{i\w_n} \,,
}
where we used that:
\es{}{
\rho(x)\equiv\sum_{i=1}^N \delta(x-x_i) \,.
}
This shows that the Mellin amplitude is an integral over random matrix eigenvalues, involving the microcanonical RMT spectral correlator. 

Applying this result to the Poincar\'e sum definition of RMT${}_2$, \eqref{eq:RMT2defPmRep}, we can write:
\begin{equation}
    \begin{split}
        {\cal Z}^{(n)}_{\text{RMT}_2}(\btau) 
        &= \sum_{{m \geq 0}}\,d_m^{1-n} \left\langle \tr\left(e^{-H}\big|_{P_{m}(\tau_1,\,\cdot\,)} \right) \cdots \tr\left(e^{-H}\big|_{P_{m}(\tau_n,\,\cdot\,)} \right)\right\rangle_c
        \;+\; {\text{\tt [zero modes]}}
    \end{split}
\end{equation}
with the following definition of the {\it spin $m$ modular Boltzmann factor}
\begin{equation}
    e^{-H}\big|_{P_{m}(\tau,\,\cdot\,)}
    \equiv \int_{\mathbb{R}} \frac{d\omega}{2\pi} \, \Gamma(-i\omega) H^{i\omega} \, P_m(\tau,\tfrac{1}{2}+i\omega) \,.
\end{equation}

%%%%%%%%%%%%%%%%%%%%%%%%%%%%%%%%%%%%%%%%%%%%%%%%
%%%%%%%%%%%%%%%%%%%%%%%%%%%%%%%%%%%%%%%%%%%%%%%%
\section{Numerical checks}
\label{app:numerics}
%%%%%%%%%%%%%%%%%%%%%%%%%%%%%%%%%%%%%%%%%%%%%%%%
%%%%%%%%%%%%%%%%%%%%%%%%%%%%%%%%%%%%%%%%%%%%%%%%

In this appendix, we provide some numerical checks of spectral decompositions.

\paragraph{Numerics of Maass cusp forms.}
When we evaluate sums over Maass cusp forms below, we use a set of 5832 even-parity Maass cusp forms: at the time of writing, the eigenvalues and Fourier coefficients for the first 1110 parity-even cusp forms can be found in the {\it L-functions and modular forms database} \cite{lmfdb}; the remaining ones up to $n=5832$ were provided by H.\ Then \cite{Then_2004}. 

\begin{figure}
\begin{center}
\includegraphics[width=.99\textwidth]{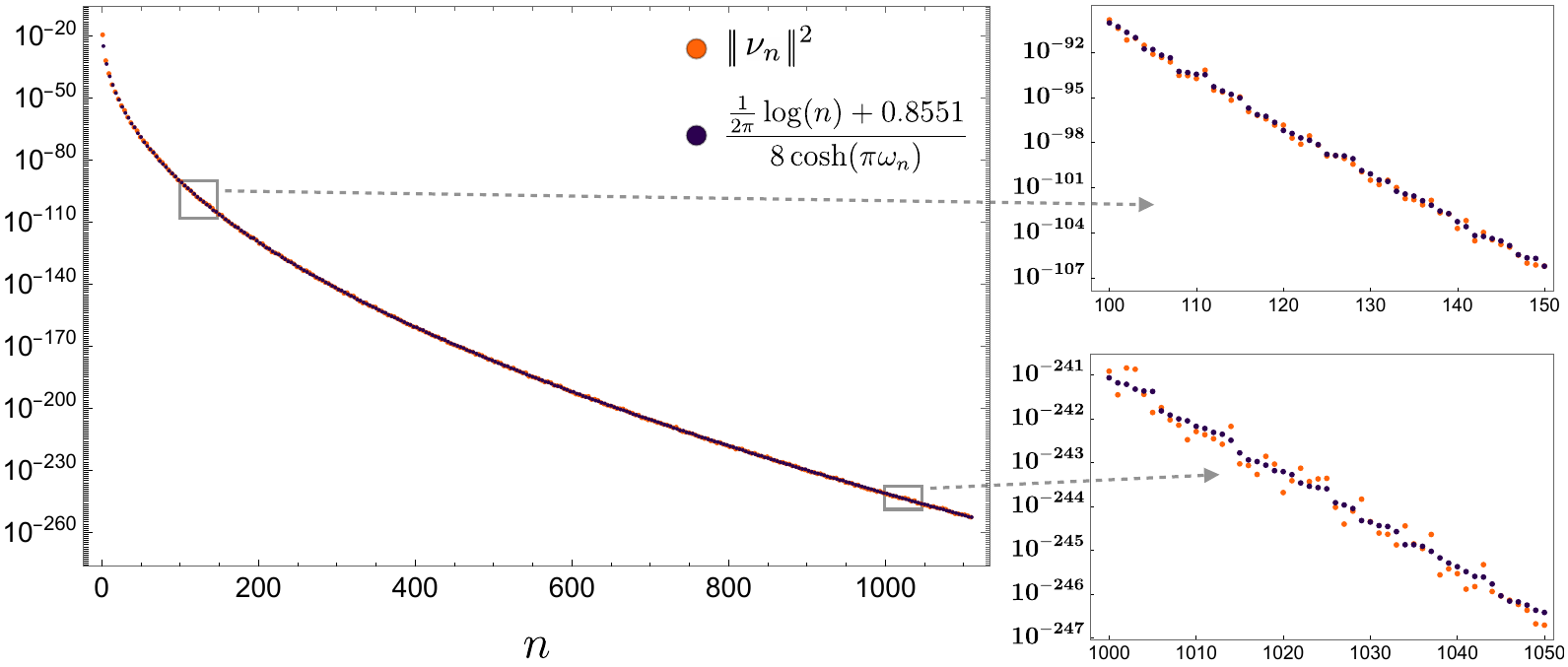}
\end{center}
\caption{Norms of (even-parity) Maass cusp forms in the Hecke normalization, compared to the approximate expression \eqref{eq:CJcusps}. The difference only becomes visible upon zooming in (right panels).}
\label{fig:CuspFormNorms}
\end{figure}

These databases use Maass cusp forms in the Hecke normalization. We denote these as $\nu_n(\tau)$ and their Fourier coefficients as $a_j^{(n)}$. The eigenvalues $\omega_n$ are of course the same in both conventions. The Hecke normalization is defined by $a_1^{(n)}=1$. The relation to the Maass cusp forms in the main text (defined by $|\!|\phi_n|\!|=1$) is\footnote{The Hecke normalized cusp forms are explicitly $\nu_n(\tau)=\sum_{j\geq 1} a_j^{(n)} \sqrt{y} K_{i\omega_n}(2\pi j y) \cos(2\pi j x)$.}
\begin{equation}
    \phi_n(\tau) = \frac{\nu_n(\tau)}{|\!|\nu_n|\!|} \,,\qquad {\tt b}_j^{(n)} = a_j^{(n)} \, {\tt b}_1^{(n)}  \,,\qquad  {\tt b}_1^{(n)} 
    = \frac{1}{2|\!|\nu_n|\!|}
    \,,
\end{equation}
where the Petersson norm of the cusp forms is
\begin{equation}
    |\!|\nu_n|\!|^2 = \int_{{\cal F}} \frac{dx dy}{y^2} \, |\nu_n(\tau=x+iy)|^2 \,.
\end{equation}
For the first 1110 cusp forms, we computed the norms $|\!|\nu_n|\!|$ by explicit integration to several digits of accuracy. For the remaining cusp forms ($n=1111,\ldots,5832$), we use the following approximation for the norm \cite{Haehl:2023xys}:
\begin{equation}
\label{eq:CJcusps}
    |\!|\nu_n|\!|^2 = \frac{L_{\text{sym}^2\nu_n}(1)}{8\cosh(\pi \omega_n)} \approx \frac{1}{8\cosh(\pi \omega_n)} \left[ \frac{1}{2\pi} \, \log(n) + 0.8551 + \ldots \right] \,,
\end{equation}
where $L_{\text{sym}^2\nu_n}(s)$ is the symmetric-square L-function associated with the $n$-th Maass cusp form.\footnote{One can also compute the norms by evaluating the series representation of the symmetric-square L-function,
\begin{equation}
    L_{\text{sym}^2\nu_n}(1) = \frac{\pi^2}{6} \sum_{m\geq 1} \frac{a_{m^2}^{(n)}}{m} \,.
\end{equation}
However, to achieve reasonable precision, this requires knowledge of the cusp form Fourier coefficients for very high spin, which is not practical.} The symbol `$\approx$' should be understood in the sense of averaging: the L-function takes a slightly different value for each $n$, and these exact values fluctuate around $\frac{1}{2\pi} \, \log(n) + 0.8551 + \ldots$ with a small variance. The coefficients in this expansion were extracted from a numerical fit and it would be interesting to understand if the empirical coefficient $\frac{1}{2\pi}$ is correct analytically. The approximation is illustrated in Fig.\ \ref{fig:CuspFormNorms}. 

Since cusp forms with larger values of $n$ contribute less to the sums we are interested in, the approximation \eqref{eq:CJcusps} is good for our purposes. The advantage of the approximation is that the spectral parameters $\omega_n$ are easier to compute and available to us for more cusp forms than the norms $|\!|\nu_n|\!|^2$.

%%%%%%%%%%%%%%%%%%%%%%%%%%%%%%%%%%%%%%%%%%%%%%%%
\subsection{Two-boundary wormhole: Maass cusp forms}
%%%%%%%%%%%%%%%%%%%%%%%%%%%%%%%%%%%%%%%%%%%%%%%%

The universal two-boundary amplitude contains the following cusp form contribution:
\begin{equation}
\label{G4}
    {\cal Z}^{(0,2)}_{\text{RMT}_2}(\tau_1,\tau_2) \supset \sum_{n\geq 1} \frac{{\tt C}_\text{RMT}}{2\cosh(\pi\omega_n)} \phi_{n}(\tau_1)\phi_{n}(\tau_2) \,.
\end{equation}

\begin{figure}
\begin{center}
\includegraphics[width=.87\textwidth]{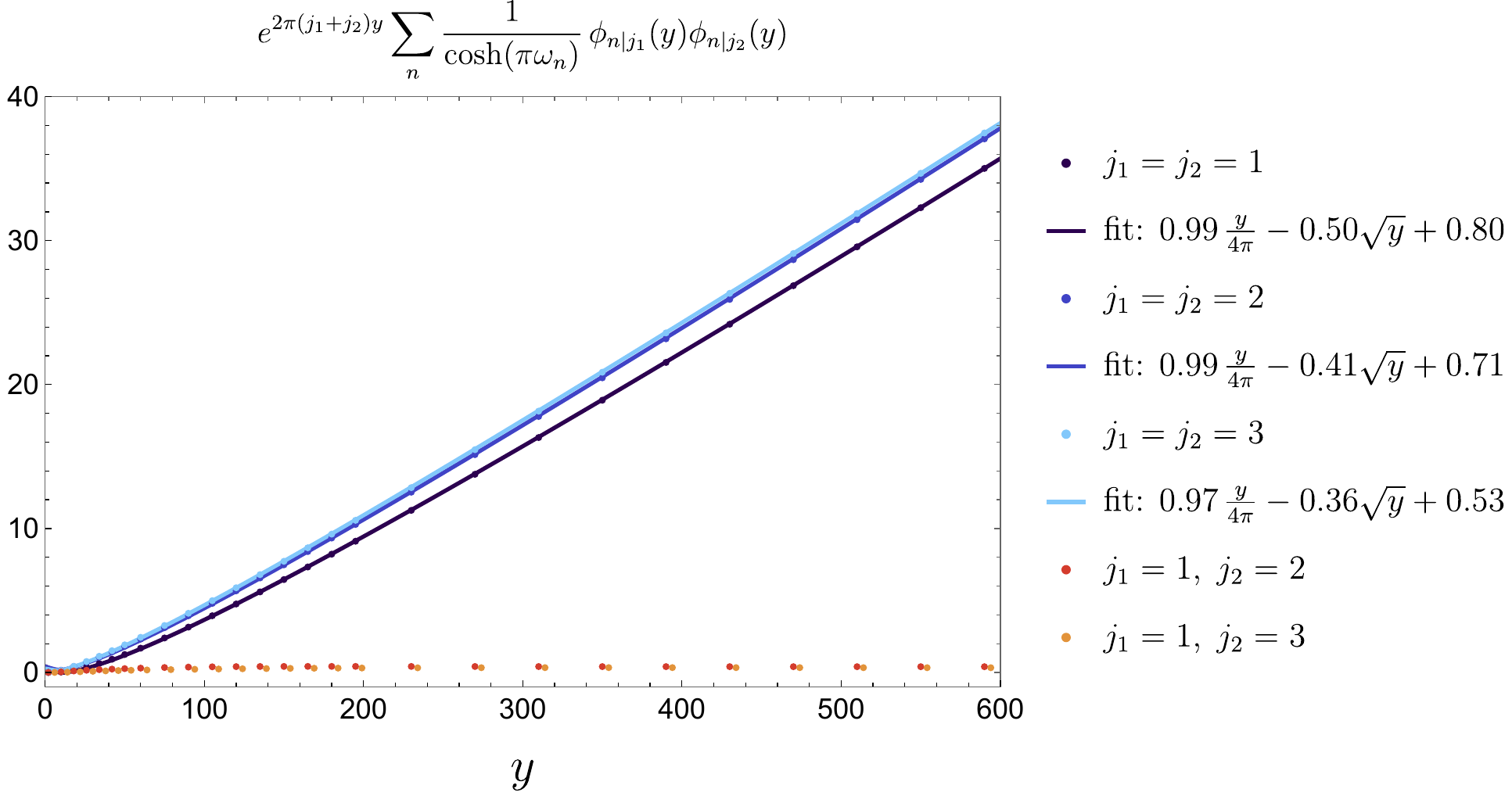}
\end{center}
\caption{Numerical evaluation of the sum over cusp forms in the $(g,n)=(0,2)$ amplitude \eqref{G4}. Aligned spins $j_1=j_2$ follow the expected leading behavior of the linear ramp with excellent accuracy. Unaligned spins $j_1\neq j_2$ give negligible contributions.}
\label{fig:CJnumerics1}
\end{figure}

In Fig.\ \ref{fig:CJnumerics1}, we evaluate this sum numerically in different spin sectors $(j_1,j_2)$. We choose the section $y_1=y_2\equiv y$ and even cusp forms. The numerical result is consistent with
\begin{equation}
    \sum_{n\geq 1} \frac{{\tt C}_\text{RMT}}{2\cosh(\pi\omega_n)} \,\phi_{n|j_1}(y_1)\phi_{n|j_2}(y_2)
    = \frac{1}{2}\,e^{-2\pi(j_1y_1+j_2y_2)}\,
     \frac{{\tt C}_\text{RMT}}{2\pi} \frac{y_1y_2}{y_1+y_2} \, \delta_{j_1j_2} + \ldots\,,
\end{equation}
where `$\ldots$' denotes terms subleading at large $y_i$, and the spin-$j$ components are $\phi_{n|j}(y) = {\tt b}_j^{(n)} \sqrt{y} K_{i\omega_n}(2\pi j y)$. This is consistent with the general axiom \eqref{eq:NErequirement}.

%%%%%%%%%%%%%%%%%%%%%%%%%%%%%%%%%%%%%%%%%%%%%%%%
\subsection{Three-boundary wormhole}
%%%%%%%%%%%%%%%%%%%%%%%%%%%%%%%%%%%%%%%%%%%%%%%%

In this section, we evaluate the triple Eisenstein and triple cusp form overlap for the universal $(g,n)=(0,3)$ amplitude numerically. 

%%%%%%%%%%%%%%%%%%%%%%%%%%%%%%%%%%%%%%%%%%%%%%%%
\subsubsection{Triple Eisenstein overlap}
\label{sec:appEEE}
%%%%%%%%%%%%%%%%%%%%%%%%%%%%%%%%%%%%%%%%%%%%%%%%

\begin{figure}
\begin{center}
\includegraphics[width=.85\textwidth]{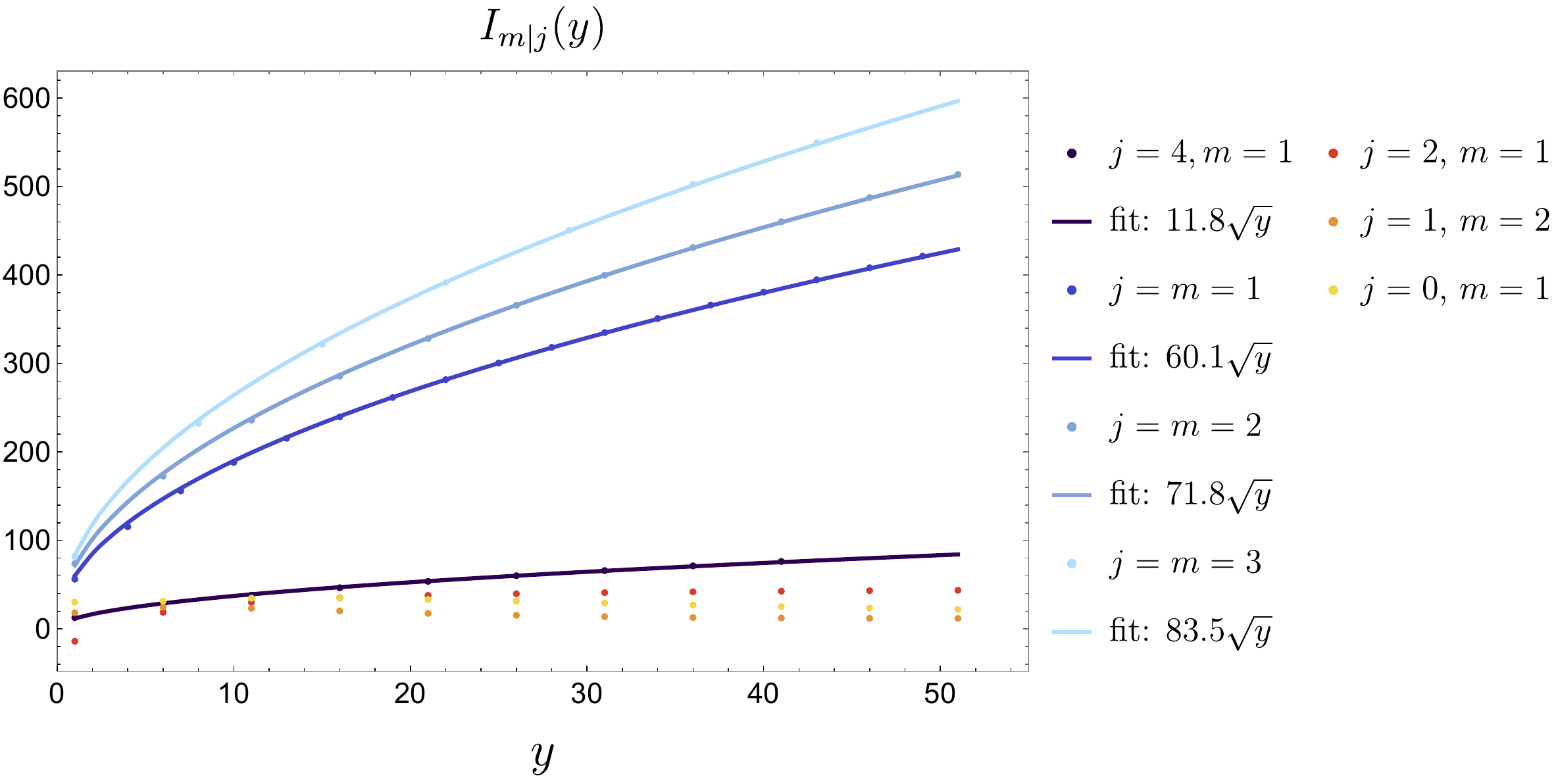}
\end{center}
\caption{Numerical evaluation and fit of the integral $I_{m|j}(y)$ computing the imprint of the base spin-$m$ Eisenstein contribution onto the physical spin-$j$ spectrum. Aligned spins ($j=m$) are enhanced by $\sqrt{y}$, as are combinations where $mj$ is a perfect square (such as $j=4$, $m=1$). Other combinations of $j$ and $m$ tend to a constant for large $y$. Both cases are still subleading compared to the RMT expectation, which would scale as $y$.}
\label{fig:spinningImprint}
\end{figure}

The triple Eisenstein overlap for $m>0$ was described in Section \ref{sec:EEEmneq0}. It is determined by three factors of the following integral over Eisenstein series basis functions:
\begin{equation}
\begin{split}
& \int_{\mathbb{R}} \frac{d\omega}{4\pi} \, E_{\frac{1}{2}+i\omega|j}(y) \, \frac{\Gamma(\tfrac{1}{2}+i\omega)\Gamma(\tfrac{1}{2}-i\omega)}{\sqrt{8\pi m \gamma}} \, e^{-\frac{1}{3}S_0(m)}\, {\tt{a}}_m^{(\frac{1}{2}-i\omega)}
\\
&\qquad = \frac{e^{-\frac{1}{3}S_0(m)}}{\sqrt{8\pi m \gamma}} \int_{\mathbb{R}} \frac{d\omega}{4\pi} \,  
 {\tt{a}}_m^{(\frac{1}{2}-i\omega)} \, {\tt{a}}_j^{(\frac{1}{2}+i\omega)} \,\frac{\pi}{\cosh(\pi \omega)}\, \sqrt{y} K_{i\omega}(2\pi j y) 
 \\
 &\qquad = \frac{e^{-\frac{1}{3}S_0(m)}}{\sqrt{8\pi m \gamma}} \int_{\mathbb{R}} \frac{d\omega}{4\pi} \, \frac{4\omega\tanh(\pi\omega)}{\zeta(2i\omega)\zeta(-2i\omega)} \, \left(\frac{m}{j}\right)^{i\omega} \sigma_{-2i\omega}(m) \sigma_{2i\omega}(j)
 \, \sqrt{y} K_{i\omega}(2\pi j y)
 \\
 &\qquad \equiv  \frac{e^{-\frac{1}{3}S_0(m)}}{\sqrt{8\pi m \gamma}}\times  \frac{1}{4\pi}\,e^{-2\pi j y} \, I_{m|j}(y)
\end{split}
\end{equation}
Recall the meaning of the integral $I_{m|j}(y)$: it exists due to the RMT constraints near extremality in the (base) spin $m$ sector. The above is the imprint of this piece of the Eisenstein overlaps onto the spin-$j$ sector (not necessarily near extremality). We analyze this integral numerically by deforming the $\omega$ contour and collecting residues due to poles of $\tanh(\pi\omega)$ and zeros of $\zeta(2i\omega)\zeta(-2i\omega)$. The result is shown in Fig.\ \ref{fig:spinningImprint}. 

We observe that for $j=m$, the imprint of the base spin $m$ is largest on the same physical spin sector, $j=m$. For $j\neq m$, the imprint either tends to a constant, or is also of order ${\cal O}(\sqrt{y})$ for special values of spins (such as $j=4$, $m=1$). In summary:
\begin{equation}
\label{eq:Imjscaling}
    I_{m|j}(y) = \text{const.} \times \sqrt{y}  + o(\sqrt{y}) \,.
\end{equation}
The value of the $(j,m)$-dependent constant is not too important (and it can be zero), as this contribution is subleading anyway compared to \eqref{eq:n3limit}. The interplay of these asymptotics onto the RMT${}_2$ amplitude is summarized in \eqref{eq:ZjjjmmmEisenstein}.

The behavior \eqref{eq:Imjscaling} is also easily confirmed analytically in a large $y$ approximation. For large $y$, the Bessel function in $I_{m|j}(y)$ can be approximated as
\begin{equation}
    \sqrt{y} K_{i\omega}(2\pi j y) \approx \frac{1}{2\sqrt{j}} \, e^{-2\pi j y - \frac{\omega^2}{4\pi j y}} \qquad (\omega = {\cal O}(\sqrt{y}))\,.
\end{equation}
In this approximation, the integral over $\omega$ becomes a Gaussian of width ${\cal O}(\sqrt{y})$, which we can factor out and evaluate. Using $\frac{\omega \tanh(\pi\omega)}{|\zeta(2i\omega)|^2} = \frac{\pi}{|\zeta(1+2i\omega)|^2}$, we obtain
\begin{equation}
\label{eq:Imjapprox}
    I_{m|j}(y)
    = \left( \lim_{W\rightarrow\infty}\, \frac{1}{2W}\int_{-W}^W d\omega \, \frac{4\pi^2}{|\zeta(1+2i\omega)|^2} \frac{\sigma_{2i\omega}(j)\sigma_{2i\omega}(m)}{(jm)^{i\omega}} \right) \sqrt{y} + o(\sqrt{y})
\end{equation}
The numerical coefficient in the bracket can be evaluated and it may or may not vanish, depending on $j$ and $m$. The largest scaling in $y$ in any case is $\sqrt{y}$, which is subleading to the ramp. 

In more detail, one can show that the average in \eqref{eq:Imjapprox} vanishes when $jm$ is not a perfect square. To see this, use Hecke relations to write \eqref{eq:Imjapprox} as follows:
\begin{equation}
I_{m|j}(y)
    = \sum_{d|\text{gcd}(j,m)} A_{\frac{jm}{d^2}} \sqrt{y} + o(\sqrt{y})\,,\qquad 
    A_n = \lim_{W\rightarrow\infty}\, \frac{1}{2W}\int_{-W}^W d\omega \, \frac{4\pi^2}{|\zeta(1+2i\omega)|^2} \frac{\sigma_{2i\omega}(n)}{n^{i\omega}} \,.
\end{equation} 
Note that $\frac{jm}{d^2}$ is a square if and only if $jm$ is a square. We thus need to show that $A_n=0$ except when $n$ is a square. Write the prime decomposition $n=p_1^{k_1}\cdots p_r^{k_r}$ and introduce prime phases $\theta_p= \omega \log(p)$. Then a prime factor appearing in the prime factorization of the divisor sum takes the form
\begin{equation}
\label{eq:divFac}
    \frac{\sigma_{2i\omega}(p^k)}{p^{i\omega k}} = \sum_{j=0}^k e^{i(2j-k)\theta_p} \,.
\end{equation}
Now observe that the inverse $\zeta$ factor in \eqref{eq:Imjapprox} has Euler factors that are invariant under $\theta_p \rightarrow \theta_p + \pi$, whereas the divisor factors in \eqref{eq:divFac} transform with a sign $(-1)^{k}$. The local phase average therefore vanishes whenever $k$ is odd. For $A_n$ to be non-zero, we thus need all $k_i$ to be even, i.e., $n$ is a square.

%%%%%%%%%%%%%%%%%%%%%%%%%%%%%%%%%%%%%%%%%%%%%%%%
\subsubsection{Triple cusp form overlap}
\label{app:cuspcuspcusp}
%%%%%%%%%%%%%%%%%%%%%%%%%%%%%%%%%%%%%%%%%%%%%%%%

\begin{figure}
\begin{center}
\includegraphics[width=.86\textwidth]{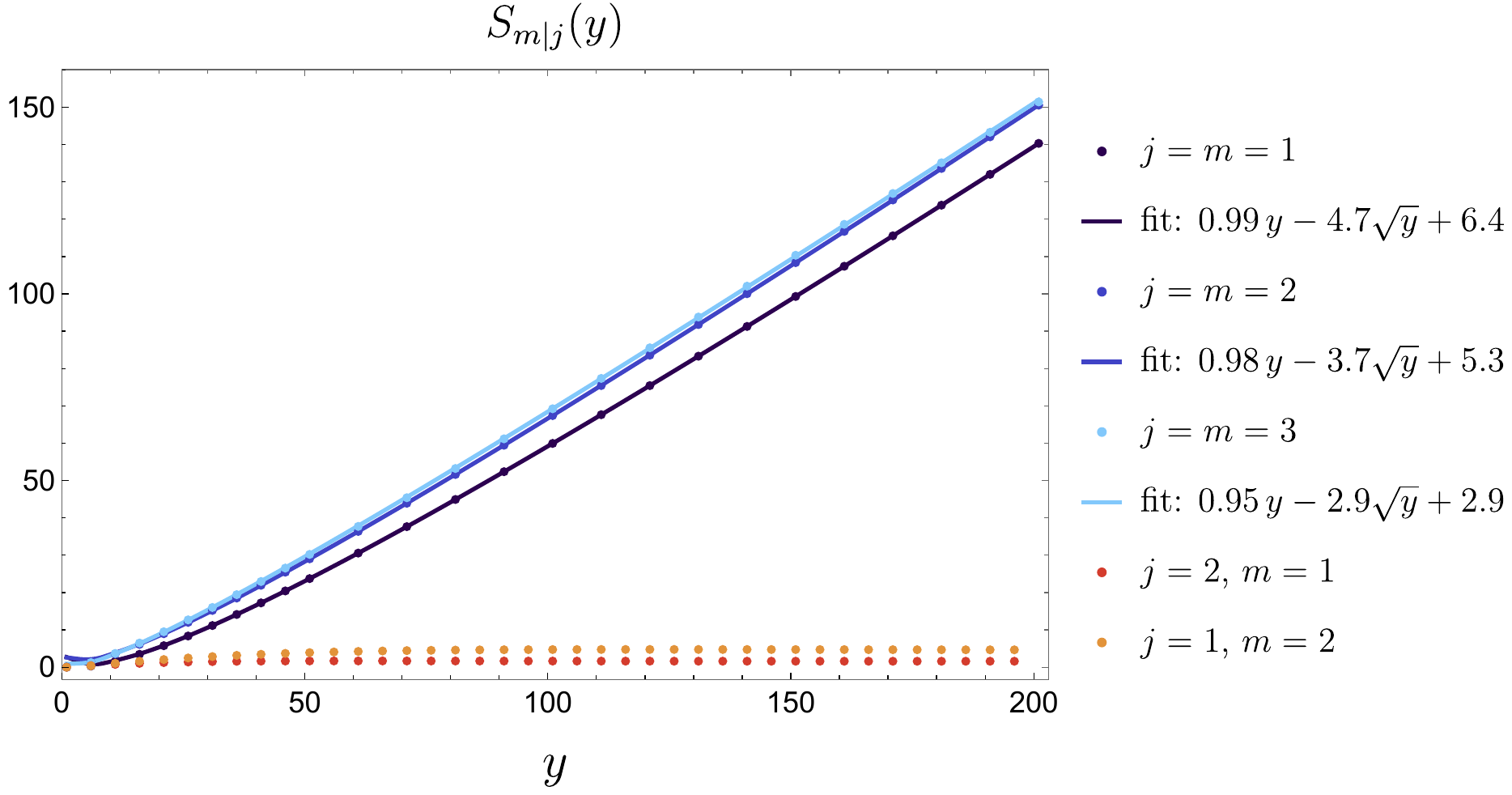}
\end{center}
\caption{Numerical evaluation of the auxiliary sum $S_{m|j}(y)$, defined in \eqref{eq:G13}, that determines the cusp form sector of the $(g,n)=(0,3)$ amplitude. For $j=m$, the sum asymptotes to a linear $y$-dependence. For $jm$ not a square it approaches a subleading constant.}
\label{fig:CJnumerics}
\end{figure}

Consider now the contribution of three Maass cusp forms, which takes the form
\begin{equation}
\label{eq:ZjjjmmmMaass}
{\cal Z}^{(0,3)}_{\text{RMT}_2|j_1,j_2,j_3}(\tau_1,\tau_2,\tau_3 | m,m,m)
\supset \frac{e^{-S_0(m)-2\pi (j_1y_1+j_2y_2+j_3y_3)} }{(2\pi  \gamma)^{3/2}} \,\frac{1}{4}\, S_{m|j_1}(y_1) \, S_{m|j_2}(y_2)  \, S_{m|j_3}(y_3) \,,
\end{equation}
where 
\begin{equation}
\label{eq:G13}
S_{m|j}(y) \equiv  \frac{e^{2\pi j y}}{\sqrt{m}}  \,\sum_{n} \frac{\pi}{\cosh(\pi \omega_n)}  \, {\tt b}^{(n)}_m\,\phi_{n|j}(y)\,.
\end{equation}
The function $S_{m|j_i}(y_i)$ computes the (rescaled) base spin-$m$ contribution to the spin-$j_i$ spectrum on the $i$-th torus. Numerically, we find good agreement with (see Fig.\ \ref{fig:CJnumerics})
\begin{equation}
\label{eq:SmjEval}
    S_{m|j}(y) = y \, \delta_{mj} + {\cal O}\big(\sqrt{y}\big) \qquad (y\rightarrow\infty) \,.
\end{equation}
This is consistent with universal RMT scaling (modulo the usual conventional overall factor of $\sqrt{y}$).

%

%%%%%%%%%%%%%%%%%%%%%%%%%%%%%%%%%%%%%%%%%%%%%%%%
%%%%%%%%%%%%%%%%%%%%%%%%%%%%%%%%%%%%%%%%%%%%%%%%
\section{Two-point function: from Poincar\'e to the harmonic decomposition}
\label{sec:CJnewderivation}
%%%%%%%%%%%%%%%%%%%%%%%%%%%%%%%%%%%%%%%%%%%%%%%%
%%%%%%%%%%%%%%%%%%%%%%%%%%%%%%%%%%%%%%%%%%%%%%%%

In this appendix we derive \eqref{eq:CJoverlaps}. 

From \eqref{eq:KLgeneral}, the (double) Kontorovich-Lebedev transform of the RMT input \eqref{eq:rampGeneral} is\footnote{The derivation given here can be applied to $m_i\rightarrow 0$ using the limits explained in Appendix C of \cite{DiUbaldo:2023qli}.}
\begin{align}
\label{2overlaps}
  &h_{\text{RMT}|m_1,m_2}^{(0,2)}(\sGen_1,\sGen_2) \notag\\
  &= d_{m_1}^2 \int_0^\infty \frac{dy_1}{y_1}\, K_{\sGen_1-\frac{1}{2}}(2\pi m_1 y_1)\,e^{-2\pi m_1 y_1}
     \int_0^\infty \frac{dy_2}{y_2}\, K_{\sGen_2-\frac{1}{2}}(2\pi m_2 y_2) \,e^{-2\pi m_2 y_2}\left(\frac{{\tt C}_\text{RMT}}{2\pi}\,\frac{\sqrt{y_1 y_2}}{y_1+y_2}\,\delta_{m_1m_2}\right)
\end{align}
Using the standard integral \cite{Gradshteyn}
\begin{equation}
  \int_0^\infty \frac{dy_1}{\sqrt{y_1}}\,\frac{e^{-2\pi m_1 y_1}}{y_1+y_2}\,K_{i\omega}(2\pi m_1 y_1)
  = \frac{\pi\, e^{2\pi m_1 y_2}}{\sqrt{y_2}\,\cosh(\pi\omega)}\,K_{i\omega}(2\pi m_1 y_2)\,,
\end{equation}
we obtain
\begin{equation}
 h_{\text{RMT}|m_1,m_2}^{(0,2)}(\sGen_1,\sGen_2)
  = \frac{d_{m_1}^2\,{\tt C}_\text{RMT}}{2\cosh(i\pi(\tfrac{1}{2}-\sGen_1))}\int_0^\infty \frac{dy}{y}\,
    K_{\sGen_1-\frac{1}{2}}(2\pi m_1 y)\,K_{\sGen_2-\frac{1}{2}}(2\pi m_2 y)\,\delta_{m_1m_2} \,.
\end{equation}
This integral is logarithmically divergent as $y\rightarrow 0$. It is natural to regularize this divergence as follows:
\begin{equation}
 h_{\text{RMT}|m_1,m_2}^{(0,2),\text{reg}}(\sGen_1,\sGen_2;s)
  \equiv \frac{d_{m_1}^2\,{\tt C}_\text{RMT}}{2\cosh(i\pi(\tfrac{1}{2}-\sGen_1))}   \int_0^\infty \frac{dy}{y^{1-s}}\,
    K_{\sGen_1-\frac{1}{2}}(2\pi m_1 y)\,K_{\sGen_2-\frac{1}{2}}(2\pi m_2 y)\,\delta_{m_1m_2} \,,
\end{equation}
where the regulator $s\rightarrow 0^+$ will be taken later.
This expression needs to be plugged into \eqref{eq:spec11} and summed over $m_1$, $m_2$. To decouple the spins, we write $\delta_{m_1m_2}$ in terms of orthogonal Fourier waves,
\be
\delta_{m_1m_2} = d_{m_1}\int_{-\frac{1}{2}}^{\frac{1}{2}}dx \, \cos(2\pi m_1 x) \cos(2\pi m_2 x)\,,
\ee
which yields
\begin{equation}
\begin{split}
&h_{\text{RMT}|m_1,m_2}^{(0,2),\text{reg}}(\sGen_1,\sGen_2;s)\\
  & = \frac{d_{m_1}^3\,{\tt C}_\text{RMT}}{2\cosh(i\pi(\tfrac{1}{2}-\sGen_1))}\int_{-\frac{1}{2}}^{\frac{1}{2}} dx \int_0^\infty \frac{dy}{y^{1-s}}\,
    K_{\sGen_1-\frac{1}{2}}(2\pi m_1 y)K_{\sGen_2-\frac{1}{2}}(2\pi m_2 y)\cos(2\pi m_1 x)\cos(2\pi m_2 x)
\end{split}
\end{equation}
Summing over spins with the (complex conjugate) Fourier coefficients as in \eqref{eq:spec11}, leads to a reconstruction of the two harmonic basis vectors (cf.\ \eqref{eq:PsiDef}):
\begin{equation}
\begin{split}
&\lim_{s\rightarrow 0^+}\sum_{m_1,m_2 \geq 0} d_{m_1}^{-1}\,  \overline{{\tt c}}_{m_1}^{(\sGen_1)}\, \overline{{\tt c}}_{m_2}^{(\sGen_2)}\,  h_{\text{RMT}|m_1,m_2}^{(0,2),\text{reg}}(\sGen_1,\sGen_2;s) \\
&\qquad = \frac{{\tt C}_\text{RMT}}{2\cosh(i\pi(\tfrac{1}{2}-\sGen_1))}\,\lim_{s\rightarrow 0^+} \int_{-\frac{1}{2}}^{\frac{1}{2}} dx \int_0^\infty \frac{dy}{y^{2}}\; \, y^{s} \, \overline{\Psi_{\sGen_1}(\tau) \, \Psi_{\sGen_2} (\tau)} \,,
\end{split}
\end{equation}
where the sum converges for $\text{Re}(s)>1$ and is continued to $s \to 0$.
Without the regulator (setting $s=0$) this is almost of the form of an inner product of the two basis functions and logarithmically divergent.  
To regulate this divergence, we keep track of the regulator $s$. Indeed, we can now turn the integral over the strip $\tau \in [-\frac{1}{2},\frac{1}{2}] \times [0,\infty)$ into an integral over the fundamental domain ${\cal F}=\mathbb{H}/\text{SL}(2,\mathbb{Z})$, using the inverse unfolding trick. The harmonic basis functions are already modular-invariant, while the factor $y^{s}$, upon summing over fundamental domains, turns into an Eisenstein series $E_{s}$:
\begin{equation}
\begin{split}
 \int_{-\frac{1}{2}}^{\frac{1}{2}} dx \int_0^\infty \frac{dy}{y^{2}}\; \, y^{s} \, \overline{\Psi_{\sGen_1}(\tau) \, \Psi_{\sGen_2} (\tau)} 
 =
 \int_{{\cal F}} \frac{dx dy}{y^2} \, E_s\, \overline{\Psi_{\sGen_1}(\tau) \, \Psi_{\sGen_2} (\tau)} \,.
\end{split}
\end{equation}
The integral can be interpreted as a triple Petersson overlap of $E_s$ with two basis functions $\Psi_{\sGen_1}\Psi_{\sGen_2}$:
\begin{equation}
\begin{split}
\lim_{s\rightarrow 0^+}\sum_{m_1,m_2\geq 0} d_{m_1}^{-1}\,\overline{{\tt c}}_{m_1}^{(\sGen_1)}\,  \overline{{\tt c}}_{m_2}^{(\sGen_2)}\,h_{\text{RMT}|m_1,m_2}^{(0,2),\text{reg}}(\sGen_1,\sGen_2;s)  
&= \frac{{\tt C}_\text{RMT}}{2\cosh(i\pi(\tfrac{1}{2}-\sGen_1))}\lim_{s\rightarrow 0^+} \left( E_s ,\, \Psi_{\sGen_1} \Psi_{\sGen_2} \right) \,.
\end{split}
\label{eq:EPsiPsi}
\end{equation}
Note that the analytic continuation of $E_s$ to $s=0$ regulates the expression for $E_s$ as an infinite sum, which would diverge for $s=0$. This mechanism is similar to zeta function regularization of divergent sums. After the analytic continuation, we can safely take the limit $s\rightarrow 0$, recalling $E_0(\tau) = 1$. Noting further $\overline{E_{s}(\tau)} = E_{1-s}(\tau)$ and $\overline{\phi_{n}(\tau)} = \phi_{n}(\tau)$, the inner product in \eqref{eq:EPsiPsi} becomes:
\begin{equation}
\begin{split}
    \lim_{s\rightarrow 0^+} \left( E_s ,\, E_{s_1} E_{s_2} \right) 
    &= \left( E_{1-s_1} ,\, E_{s_2} \right)
    = 2\pi\big(\delta(\omega_1+\omega_2) + \varphi(1-s_2) \delta(\omega_1-\omega_2) \big) \,, \\
    \lim_{s\rightarrow 0^+} \left( E_s ,\, \phi_{n_1}\phi_{n_2} \right) &= \left( \phi_{n_1},\,\phi_{n_2} \right) =  \delta_{n_1,n_2} \,,\\
    \lim_{s\rightarrow 0^+} \left( E_s ,\, E_{s_1}\,\phi_{n_2} \right) &= 0 \,.
\end{split}
\end{equation}
Here we used the orthogonality of Eisenstein series and cusp forms. Inside spectral integrals, the reflected term involving $\varphi(1-s_2)$ simply produces a factor 2. This yields \eqref{eq:CJoverlaps}, confirming the consistency of the Poincar\'e sum expression \eqref{eq:CJpoincare}.

%%%%%%%%%%%%%%%%%%%%%%%%%%%%%%%%%%%%%%%%%%%%%%%%
%%%%%%%%%%%%%%%%%%%%%%%%%%%%%%%%%%%%%%%%%%%%%%%%
\section{Calculations for the GOE Airy model}
\label{app:GOE}
%%%%%%%%%%%%%%%%%%%%%%%%%%%%%%%%%%%%%%%%%%%%%%%%
%%%%%%%%%%%%%%%%%%%%%%%%%%%%%%%%%%%%%%%%%%%%%%%%

In this appendix, we give more details about the $\uptau$-scaled Airy model in the GOE universality class.

%%%%%%%%%%%%%%%%%%%%%%%%%%%%%%%%%%%%%%%%%%%%%%%%
\subsection{Verification of the Mellin amplitude}
\label{app:mellinGOEAiry}
%%%%%%%%%%%%%%%%%%%%%%%%%%%%%%%%%%%%%%%%%%%%%%%%

The Eisenstein sector, as well as the RMT${}_2$ in the Poincar\'e decomposition, is fixed by the Mellin transform of the $\uptau$-scaled GOE Airy SFF \eqref{eq:GOEAiryMellin}, which we recall here:
\begin{equation}
\label{eq:GOEAiryMellinApp}
    \begin{split}
    f^{(2)}_{\uptau\text{-Airy-GOE}}(S_0;\bar{s}_1,\bar{s}_2) 
    &= e^{-2i\omega_+S_0} \, \Gamma\left(\frac{1}{2}-\frac{2i\omega_1}{3}+\frac{i\omega_2}{3}\right)\Gamma\left(\frac{1}{2}+\frac{i\omega_1}{3} - \frac{2i\omega_2}{3}\right) \mathfrak{f}^{(2)}_{\uptau\text{-Airy-GOE}} (\omega_+)  \\
   \mathfrak{f}^{(2)}_{\uptau\text{-Airy-GOE}} (\omega_+) &=
   2\times \mathfrak{f}^{(2)}_{\uptau\text{-Airy-GUE}} (\omega_+) -\frac{4^{i\omega_+}}{12\pi \omega_+^2} \bigg[ {}_2F_1(2i\omega_+,2i\omega_+,2i\omega_++1;-1)- \frac{2\pi\omega_+}{\sinh(2\pi \omega_+)} \bigg] \end{split}
\end{equation}
with $\bar s_i=\frac{1}{2}-i\omega_i$.
The $\omega_-$-dependence has the same universal structure
already present in the GUE amplitude. All of the GOE data is carried by the single-variable function $\mathfrak{f}^{(2)}_{\uptau\text{-Airy-GOE}} (\omega_+)$.

Rather than deriving \eqref{eq:GOEAiryMellinApp} via Mellin transform, it is easier to confirm its correctness by performing the inverse Mellin transform and reproducing the topological expansion of the RMT SFF. The inverse Mellin transform is
\begin{equation}
\begin{split}
\text{K}^{(2)}_{\uptau\text{-Airy-GOE}}(S_0;y_1,y_2) &= \int_{\mathbb{R}^2}\frac{d\omega_1}{2\pi}\frac{d\omega_2}{2\pi}\,y_1^{i\omega_1}y_2^{i\omega_2}\,f^{(2)}_{\uptau\text{-Airy-GOE}}(S_0;\bar s_1,\bar s_2)
\\
 &=\frac{9}{2}\int_{\mathbb{R}^2} \frac{d\omega_+}{2\pi} \frac{d\omega_-}{2\pi}\, y_1^{\frac{3i}{2}(\omega_++\omega_-)} y_2^{\frac{3i}{2}(\omega_+-\omega_-)}\, e^{-2i\omega_+ S_0} \, \\
 &\qquad\qquad\times \Gamma\left(\frac{1-i\omega_+-3i\omega_-}{2}\right)\Gamma\left(\frac{1-i\omega_++3i\omega_-}{2}\right)\mathfrak{f}^{(2)}_{\uptau\text{-Airy-GOE}} (\omega_+)\,.
\end{split}
\end{equation}
The $\omega_-$ integral is a Barnes
integral and is done in closed form,
\begin{equation}
\int\frac{d\omega_-}{2\pi}\left(\frac{y_1}{y_2}\right)^{\frac{3i}2\omega_-}
 \Gamma\left(\frac{1-i\omega_+-3i\omega_-}{2}\right)\Gamma\left(\frac{1-i\omega_++3i\omega_-}{2}\right)
 =\frac{2}{3}\,\Gamma(1-i\omega_+)\,\frac{(y_1y_2)^{\frac12-\frac{i\omega_+}2}}{(y_1+y_2)^{1-i\omega_+}} \,.
\label{eq:barnesw}
\end{equation}
This collapses the inversion to a single $\omega_+$ contour integral:
\begin{equation}
\text{K}^{(2)}_{\uptau\text{-Airy-GOE}}(S_0;y_1,y_2) =3\,\frac{\sqrt{y_1y_2}}{y_1+y_2}\int\frac{d\omega_+}{2\pi}\,
 \Gamma(1-i\omega_+)\,\left(e^{-2S_0}(y_1y_2)(y_1+y_2)\right)^{i\omega_+}\,\mathfrak{f}^{(2)}_{\uptau\text{-Airy-GOE}} (\omega_+)\,.
\label{eq:wp}
\end{equation}
Closing in the lower half plane (as required by the factor $e^{-2i\omega_+ S_0}$) and summing residues reproduces the SFF \eqref{eq:GOEairyResult} genus by genus:
\begin{itemize}
\item \textbf{Universal $\Gamma$-function poles:} The factor $\Gamma(1-i\omega_+)$ has the following
\begin{equation}
    \text{universal } \Gamma \text{ poles:} \qquad \omega_+ = -ig \,,\qquad g=1,2,3,\ldots 
\end{equation}
Except for the double poles discussed below, these are simple poles and give rise to two times the GUE result plus the hypergeometric function in the $A_g$ coefficient of \eqref{eq:AresultAiry}.\footnote{A conventional relative factor of $e^{S_0}$ is absorbed in the definition $\text{K}^{(2)}_{\uptau\text{-Airy-GOE}} = e^{S_0} {\cal K}^{(2)}_{\uptau\text{-Airy-GOE}}$; see also the discussion around \eqref{eq:tauAiryGUE2}.}
\item \textbf{Crosscaps:} The function $\mathfrak{f}^{(2)}_{\uptau\text{-Airy-GOE}} (\omega_+)$ has simple poles due to the term $1/\sinh(2\pi\omega_+)$, located at the half-integer
\begin{equation}
    \text{crosscap poles:} \qquad \omega_+=-i\tilde g \,,\qquad \tilde{g} = \tfrac{1}{2},\,\tfrac{3}{2},\,\tfrac{5}{2},\ldots\,.
\end{equation}
These give a contribution proportional to $(y_1+y_2)^{\tilde g-1}(y_1y_2)^{\tilde g+\frac12}$, i.e., the even
powers of $\uptau$ associated with the ``crosscap'' coefficients $C_{\tilde g}$ of \eqref{eq:GOEairyResult}.
\item \textbf{Double poles:} At $\omega_+=-ig$ the pole of $1/\sinh(2\pi\omega_+)$ collides with the universal $\Gamma$-function pole to produce a double pole.
Its residue thus involves a logarithmic derivative, which gives rise to the $\log(\uptau)$ contribution, i.e., the $B_g$ coefficient in \eqref{eq:GOEairyResult}. It also produces non-logarithmic terms that contribute to $A_g$.
\end{itemize}

%%%%%%%%%%%%%%%%%%%%%%%%%%%%%%%%%%%%%%%%%%%%%%%%
\subsection{The Kontorovich-Lebedev transform and the cusp-form overlaps}
\label{app:KL}
%%%%%%%%%%%%%%%%%%%%%%%%%%%%%%%%%%%%%%%%%%%%%%%%

While \eqref{eq:GOEAiryMellinApp} fixes the Eisenstein sector, the cusp-form overlaps require the KL transform of the RMT input. The moments in the harmonic basis follow from \eqref{eq:spec11Rep}:
\begin{equation}
\left({\cal Z}^{(2)}_{\uptau\text{-Airy-GOE}},\,\Psi_{\sGen_1}\Psi_{\sGen_2}\right)=\sum_{m\ge1} d_m^{-1}\,\bar {\tt c}^{(\sGen_1)}_m \, \bar {\tt c}^{(\sGen_2)}_m\,h^{(2)}_{\uptau\text{-Airy-GOE}|m,m}(\sGen_1,\sGen_2)\,,
\end{equation}
with the double KL transform
\begin{equation}
\label{eq:hGOEairyApp}
\begin{split}
   h_{\uptau\text{-Airy-GOE}|m_1,m_2}^{(2)}(\sGen_1,\sGen_2)
  &=d_{m_1}^2 \delta_{m_1m_2}\int_0^\infty \frac{dy_1}{y_1}\, K_{\sGen_1-\frac{1}{2}}(2\pi m_1 y_1) \, e^{-2\pi m_1 y_1}\\
  &\quad\;\;\times
     \int_0^\infty \frac{dy_2}{y_2}\, K_{\sGen_2-\frac{1}{2}}(2\pi m_2 y_2) \,  e^{-2\pi m_2 y_2}\, \text{K}^{(2)}_{\uptau\text{-Airy-GOE}}(S_0(m_1);y_1,y_2) \,.
\end{split}
\end{equation}
To evaluate this transform, we recall the $\uptau$-scaling dictionary $\beta = \frac{y_1+y_2}{2}$ and $\uptau = e^{-S_0} \sqrt{y_1y_2}$.

The RMT amplitude contains three qualitatively different terms in these variables:
\begin{equation}
\label{eq:KairyToKL}
\begin{split}
    &e^{-S_0} \, \text{K}^{(2)}_{\uptau\text{-Airy-GOE}}(S_0;y_1,y_2) \\
    &\quad = [g=0] + \sum_{g\geq 1} \left[\tilde{a}_{g} + \tilde{b}_g \, \log((y_1+y_2)y_1y_2 \,e^{-2S_0})\right](y_1+y_2)^{g-1} (y_1y_2)^{g+\frac{1}{2}}\, e^{-(2g +1)S_0}
    \\
    &\qquad + \sum_{\tilde{g}=\frac{1}{2},\frac{3}{2},\ldots} \tilde{c}_{\tilde g} \, (y_1+y_2)^{\tilde g-1} \, \left(y_1y_2\right)^{\tilde g+\frac{1}{2}}\, e^{-(2\tilde g+1)S_0}
\end{split}
\end{equation}
where the genus zero term is universal as discussed earlier, and we defined the following numerical ($y_{1,2}$-independent) coefficients:
\begin{equation}
    \begin{split}
    \tilde{a}_g &= \frac{(-1)^{g}}{\pi (2g+1) g!} + \frac{(-4)^{g-1}}{\pi  g!} \left[\frac{1}{g}\;{}_2F_1(2g,2g,2g+1,-1) +  \psi(g+1)  - \log\left(4\right)  \right] 
    \,,\\
    \tilde{b}_g &= -\frac{(-4)^{g-1}}{\pi g!}\,,\qquad
    \tilde{c}_{\tilde g} = -\frac{4^{\tilde{g}-1}}{\pi \tilde{g}} \Gamma(1-\tilde{g}) \,.
    \end{split}
\label{eq:abctilde}
\end{equation}
Our task is thus to compute the KL transform of the three types of structures appearing in \eqref{eq:KairyToKL}.

We note that all terms descend from a common Mellin-Barnes integral:
\begin{equation} 
\label{eq:hGOEairyKLfinal} 
\begin{split} 
& h^{(2)}_{\uptau\text{-Airy-GOE}|m_1,m_2} (\sGen_1,\sGen_2) \\
&\quad = [g=0]+ \sum_{g\ge 1}  \left[ \tilde a_g\, \mathfrak H_{\,g-1,\;g+\frac12}\,e^{-2gS_0} + \tilde b_g \, \frac{\partial}{\partial g} \left(\mathfrak H_{g-1,g+\frac12}\,e^{-2gS_0}\right) \right] +  \sum_{\tilde g=\frac12,\frac32,\ldots} \tilde c_{\tilde g} \, \mathfrak H_{\, \tilde g-1,\; \tilde g+\frac12}\, e^{-2\tilde g S_0}\,,
\end{split} 
\end{equation}
where 
\begin{equation} 
\label{eq:masterKLexplicit} 
\begin{split} 
&\mathfrak H_{\lambda,\mu} (\sGen_1,\sGen_2|m_1,m_2) \\ 
&\quad \equiv d_{m_1}^2\delta_{m_1m_2}\int_0^\infty\frac{dy_1}{y_1} K_{\sGen_1-\frac{1}{2}}(2\pi m_1y_1)\,e^{-2\pi m_1y_1} \int_0^\infty\frac{dy_2}{y_2} K_{\sGen_2-\frac{1}{2}}(2\pi m_2y_2)\,e^{-2\pi m_2y_2} \,(y_1+y_2)^\lambda(y_1y_2)^\mu \\
&\quad= \frac{d_{m_1}^2\delta_{m_1m_2}}{\Gamma(-\lambda)} \int_{\cal C}\frac{ds}{2\pi i}\, \Gamma(-s)\Gamma(s-\lambda)\, \mathcal I_{\mu+s}(\sGen_1,m_1)\, \mathcal I_{\mu+\lambda-s}(\sGen_2,m_2)
\\
&\qquad {\cal I}_{a}(\sGen,m) \equiv \frac{\sqrt{\pi}\, \Gamma\left(a+\sGen-\frac12\right) \Gamma\left(a-\sGen+\frac12\right)} {(4\pi m)^{a}\,\Gamma\left(a+\frac12\right)}
\end{split} 
\end{equation}
Explicit evaluation yields the hypergeometric representation 
\begin{equation}
\begin{split}
\mathfrak H_{\lambda,\mu} &= \frac{\pi d_{m_1}^2\delta_{m_1m_2}}{(4\pi m_1)^\mu(4\pi m_2)^{\mu+\lambda}}\,\widetilde{\mathfrak H}_{\lambda,\mu}\,,\\
\widetilde{\mathfrak H}_{\lambda,\mu}
&=\frac{\Gamma(\mu+\sGen_1-\frac12)\Gamma(\mu-\sGen_1+\frac12)\Gamma(\mu+\lambda+\sGen_2-\frac12)\Gamma(\mu+\lambda-\sGen_2+\frac12)}{\Gamma(\mu+\frac12)\Gamma(\mu+\lambda+\frac12)}\\
&\qquad\times {}_4F_3\left(
\begin{matrix}
-\lambda,\,\mu+\sGen_1-\frac12,\,\mu-\sGen_1+\frac12,\,\frac12-\mu-\lambda
\\
\mu+\frac12,\,\frac32-\mu-\lambda-\sGen_2,\,\frac12-\mu-\lambda+\sGen_2
\end{matrix}
\,;\,\frac{m_2}{m_1}\right)
\\[4pt]
&\quad+\left(\frac{m_2}{m_1}\right)^{\mu+\lambda+\sGen_2-\frac12} \frac{\Gamma(\frac12-\mu-\lambda-\sGen_2)\Gamma(\mu+\sGen_2-\frac12)\Gamma(1-2\sGen_2)}{\Gamma(-\lambda)\Gamma(1-\sGen_2)}
\\
&\qquad\times\frac{\Gamma(2\mu+\lambda+\sGen_1+\sGen_2-1)\Gamma(2\mu+\lambda-\sGen_1+\sGen_2)}{\Gamma(2\mu+\lambda+\sGen_2)}
\\
&\qquad\times {}_4F_3\left(
\begin{matrix}
\mu+\sGen_2-\frac12,\,2\mu+\lambda+\sGen_1+\sGen_2-1,\,2\mu+\lambda-\sGen_1+\sGen_2,\,\sGen_2
\\
2\mu+\lambda+\sGen_2,\,\mu+\lambda+\sGen_2+\frac12,\,2\sGen_2
\end{matrix}
\,;\,\frac{m_2}{m_1}\right)
\\[4pt]
&\quad+\left(\frac{m_2}{m_1}\right)^{\mu+\lambda-\sGen_2+\frac12}\frac{\Gamma(\sGen_2-\mu-\lambda-\frac12)\Gamma(\mu-\sGen_2+\frac12)\Gamma(2\sGen_2-1)}{\Gamma(-\lambda)\Gamma(\sGen_2)}
\\
&\qquad\times\frac{\Gamma(2\mu+\lambda+\sGen_1-\sGen_2)\Gamma(2\mu+\lambda-\sGen_1-\sGen_2+1)}{\Gamma(2\mu+\lambda-\sGen_2+1)}
\\
&\qquad\times {}_4F_3\left(
\begin{matrix}
\mu-\sGen_2+\frac12,\,2\mu+\lambda+\sGen_1-\sGen_2,\,2\mu+\lambda-\sGen_1-\sGen_2+1,\,1-\sGen_2
\\
2\mu+\lambda-\sGen_2+1,\,\mu+\lambda-\sGen_2+\frac32,\,2-2\sGen_2
\end{matrix}
\,;\,\frac{m_2}{m_1}\right)\,.
\end{split}
\end{equation}
The expression is understood by analytic continuation in its parameters.
On the diagonal $m_1=m_2$, the ratio $m_2/m_1$ is treated as an auxiliary
continuous variable, and the limit $m_2/m_1\to1^{-}$ is taken after
combining all three terms in $\widetilde{\mathfrak H}_{\lambda,\mu}$.
Coincident-pole cases, including $\sGen_2=\frac12$, are likewise
evaluated by taking the limit of the complete sum.

Note that only the first term in $\widetilde{\mathfrak H}_{\lambda,\mu}$ contributes for $\lambda = g-1 \in \mathbb{Z}$ (integer genus), due to the factor $1/\Gamma(-\lambda)$ in the remaining terms. The second and third hypergeometric functions only contribute for half-integer genus.

We now use base spin diagonality $m_1=m_2\equiv m$ to simplify the three terms in \eqref{eq:hGOEairyKLfinal}. All $m$-dependence is then encoded in the prefactor of $\mathfrak{H}_{\lambda,\mu}$, with $\widetilde{\mathfrak{H}}_{\lambda,\mu}$ being $m$-independent. We find the following structures:
\begin{itemize}
    \item First term ($\tilde{a}_g$): for integer genus the hypergeometric sum terminates and reproduces the structure we discussed in the GUE Airy model in Section \ref{sec:GUEairySpectral}:
\begin{equation}
\label{eq:Hsimple}
\begin{split}
    &\widetilde{\mathfrak{H}}_{g-1,g+\frac{1}{2}}(\sGen_1,\sGen_2)\\
    &\quad =  \sum_{k=0}^{g-1} \binom{g-1}{k} \frac{\Gamma(g+k+\sGen_1)\Gamma(g+k+1-\sGen_1)\Gamma(2g-k-\sGen_2)\Gamma(2g-k-1+\sGen_2)}{\Gamma(g+k+1)\Gamma(2g-k)}
\end{split}
\end{equation}
\item  Second term ($\tilde{b}_g$): for the coefficient $\tilde{b}_g$ in \eqref{eq:hGOEairyKLfinal}, we need to take the $g$-derivative before evaluating for integer genus. We observe 
\begin{equation}
\begin{split}
    &\frac{\partial}{\partial g} \left(\mathfrak{H}_{g-1,g+\frac{1}{2}}(\sGen_1,\sGen_2)\, e^{-2gS_0} \right)
    \\
    &\quad = \frac{\pi d_m^2 \, e^{-2gS_0}}{(4\pi m)^{3g}} \left[-(3 \log(4\pi m)+2S_0) \, \widetilde{\mathfrak{H}}_{g-1,g+\frac{1}{2}}(\sGen_1,\sGen_2|m,m) + \widetilde{\mathfrak{H}}_{g-1,g+\frac{1}{2}}^\text{reg.}(\sGen_1,\sGen_2)\right] \,.
\end{split}
\end{equation}
The first term takes the same form as \eqref{eq:Hsimple}, leading to a combined contribution $(\tilde{a}_g - (3 \log(4\pi m)+2S_0) \, \tilde{b}_g) \mathfrak{H}_{g-1,g+\frac{1}{2}}$. The `regular' term involves a complicated genus derivative, which is however entirely independent of $m$:
\begin{equation}
    \begin{split}
        \widetilde{\mathfrak{H}}_{g-1,g+\frac{1}{2}}^\text{reg.}
        &=\left( \frac{\partial}{\partial \mu}+\frac{\partial}{\partial \lambda}\right) \left[ \widetilde{\mathfrak{H}}_{\lambda,\mu} \right]_{\lambda=g-1,\mu=g+\frac{1}{2}}
    \end{split}
\end{equation}
We refrain from simplifying this expression further.
\item  Third term ($\tilde{c}_{\tilde g}$): finally, for crosscap terms the hypergeometric sum also simplifies. 
For half-integer genus $\tilde g \equiv g+\frac12$, with $g=0,1,2,\ldots$,
write $\sGen_i=\frac12+i\omega_i$. The complete result is then
\begin{equation}
\begin{split}
&\widetilde{\mathfrak H}_{\tilde g-1,\tilde g+\frac12}
  (\sGen_1,\sGen_2)=
\frac{
\cosh^2(\pi\omega_2)\,
T_{\tilde g}(\sGen_1,\sGen_2)
+
\cosh^2(\pi\omega_1)\,
T_{\tilde g}(\sGen_2,\sGen_1)
- P_{\tilde g}(\sGen_1,\sGen_2)}
{
\cosh\bigl(\pi(\omega_1+\omega_2)\bigr)
\cosh\bigl(\pi(\omega_1-\omega_2)\bigr)
}\,,
\end{split}
\end{equation}
where
\begin{equation}
\begin{split}
T_{\tilde g}(\sGen_1,\sGen_2)
& =
\Gamma\left(\tilde g\right)\sum_{k=0}^{2\tilde g-1}
\frac{
\Gamma(\tilde g+k+\sGen_1)
\Gamma(\tilde g+k+\bar \sGen_1)\Gamma(2\tilde g-k-1+\sGen_2)
\Gamma(2\tilde g-k-1+\bar\sGen_2)}
{
\Gamma(k+1)\Gamma(\tilde g-k)\Gamma(\tilde g+k+1)\Gamma(2\tilde g-k)
}\,,
\\
P_{\tilde g}(\sGen_1,\sGen_2)
& = \Gamma(\tilde g)^3
\sum_{r=0}^{2\tilde g -1}\sum_{s=0}^{2\tilde g-1-r}
\frac{
\left(\tilde g\right)_{2\tilde g-1-r-s}}
{(2\tilde g -1-r-s)!}
\frac{
\left(\sGen_1\right)_r
\left(\bar \sGen_1\right)_r
\left(\sGen_2\right)_s
\left(\bar \sGen_2\right)_s}
{
r!\,s!\,
\left(1-\tilde g\right)_r
\left(1-\tilde g\right)_s
}\,.
\end{split}
\end{equation}
The latter is a symmetric polynomial in $\omega_1^2$ and $\omega_2^2$, for example,
\begin{equation}
P_{\frac12}=\pi^{\frac{3}{2}}\,,
\qquad
P_{\frac32}
=
-\frac{\pi^{\frac{3}{2}}}{64} \left[
16(\omega_1^2-\omega_2^2)^2
+56(\omega_1^2+\omega_2^2)+13\right]\,.
\end{equation}
This determines the ``crosscap'' contribution in \eqref{eq:hGOEairyKLfinal}.
\end{itemize}

%%%%%%%%%%%%%%%%%%%%%%%%%%%%%%%%%%%%%%%%%%%%%%%%

\newpage
\bibliographystyle{utphys2}
{\small \bibliography{Biblio}{}}
 
\end{document}